\documentclass[onecolumn,prc,floatfix,showpacs,preprintnumbers,nofootinbib,%
superscriptaddress]{revtex4}
\usepackage{mathrsfs}
\usepackage{amssymb}	
\usepackage{amsmath}
\usepackage{graphicx}
\usepackage{subdepth}
\usepackage{xcolor}
\usepackage[normalem]{ulem} 
\definecolor{lcolor}{rgb}{0.5,0,0}
\definecolor{citcolor}{rgb}{0,0.3,0.0}

\usepackage[latin1]{inputenc}

\usepackage[breaklinks,colorlinks,urlcolor=blue,citecolor=citcolor,linkcolor=lcolor]{hyperref}

\usepackage{tikz}

\usepackage{mciteplus}

\newcommand{\Pt}{{\mathbf{P}}}

\newcommand{\ktp}{{\mathbf{k}'}}
\newcommand{\ktpp}{{\mathbf{k}''}}
\newcommand{\ptp}{{\mathbf{p}'}}

\newcommand{\xtp}{{\mathbf{x}'}}

\newcommand{\ztp}{{\mathbf{z}'}}

\newcommand{\rt}{{\mathbf{r}}}
\newcommand{\xt}{{\mathbf{x}}}
\newcommand{\bt}{{\mathbf{b}}}
\newcommand{\yt}{{\mathbf{y}}}
\newcommand{\zt}{{\mathbf{z}}}
\newcommand{\pt}{{\mathbf{p}}}
\newcommand{\qt}{{\mathbf{q}}}
\newcommand{\kt}{{\mathbf{k}}}
\newcommand{\nt}{{\mathbf{n}}}
\newcommand{\mt}{{\mathbf{m}}}
\newcommand{\Rt}{{\mathbf{R}}}
\newcommand{\ot}{\mathbf{0}}

\newcommand{\hht}{\mathbf{h}}
\newcommand{\Kt}{\mathbf{K}}

\newcommand{\lt}{\mathbf{l}}

\newcommand{\epst}{\boldsymbol{\varepsilon}}

\newcommand{\kvec}{{\vec{k}}}
\newcommand{\pvec}{{\vec{p}}}
\newcommand{\ppvec}{{{\vec{p}}{\, '}}}
\newcommand{\pppvec}{{{\vec{p}}{\, ''}}}
\newcommand{\kpvec}{{{\vec{k}}{\, '}}}
\newcommand{\kppvec}{{{\vec{k}}{\, ''}}}
\newcommand{\qvec}{{\vec{q}}}
\newcommand{\qpvec}{{{\vec{q}}{\, '}}}

\newcommand{\lo}{{\textnormal{LO}}}
\newcommand{\nlo}{{\textnormal{NLO}}}

\newcommand{\epsl}{{\varepsilon\!\!\!/}}

\newcommand{\psl}{{p\!\!\!/}}
\newcommand{\ppsl}{{p\!\!\!/}'}

\newcommand{\dk}{{\widetilde{\mathrm{d} k}}}
\newcommand{\dkp}{{\widetilde{\mathrm{d} k}{'}}}
\newcommand{\dkpp}{{\widetilde{\mathrm{d} k}{''}}}
\newcommand{\ddp}{{\widetilde{\mathrm{d} p}}}
\newcommand{\dpp}{{\widetilde{\mathrm{d} p}{'}}}
\newcommand{\dppp}{{\widetilde{\mathrm{d} p}{''}}}

\newcommand{\qemit}{V}
\newcommand{\qemitas}{\mathcal{V}}
\newcommand{\qbemit}{\overline{V}}
\newcommand{\qbemitas}{\overline{\mathcal{V}}}
\newcommand{\paircr}{A}
\newcommand{\paircras}{\mathcal{A}}
\newcommand{\annih}{\overline{A}}
\newcommand{\annihas}{\overline{\mathcal{A}}}

\newcommand{\ud}{\, \mathrm{d}}

\newcommand{\R}{\mathrm{Re}}
\newcommand{\nc}{{N_\mathrm{c}}}

\newcommand{\cf}{C_\mathrm{F}}

\newcommand{\nr}[1]{(\ref{#1})}

\newcommand{\as}{\alpha_{\mathrm{s}}}

\newcommand{\fig}{Fig.~}
\newcommand{\figs}{Figs.~}
\newcommand{\eq}{Eq.~}

\newcommand{\eqs}{Eqs.~}

\newcounter{diag}
\newcommand{\namediag}[1]{\refstepcounter{diag} \thediag \label{#1}}
\renewcommand{\thediag}{(\alph{diag})}

\newcommand{\epsmsbar}{\varepsilon_{\overline{\textnormal{MS}}}}

\begin{document}

\author{H. H\"anninen}
\affiliation{
Department of Physics, %
 P.O. Box 35, 40014 University of Jyv\"askyl\"a, Finland}

 \author{T. Lappi}
\affiliation{
Department of Physics, %
 P.O. Box 35, 40014 University of Jyv\"askyl\"a, Finland}
\affiliation{
Helsinki Institute of Physics, P.O. Box 64, 00014 University of Helsinki,
Finland}

\author{R. Paatelainen}
\affiliation{
Department of Physics, %
 P.O. Box 64, 00014 University of Helsinki, Finland}
\affiliation{
Helsinki Institute of Physics, P.O. Box 64, 00014 University of Helsinki,
Finland}
\affiliation{
Department of Physics, %
 P.O. Box 35, 40014 University of Jyv\"askyl\"a, Finland}

\title{
One-loop corrections to light cone wave functions: the dipole picture DIS cross section
}

\preprint{HIP-2017-32/TH}

\begin{abstract}
We develop methods needed to perform loop calculations in light cone perturbation theory using a helicity basis, refining the method introduced in our earlier work. In particular this includes implementing a consistent way to contract the four-dimensional tensor structures from the helicity vectors with $d$-dimensional tensors arising from loop integrals, in a way that can be fully automatized. We demonstrate this explicitly by calculating the one-loop correction to the virtual photon to quark-antiquark dipole light cone wave function. This allows us to calculate the deep inelastic scattering cross section in the dipole formalism to next-to-leading order accuracy. Our results, obtained using the  four dimensional helicity scheme, agree with the recent calculation by Beuf using conventional dimensional regularization, confirming the regularization scheme independence of this cross section.
\end{abstract}

\maketitle

\section{Introduction}

Light cone perturbation theory (LCPT), the Hamiltonian formulation of field theory on the light front~\cite{Kogut:1969xa,Bjorken:1970ah,Lepage:1980fj,Brodsky:1997de} is a widely used calculational tool in particle and hadronic physics. Its added calculational complexity compared to covariant perturbation theory is balanced by  several advantages in the description of bound states or other multiparton systems. The light cone wave functions (LCWF's) and operators  have a simpler behavior under transverse Lorentz boosts than covariant ones. The perturbative expansion is organized in terms of a Fock state expansions involving only physical degrees of freedom with definite helicities. This gives a natural physical interpretation for the factorization of scattering processes into the properties of the incoming and outgoing hadronic states on one hand, and the short distance partonic scatterings between elementary constituents on the other.

Modern hadronic and nuclear scattering experiments probe QCD with increasing accuracy in the high energy or small-$x$ regime. Here the large  available phase space for gluon radiation enables the generation of a dense system of gluons with  nonperturbatively large gluon fields. One the other hand, balancing the complication arising from the nonlinear dynamics of the gluons, the high collision energy simplifies the treatment of the scattering by allowing an eikonal approximation for the interactions of individual partons with the color field. Typically this situation is described using the effective theory of QCD known as the Color Glass Condensate (CGC)~\cite{Gelis:2010nm}. In this picture, the scattering of a dilute probe off the dense color field is factorized into the partonic structure of the ``simple'' probe (virtual photon, or an individual quark or gluon in the case of forward rapidities in proton-nucleus collisions), and the eikonal scattering of the partons of the probe with the target color field. This allows for a treatment that includes nonlinear interactions in the dense target color field to all orders, while the simple probe can be treated exactly. This picture is advantageous in particular for understanding exclusive processes. Light cone perturbation theory is the method of choice for understanding the structure of the probe.

In order to develop a more quantitative description of several scattering processes in the high energy limit, CGC calculations have recently been advancing to next-to-leading order (NLO) accuracy for several different processes. The NLO corrections to the small-$x$ evolution equations (in particular the Balitsky-Kovchegov (BK) equation~\cite{Balitsky:1995ub,Kovchegov:1999ua,Kovchegov:1999yj}) have been derived and the required resummations of collinear logarithms studied in several papers~\cite{Balitsky:2008zza,Balitsky:2013fea,Kovner:2013ona,Balitsky:2014mca,Beuf:2014uia,Lappi:2015fma,Iancu:2015vea,Iancu:2015joa,Albacete:2015xza,Lappi:2016fmu,Lublinsky:2016meo}. 
There have been several caluclations of single~\cite{Altinoluk:2011qy,JalilianMarian:2011dt,Chirilli:2012jd,Stasto:2013cha,Kang:2014lha,Altinoluk:2014eka} and double~\cite{Ayala:2016lhd} inclusive parton production at forward rapidity in high energy proton-nucleus collisions. In the context of deep inelastic scattering, both inclusive~\cite{Beuf:2011xd,Balitsky:2012bs,Beuf:2016wdz,Beuf:2017bpd,Ducloue:2017ftk} and exclusive~\cite{Boussarie:2014lxa,Boussarie:2016bkq,Boussarie:2016ogo} processes have been studied at the NLO order. 

Our present paper is a follow-up of our recent work~\cite{Lappi:2016oup}, where we introduced the idea of performing loop calculations in LCPT using a helicity basis for the elementary vertices.
In this paper we will present a better formulation of the calculational scheme introduced in~\cite{Lappi:2016oup}, correcting a partially incorrect formulation used in that paper. As a demonstration, we will calculate the one-loop correction to the virtual photon to quark-antiquark dipole light cone wave function. We will then use this to derive the NLO cross section for inclusive DIS in the dipole factorization picture. We perform the calculation using the four-dimensional helicity (FDH) scheme, where polarization sums are calculated in four dimensions, and ultraviolet divergences are regularized by performing momentum integrals in $d$ dimensions. Our results recover the ones obtained in~\cite{Beuf:2016wdz,Beuf:2017bpd} after a lengthy manual calculation, in what we would argue to be a more systematical and economical way. Also, although intermediate results are different in the FDH scheme used here and the conventional dimensional regularization (CDR) used in~\cite{Beuf:2016wdz,Beuf:2017bpd}, we see that these scheme dependent terms cancel in the final result. As a separate small difference to the calculation in~\cite{Beuf:2016wdz,Beuf:2017bpd}, we implement the cancellation between UV divergences in the real and virtual corrections to the cross section by adding and subtracting a slightly different subtraction term, leading to a numerically smoother expression for the cross section. We verify both analytically and numerically that our results are equivalent to those in~\cite{Beuf:2016wdz,Beuf:2017bpd}.

The rest of the paper is structured as follows. We will first discuss the technical aspects of the calculation in Sec.~\ref{sec:nlolcpt}, concentrating on the differences compared to our earlier work \cite{Lappi:2016oup}. We then recall how one calculates cross sections by combining eikonal interactions with a color field target with light cone wave functions in Sec.~\ref{sec:sigma}. After briefly rederiving the leading order virtual photon wave functions in Sec.~\ref{sec:lo} we calculate the corresponding one-loop corrections in Sec.~\ref{sec:loop}. After calculating also the real corrections, i.e. the wave functions for gluon emission from the dipole, in Sec.~\ref{sec:real} we combine our results into the NLO DIS cross sextion in Sec.~\ref{sec:nlodis}, before concluding in Sec.~\ref{sec:conc}. Technical details of the calculation have been spelled out in the Appendices.

\section{Higher order  LCPT computations}
\label{sec:nlolcpt}

The purpose of this paper is to develop and demonstrate techniques for the perturbative calculation of \emph{light cone wave functions}. These are formed from interaction vertices, where spatial momentum ($\pvec = (p^+,\pt)$) is conserved but light cone energy $p^-$ is not, and energy denominators which depend on the energy differences between intermediate states. In loop diagrams, the spatial momentum circulating in the loop must be integrated over. In these momentum integrals we encounter $p^+\to 0$ divergences, $\pt\to0$ divergences and 
ultraviolet (UV) divergences ($\pt\to\infty$). Combinations of the first two divergences encode soft, collinear and spurious gauge divergences. All of these divergences have very different physical interpretations, and it makes therefore sense to reguralize them all by different means. In particular we will regularize, if needed, the soft divergence with a cutoff, and regulate the UV divergences by integrating over the transverse momenta in $2-2\varepsilon$ dimensions. The basic normalizations, notations etc are explained in much more detail in Ref.~\cite{Lappi:2016oup}, and we will here concentrate only on the differences with respect to the formulation used there. We first discuss the different flavors of dimensional  regularization in Sec.~\ref{sec:dimreg}, and the required modifications to the formulation of the elementary vertices compared to the explicitly 2-dimensional one used in~\cite{Lappi:2016oup} in Sec.~\ref{sec:vertices}. We will then, as an explicit demostration, calculate two helicity sums appearing in the calculation of the NLO DIS impact factor in Sec.~\ref{sec:numerators}, and briefly write down the instantaneous vertices needed in our calculation in 
Sec.~\ref{sec:otherdiags}.

\subsection{Dimensional regularization schemes in gauge field theories}
\label{sec:dimreg}

In the evaluation of loop and phase space momentum integrals one encounters divergences which have to be properly regularized. In gauge field theories a satisfactory regulator has to respect gauge invariance and unitarity which requires that one treats the momenta and helicities equally. For practical computations, the only choice is a form of dimensional regularization.

For the discussion of different versions of dimensional regularization schemes it is useful to define unobserved and observed particles. Unobserved particles are either virtual ones which circulate in internal loops or particles which are external but soft or collinear with other external particles. All the rest are observed particles. The common feature in all dimensional regularization schemes is the continuation of the momenta of the unobserved particles into $d \neq 4$.  Once this is done, there is still some freedom regarding the dimensionality of the momenta of the observed particles as well as the treatment of polarization vectors (or helicities) of the unobserved and observed particles. Thus, one can define a set of different versions of dimensional regularization schemes:

\begin{itemize}
\item The conventional dimensional regularization (CDR) scheme~\cite{Collins:1984xc}, in which both observed and unobserved polarization vectors and momenta are continued to $d$ dimension (i.e. all gluons have $d-2$ helicity states).
\item The 't Hooft-Veltman (HV) scheme~\cite{tHooft:1972tcz}, in which the unobserved particle momenta and polarization vectors are continued to $d$ dimensions (i.e. unobserved gluons have $d-2$ helicity states), but the momenta and polarization vectors of observed particles are kept in four dimensions (i.e. observed gluons have 2 helicity states)
\item The dimensional reduction (DR) scheme~\cite{Siegel:1979wq}, in which the momenta of unobserved particles are continued to $d<4$ dimensions, but polarization vectors of unobserved and observed particles are kept in four dimensions (i.e. all gluons have 2 helicity states).
\item The four dimensional helicity (FDH) scheme~\cite{Bern:1991aq,Bern:2002zk}, in which the momenta of unobserved particles is continued to $d>4$ dimensions, and all observed particles are kept in four dimensions (i.e. observed gluons have 2 helicity states). All unobserved internal states are treated as $d_s$-dimensional, where $d_s>d$ in all intermediate steps. Any factor of dimension arising from the numerator Lorentz and Dirac algebra should be labeled as $d_s$, and should be distinct from the dimension $d$. Once the spin and tensor algebra is done one analytically continues the result to $d<4$ and takes the limit $d_s\rightarrow 4$ for the spins of the internal particles.
\end{itemize}
Typically the dimensionality is parametrized as $d=4-2\varepsilon$. We will also use the notation $d_\perp\equiv d-2 = 2-2\varepsilon$ for the number of transverse dimensions in light cone coordinates.
Within the DR and FDH schemes one can still choose the momentum of observed particles to be either $d$-dimensional or 4-dimensional. At one-loop order, however, these choices lead to difference of $\mathcal{O}(\varepsilon)$ and thus one can set the observed particles momenta to be 4-dimensional.

The question of which regularization scheme is most efficient for a given calculation is of course very subjective. We would like to argue in this paper that for one-loop LCPT calculations the helicity basis supplemented with the FDH regularization scheme is in fact the most efficient one. However, as we will show below, the helicity basis approach can also be combined with other dimensional regularization scheme choices, and in particular with the CDR scheme. 
Our overall motivation for using the FDH scheme is the following.
The one-loop results for physical observables arise from a product of a one-loop tensorial loop integral and another tensor  from the spin/helicity structure of the vertices. The resulting contributions can be classified into three kinds of terms. The most divergent part is obtained by taking the divergent $1/\varepsilon$-term from the integral, and evaluating the helicity structure in 4 spacetime dimensions. This part has no scheme dependence. The scheme dependent finite part comes from taking  a $\sim \varepsilon$ term from the helicity structure and multiplying it by the $1/\varepsilon$-term from the loop integral. The scheme independent finite part, on the other hand, involves the finite part of the integral and a helicity structure which can, at one-loop accuracy, be evaluated in $d_s=4$ dimensions. Out of these three, the scheme independent finite part is by far the most complicated one, because in many cases the tensorial structure in the finite part of the loop integral is much more complicated than in the pole part. Thus being able to calculate the finite scheme independent part as efficiently as possible is a priority.

Our strategy is to write the elementary vertices of the theory in a way which, in 4 dimensions, has a very practical structure in terms of the helicities of the particles. These structures are, when the helicities are evaluated in 4 dimensions, written in terms of Levi-Civita tensors in 2 transverse dimensions. This leads to a very easy way to calculate the most complicated scheme independent finite part. The price to pay, however, is that calculating the scheme dependent  $\varepsilon/\varepsilon$-part becomes more complicated, because to evaluate the helicity sums accurately up to order $\varepsilon$ the Levi-Civita structure cannot be used any more. In stead, one must carefully evaluate contractions involving both $d_s$-dimensional structures from the spin sums and $d$-dimensional ones from the loop integrals.  Here, however, one is dealing with the simpler tensorial structure  of the $1/\varepsilon$-part of the loop integral, and this represents a relatively small part of the calculation.

\subsection{Decomposition of quark vertices}
\label{sec:vertices}

\begin{figure*}[t]
\centerline{
\includegraphics[width=6.00cm]{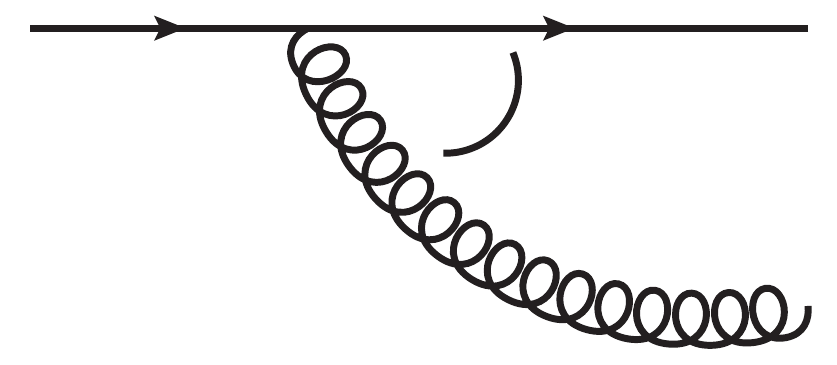}
\begin{tikzpicture}[overlay]
\node[anchor=south east] at (-5cm,2.6cm) {$\pvec,h,\alpha$};
\node[anchor=south west] at (-2cm,2.6cm) {$\ppvec \equiv \pvec-\kvec,h',\beta$};
\node[anchor=south west] at (-2cm,0.7cm) {$\kvec,\lambda,a; \quad k^+ = z p^+$};
\node[anchor=north west] at (-2.5cm,2cm) {$\qt \equiv \kt - z\pt$};
\end{tikzpicture}
\rule{8em}{0pt}
\reflectbox{\includegraphics[width=6.00cm]{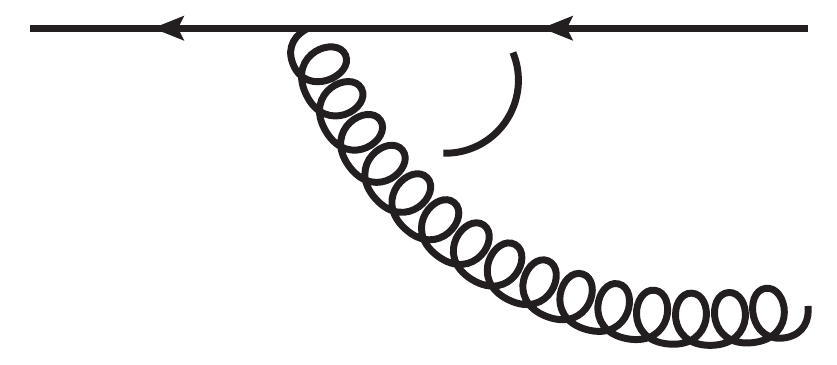}}
\begin{tikzpicture}[overlay]
\node[anchor=south west] at (-1.5cm,2.6cm) {$\pvec,h,\alpha$};
\node[anchor=south west] at (-5cm,2.6cm) {$\ppvec \equiv \pvec-\kvec,h',\beta$};
\node[anchor=south west] at (-7cm,0.7cm) {$\kvec,\lambda,a; \quad k^+ = z p^+$};
\node[anchor=north east] at (-4.1cm,2cm) {$\qt \equiv \kt - z\pt$};
\end{tikzpicture}
}
\caption{Left: Gluon emission vertex from a quark 
$\qemit^{\alpha;\beta,a}_{h;h',\lambda}(\qt,z)$
\eq\nr{eq:vertexqtoqgd}, 
where $\alpha,\beta$ are quark colors, $h,h'$ the quark helicities before and after the emission, $a$ the gluon color and $\lambda$ the gluon helicity. Right: Gluon absorption vertex into quark
 $\qemit^{\beta,a;\alpha}_{h',\lambda;h}(\qt,z)$ \eq\nr{eq:vertexqgtoqd}. 
 }
\label{fig:vertexqtoqg}
\end{figure*}
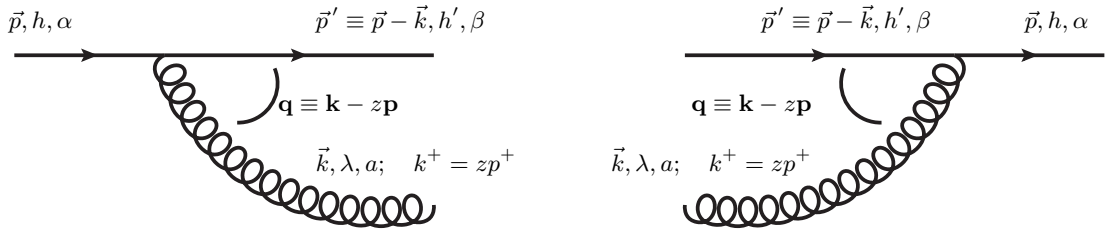

\begin{figure*}[t]
\centerline{\includegraphics[width=6.28cm]{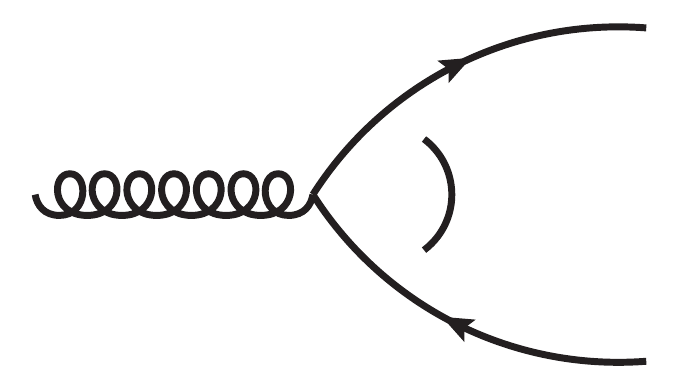}
\begin{tikzpicture}[overlay]
\node[anchor=south west] at (-5cm,2.1cm) {$\pvec,\lambda,a$};
\node[anchor=south west] at (-2.2cm,2.4cm) {$\kvec,h,\alpha; \quad k^+ = z p^+$};
\node[anchor=south west] at (-2cm,0.7cm) {$\pvec-\kvec,h',\beta$};
\node[anchor=north east] at (-0cm,2cm) {$\qt \equiv \kt - z\pt$};
\end{tikzpicture}
\rule{8em}{0pt}
\reflectbox{\includegraphics[width=6.28cm]{diags/vertexgtoqqb}}
\begin{tikzpicture}[overlay]
\node[anchor=south west] at (-2cm,2.1cm) {$\pvec,\lambda,a$};
\node[anchor=south east] at (-4cm,0.7cm) {$\kvec,h,\alpha; \quad k^+ = z p^+$};
\node[anchor=south east] at (-4cm,2.4cm) {$\pvec-\kvec,h',\beta$};
\node[anchor=north east] at (-4.5cm,2cm) {$\qt \equiv \kt - z\pt$};
\end{tikzpicture}
}
\caption{Left: Gluon splitting vertex into quark-antiquark pair,
$A_{\lambda;h,h'}^{a;\beta,\alpha}(\qt,z)$ \eq\nr{eq:vertexgtoqqb}.
Right: Quark-antiquark annihilation vertex into gluon
$A_{h',h;\lambda}^{\beta,\alpha;a}(\qt,z)$ \eq\nr{eq:vertexqqbtog}.
}\label{fig:vertexgtoqqb}
\end{figure*}

Let us first consider the simplest light cone vertex shown in \fig\ref{fig:vertexqtoqg} (left), where a gluon with momentum $\kvec$ and helicity $h$ is emitted from a quark with momentum $\pvec$ and helicity $h$. For simplicity of notation we denote the two quark spin states with spin $\pm 1/2$ by $h=\pm1$, i.e. the actual helicity of the quark is $h/2$. As discussed in \cite{Lappi:2016oup}, we denote this vertex\footnote{Our sign convention for the covariant derivative is $D_\mu = \partial_\mu -igA_\mu$, which is the opposite to that of Refs.~\cite{Beuf:2016wdz,Beuf:2017bpd}.} as 
\begin{equation}
\label{eq:gluonemission}
\qemit_{h;h',\lambda}^{\alpha;\beta,a}
= -gt^{a}_{\beta\alpha}\biggl [ \bar{u}_{h'}(p') \epsl^*_\lambda(k) u_h(p)\biggr ]. 
\end{equation}

Using the Dirac equation satisfied by the spinors, 3-momentum conservation $\pvec=\ppvec+\kvec$  and some Dirac algebra (see Appendix~\ref{appendix:decomplcvertices} for the details), the tensoral structure of matrix element in \eq\nr{eq:gluonemission} can be decomposed to the symmetric and antisymmetric parts as
\begin{equation}
\qemit_{h;h',\lambda}^{\alpha;\beta,a}(\qt,z) = \frac{-gt^{a}_{\beta\alpha}}{z(1-z)p^+} \biggl [\left (1 - \frac{z}{2}\right )\delta^{ij}\bar{u}_{h'}(p')\gamma^{+}u_h(p) - \frac{z}{4} \bar{u}_{h'}(p')\gamma^{+}[\gamma^i,\gamma^j]u_h(p) \biggr ] \qt^i\epst_{\lambda}^{\ast j}.
\end{equation}
Here we have  expressed the  vertex in terms of the momentum fraction $z = k^+/p^+$, $0\leq z\leq 1$ and the center-of-mass transverse momentum $\qt = \kt-z\pt$.
The first matrix element is simple
\begin{equation}
\bar{u}_{h'}(p')\gamma^{+}u_h(p) =  2p^+\sqrt{1-z}\delta_{h,h'}.
\end{equation}
The antisymmetric matrix element $\bar{u}_{h'}(p')\gamma^{+}[\gamma^i,\gamma^j]u_h(p)$, on the other hand, can only be calculated simply in exactly four dimensions by relating it to the helicity operator. For the loop computations we also encounter it in situations where the indices $i,j$ have to be contracted with $d_\perp$-dimensional Kronecker deltas arising from $d_\perp$-dimensional tensorial transverse momentum integrals. For performing the numerator algebra in these cases we introduce for it  a more general  notation
\begin{equation}
\label{eq:defqemitas}
\qemitas^{ij}_{h',h} \equiv
\frac{\bar{u}_{h'}(p')\gamma^{+}[\gamma^i,\gamma^j]u_h(p)}{2p^+\sqrt{1-z}}.
\end{equation}
In exactly $d_\perp=2$ transverse dimensions this simplifies (see Appendix~\ref{appendix:decomplcvertices}) to
\begin{equation}
 \qemitas^{ij}_{h',h}
 \underset{d_\perp \to 2 } {\to} -2ih\delta_{h,h'} \epsilon^{ij}.
 \end{equation}
However, when the helicity sums (numerators of loop diagrams) are needed to order $\epsilon$ we need to remember the full definition~\nr{eq:defqemitas}.  
A similar procedure can be carried out for the gluon absorption vertex, for antiquarks and for quark-antiquark pair creation and annihilation vertices. Let us simply collect the results here, in every case parametrizing the longitudinal momentum with a splitting momentum fraction $0\leq z\leq 1$:
\begin{itemize}
 \item Gluon emission from quark \fig\ref{fig:vertexqtoqg} (left), with momentum conservation $\pvec= \ppvec+\kvec$,   $z = k^+/p^+$ and $\qt = \kt-z\pt$:
 \begin{equation}
\label{eq:vertexqtoqgd}
\qemit_{h;h',\lambda}^{\alpha;\beta,a}(\qt,z) = -gt^{a}_{\beta\alpha}\biggl [ \bar{u}_{h'}(p') \epsl^*_\lambda(k) u_h(p)\biggr ] = \frac{-2gt^{a}_{\beta\alpha}}{z\sqrt{1-z}} \biggl [\left (1 - \frac{z}{2}\right )\delta^{ij}\delta_{h',h}
- \frac{z}{4} 
\qemitas^{ij}_{h',h}
\biggr ] \qt^i\epst_{\lambda}^{\ast j},
\end{equation}
with 
\begin{equation}
\label{eq:vertexqtoqgdas}
\qemitas^{ij}_{h',h} \equiv
\frac{\bar{u}_{h'}(p')\gamma^{+}[\gamma^i,\gamma^j]u_h(p)}{2p^+\sqrt{1-z}}
\underset{d_\perp \to 2 } {\to} -2ih\delta_{h',h} \epsilon^{ij}.
\end{equation}
\item Gluon absorbtion by quark \fig\ref{fig:vertexqtoqg} (right), with momentum conservation $\pvec= \ppvec+\kvec$,   $z = k^+/p^+$ and $\qt = \kt-z\pt$:
 \begin{equation}
 \label{eq:vertexqgtoqd}
 \qemit_{h',\lambda;h}^{\beta,a;\alpha}(\qt,z)
=
-gt^{a}_{\alpha\beta}\biggl [ \bar{u}_{h}(p) \epsl_\lambda(k) u_{h'}(p')\biggr ] = \frac{-2gt^{a}_{\alpha\beta}}{z\sqrt{1-z}} \biggl [\left (1 - \frac{z}{2}\right )\delta^{ij}\delta_{h,h'} + \frac{z}{4} 
\qemitas^{ij}_{h,h'}
\biggr ] \qt^i\epst_{\lambda}^{ j},
\end{equation}
with 
\begin{equation}
 \label{eq:vertexqgtoqdas}
\qemitas^{ij}_{h,h'} \equiv
\frac{\bar{u}_{h}(p)\gamma^{+}[\gamma^i,\gamma^j]u_{h'}(p')}{2p^+\sqrt{1-z}}
\underset{d_\perp \to 2 } {\to} -2ih\delta_{h,h'} \epsilon^{ij}.
\end{equation}
\item Gluon emission from antiquark with momentum $\pvec$,  with momentum conservation $\pvec = \ppvec+\kvec$, $z=k^+/p^+$ and $\qt = \kt-z\pt$:
 \begin{equation}
\label{eq:vertexqbartoqbarg}
\qbemit_{h;h',\lambda}^{\alpha;\beta,a}(\qt,z)
=
- gt^{a}_{\alpha\beta}\biggl [ - \bar{v}_{h}(p) \epsl^*_\lambda(k) v_{h'}(p')\biggr ] 
= \frac{2gt^{a}_{\alpha\beta}}{z\sqrt{1-z}} \biggl [\left (1 - \frac{z}{2}\right )\delta^{ij}\delta_{h,h'} + \frac{z}{4} 
\qbemitas^{ij}_{h,h'}
\biggr ] \qt^i\epst_{\lambda}^{* j},
\end{equation}
with 
\begin{equation}
\label{eq:vertexqbartoqbargas}
\qbemitas^{ij}_{h,h'} \equiv
\frac{\bar{v}_{h}(p)\gamma^{+}[\gamma^i,\gamma^j]v_{h'}(p')}{2p^+\sqrt{1-z}}
\underset{d_\perp \to 2 } {\to} 2ih\delta_{h,h'} \epsilon^{ij}.
\end{equation}
\item Gluon absorption into antiquark, with momentum conservation $\pvec = \ppvec+\kvec$, $z=k^+/p^+$ and $\qt = \kt-z\pt$:
 \begin{equation}
 \label{eq:vertexqbargtotoqbar}
\qbemit_{h',\lambda;h}^{\beta,a;\alpha}(\qt,z)
=
- gt^{a}_{\beta\alpha}\biggl [ - \bar{v}_{h'}(p') \epsl_\lambda(k) v_{h}(p)\biggr ] 
= \frac{2gt^{a}_{\beta\alpha}}{z\sqrt{1-z}} \biggl [\left (1 - \frac{z}{2}\right )\delta^{ij}\delta_{h',h} - \frac{z}{4} 
\qbemitas^{ij}_{h',h}
\biggr ] \qt^i\epst_{\lambda}^{ j},
\end{equation}
with 
\begin{equation}
 \label{eq:vertexqbargtotoqbaras}
\qbemitas^{ij}_{h',h} \equiv
\frac{\bar{v}_{h'}(p')\gamma^{+}[\gamma^i,\gamma^j]v_{h}(p)}{2p^+\sqrt{1-z}}
\underset{d_\perp \to 2 } {\to} 2ih\delta_{h',h} \epsilon^{ij}.
\end{equation}
\item 
Gluon with momentum $\pvec$ splitting into quark with momentum $\kvec$ and antiquark, \fig\ref{fig:vertexgtoqqb},
with momentum conservation $\pvec = \ppvec+\kvec$, $z=k^+/p^+$ and $\qt = \kt-z\pt$:
 \begin{equation}
 \label{eq:vertexgtoqqb}
\paircr_{\lambda;h,h'}^{a;\alpha,\beta}(\qt,z)
=
- gt^{a}_{\alpha\beta}\biggl [ \bar{u}_{h}(k) \epsl_\lambda(p) v_{h'}(p')\biggr ] 
= \frac{-2gt^{a}_{\alpha\beta}}{\sqrt{z(1-z)}} \biggl [\left (z - \frac{1}{2}\right )\delta^{ij}\delta_{h,-h'} + \frac{1}{4} 
\paircras^{ij}_{h,h'}
\biggr ] \qt^i\epst_{\lambda}^{ j},
\end{equation}
with 
\begin{equation}
 \label{eq:vertexgtoqqbas}
\paircras^{ij}_{h,h'} \equiv
\frac{\bar{u}_{h}(k)\gamma^{+}[\gamma^i,\gamma^j]v_{h'}(p')}{2p^+\sqrt{z(1-z)}}
\underset{d_\perp \to 2 } {\to} -2ih\delta_{h,-h'} \epsilon^{ij}.
\end{equation}
\item Quark with momentum $\kvec$ and antiquark annihilating to gluon with momentum $\pvec$, \fig\ref{fig:vertexgtoqqb},
with momentum conservation $\pvec = \ppvec+\kvec$, $z=k^+/p^+$ and $\qt = \kt-z\pt$:
 \begin{equation}
\label{eq:vertexqqbtog}
\annih_{h',h;\lambda}^{\alpha,\beta;a}(\qt,z)
=
- gt^{a}_{\beta\alpha}\biggl [ - \bar{v}_{h'}(p') \epsl^*_\lambda(p) u_{h}(k)\biggr ] 
= \frac{2gt^{a}_{\beta\alpha}}{\sqrt{z(1-z)}} \biggl [\left (z - \frac{1}{2}\right )\delta^{ij}\delta_{h,-h'} - \frac{1}{4} 
\annihas^{ij}_{h',h}
\biggr ] \qt^i\epst_{\lambda}^{* j},
\end{equation}
with 
\begin{equation}
\label{eq:vertexqqbtogas}
\annihas^{ij}_{h',h} \equiv
\frac{\bar{v}_{h'}(p')\gamma^{+}[\gamma^i,\gamma^j]u_{h}(k)}{2p^+\sqrt{z(1-z)}}
\underset{d_\perp \to 2 } {\to} -2ih\delta_{h',-h} \epsilon^{ij}.
\end{equation}

\end{itemize}

\subsection{Evaluating helicity sums}
\label{sec:numerators}

The value of a diagram in the perturbative expansion of light cone wave functions is obtained by multiplying the factors for the vertices, integrating over internal momenta in loops and summing over the helicities of internal particles. Let us demonstrate how this procedure works in terms of the quark vertices introduced above with two concrete examples that will be needed in the calculation of the virtual photon wave function.

First, let us look at a quark propagator correction diagram such as the one shown in \fig\ref{fig:oneloopSEUPT}. The loop part involves the product of the gluon emission vertex \nr{eq:vertexqtoqgd} and the absorption of the same gluon \nr{eq:vertexqgtoqd}, summed over the helicities of the quark and gluon inside the loop
\begin{equation}
\sum_{\lambda,h'} \qemit^{\alpha;\beta,a}_{h;h',\lambda}(\qt,z)\qemit^{\beta,a;\alpha}_{h',\lambda;h}(\qt,z).
\end{equation}
The integrand in the transverse momentum integral is proportional to $q^iq^j$, thus the value of the dimensionally regulated integral is proportional to a $(d-2)$-dimensional Kronecker delta $\delta_{(d)}^{ij}$. The vertices are proportional to $(d_s-2)$-dimensional gluon polarization vectors, and summing over the helicity states of the gluon yields 
\begin{equation}
\label{eq:polsum}
 \sum_{\lambda} \epst_{\lambda}^{* k}\epst_{\lambda}^{l} = \delta_{(d_s)}^{kl}.
\end{equation}
We are then tasked with evaluating the expression
\begin{equation}
\label{eq:defnum1}
\text{num}_1=\sum_{h'}
\left[ \left (1 - \frac{z}{2}\right )\delta^{ik}\delta_{h',h}
- \frac{z}{4} 
\qemitas^{ik}_{h',h}
\right]
\left[
\left (1 - \frac{z}{2}\right )\delta^{jl}\delta_{h,h'} + \frac{z}{4} 
\qemitas^{jl}_{h,h'}
\right]\delta_{(d)}^{ij}\delta_{(d_s)}^{kl}.
\end{equation}
Now in principle, to correctly evaluate this for arbitrary $d_s>d>4$, we need to use the definitions~\nr{eq:vertexqtoqgdas} and \nr{eq:vertexqgtoqdas} and carefully perform the Dirac matrix algebra. This we will do in detail for the more complicated case of \eq\nr{eq:defnum2} below. However, let us here evaluate the sum \nr{eq:defnum1} with a simple, but less general trick that yields the same result.

In this case the most complicated structure appearing is the product of two antisymmetric tensors, not more. In fact, in such a case we can formally express the antisymmetric vertex structure in terms of a ``$(d_s-2)$-dimensional'' two-index Levi-Civita tensor $\epsilon^{ij}_{(d_s)}$. In general such an object does of course not exist, but here it can be given an explicit meaning in terms of perfectly well-defined $(d_s-2)$-dimensional Kronecker deltas using the
Fierz identity 
\begin{equation}\label{eq:fierz}
 \epsilon^{ij}_{(d_s)}\epsilon^{kl}_{(d_s)} = 
 \delta_{(d_s)}^{ik} \delta_{(d_s)}^{jl} - \delta_{(d_s)}^{il} \delta_{(d_s)}^{jk}.
\end{equation}
When there are more than two Levi-Civita tensors, there would be several inequivalent ways to get rid of them using the Fierz identity. Thus the trick we are now describing cannot be used in these more complicated cases.

To now evaluate the helicity sum \nr{eq:defnum1} we first use the fact that 
for massless quarks helicity is conserved at the emission vertex and thus the sum over the intermediate quark helicity $h'$ is trivial. This gives, promoting the 4-dimensional expressions for the antisymmetric tensors in \eqs\nr{eq:vertexqtoqgdas} and \nr{eq:vertexqgtoqdas} into $d_s$-dimensional ones,
\begin{equation}
\text{num}_1=
\left[ \left (1 - \frac{z}{2}\right )\delta^{ik}
+ih \frac{z}{2} 
 \epsilon^{ik}_{(d_s)}
\right]
\left[
\left (1 - \frac{z}{2}\right )\delta^{jl}
-ih \frac{z}{2} 
\epsilon^{jl}_{(d_s)}
\right]
\delta_{(d)}^{ij}\delta_{(d_s)}^{kl}.
\end{equation}
We then get rid of the  Levi-Civita-tensors using the  
Fierz identity~\nr{eq:fierz}
to get
\begin{equation}\label{eq:num1value}
\text{num}_1= \left (1 - \frac{z}{2}\right )^2 (d-2) + \left(\frac{z}{2}\right)^2(d-2)(d_s-3),
\end{equation}
where one must remember that $d_s>d$, i.e. 
\begin{equation}
\delta_{(d_s)}^{ij}\delta_{(d_s)}^{ij}=d_s-2, \quad \delta_{(d)}^{ij} \delta_{(d)}^{ij} = d-2, \quad \delta_{(d_s)}^{ij}\delta_{(d)}^{ij}=d-2
\end{equation}
and that $h^2=1$ in our convention. This yields the correct result for both the FDH  (taking $d_s=4,d=4-2\varepsilon$) and for the CDR schemes (taking $d_s=d=4-2\varepsilon$).
The result \nr{eq:num1value} appears in perfectly conventional QCD calculations of the $q\to qg$ splitting function in dimensional regularization. One could speculate about a physical interpretation for the two terms, independent of $d_s$ and proportional to $d_s-3$. The first one results from the part of the vertex that is independent of helicity, and therefore does not depend on the number of helicity states. The second term comes from the antisymmetric part of the vertex where the gluon and quark are constrained to have a different helicity, thus it is proportional not to the total number of gluon helicities $d_s-2$, but to the number of helicities orthogonal to that of the quark, namely $d_s-3$.

A more complicated example is provided by vertex correction diagrams, such as the one in \fig\ref{fig:vertexqbaremT}. Here (with a trivial simplification of the color structure of the gluon splitting vertex \nr{eq:vertexgtoqqb} to a virtual photon splitting), one has a structure like
\begin{equation}\label{eq:3vert}
\sum_{h',h'',\sigma}
\qemit_{h',\sigma;h}^{\bar{\alpha},a;\alpha}(\qt,z_1)
\paircr_{\lambda;h',h''}(\kt,z_2)
 \qbemit_{h'';h''',\sigma}^{\bar{\alpha};\beta,a}(\pt,z_3).
 \end{equation}
Writing this out in terms of the decompositions \nr{eq:vertexqgtoqd}, \nr{eq:vertexgtoqqb} and \nr{eq:vertexqbartoqbarg} of the vertices into symmetric and antisymmetric parts, one encounters a product of three antisymmetic vertex factors. This structure is then multiplied with a $(d_s-2)$-dimensional Kronecker delta from the sum over the internal gluon helicity $\sigma$, but also $(d-2)$-dimensional ones from the loop integrals. Now there would be three inequivalent ways to use the Fierz identity \nr{eq:fierz} to remove two of the three Levi-Civita tensors, and thus we cannot get an unambigous result in the same way as above. Thus we need to use the definitions of the antisymmetric vertex factors, \nr{eq:vertexqgtoqdas}, \nr{eq:vertexgtoqqbas} and \nr{eq:vertexqbartoqbargas}. 

In stead of working out the full expression here, let us concentrate on the most difficult part involving a product of three ansisymmetric structures in the vertices. We take as an example one of the kind of terms that arise when evaluating the structure \nr{eq:3vert}, and calculate
\begin{equation}
\label{eq:defnum2}
\text{num}_2^{lm}=
\sum_{h',h''}
\qemitas_{h,h'}^{ij}
\paircras_{h',h''}^{kl}
 \qbemitas_{h'',h'''}^{mn}
 \delta_{(d)}^{ik}
 \delta_{(d_s)}^{jn},
\end{equation}
where we have already performed the sum over the helicity $\sigma$, yielding a $(d_s-2)$-dimensional  $\delta_{(d_s)}^{jn}$, and taken one particular term of the $(d-2)$-dimensional tensor integral with indices $ikm$. Writing this out in terms of the full definitions of the antisymmetric vertex factors \nr{eq:vertexqgtoqdas}, \nr{eq:vertexgtoqqbas} and \nr{eq:vertexqbartoqbargas} we have
\begin{equation}
\text{num}_2^{lm}=
\sum_{h',h''}
\frac{\bar{u}_{h}(p_1)\gamma^{+}[\gamma^i,\gamma^j]u_{h'}(p_2)}{2\sqrt{p_1^+p_2^+}}
\frac{\bar{u}_{h'}(p_2)\gamma^{+}[\gamma^k,\gamma^l]v_{h''}(p_3)}{2\sqrt{p_2^+p_3^+}}
\frac{\bar{v}_{h''}(p_3)\gamma^{+}[\gamma^m,\gamma^n]v_{h'''}(p_4)}{2\sqrt{p_3^+p_4^+}}
 \delta_{(d)}^{ik}
 \delta_{(d_s)}^{jn}.
\end{equation}
In a massless theory helicity is conserved at the vertex, therefore we know that $h=h'=-h''=-h'''$. However, in order to evaluate this expression we do not use this, but revert to the usual procedure from covariant perturbation theory calculations and transform the sums over intermediate fermion helicities to Dirac matrices. Thus we substitute
\begin{equation}
\sum_{h}
u_{h}(p) \bar{u}_{h}(p) = p\!\!\! /
\end{equation}
to write 
\begin{equation}
\text{num}_2^{lm}=
\frac{\bar{u}_{h}(p_1)
\gamma^{+}[\gamma^i,\gamma^j]
p\!\!\!/_2
\gamma^{+}[\gamma^k,\gamma^l]
p\!\!\!/_3
\gamma^{+}[\gamma^m,\gamma^n]
v_{h'''}(p_4)
}{2\sqrt{p_1^+p_2^+} 2\sqrt{p_2^+p_3^+} 2\sqrt{p_3^+p_4^+}}
 \delta_{(d)}^{ik}
 \delta_{(d_s)}^{jn}.
\end{equation}
Now we note that $\gamma^+\gamma^+=0$ and $\gamma^+$ and $\gamma^-$ commute with 
$[\gamma^i,\gamma^j]$. Thus the only nonzero contribution to the matrix element comes from the terms where one takes from every $p\!\!\!/_i$ the term $p_i^+ \gamma^-$ in order to kill the corresponding $\gamma^+$. Using $\gamma^+\gamma^-\gamma^+ = 2\gamma^+$ it is easy to see that effectively every factor $p\!\!\!/_i\gamma^+$ is just replaced by $2p_i^+$, canceling the corresponding factor in the denominator. We are then left with
\begin{equation}
\text{num}_2^{lm}=
\frac{\bar{u}_{h}(p_1)
\gamma^{+}[\gamma^i,\gamma^j]
[\gamma^k,\gamma^l]
[\gamma^m,\gamma^n]
v_{h'''}(p_4)
}{2\sqrt{p_1^+p_4^+}}
 \delta_{(d)}^{ik}
 \delta_{(d_s)}^{jn}.
\end{equation}
Now remembering that the external momenta and polarization vectors are 2-dimensional, and $d_s>d>4$ at this stage, it is a straightforward task to evaluate either manually or, most importantly, using a symbolic calculation program:
\begin{equation}
[\gamma^i,\gamma^j]
[\gamma^k,\gamma^l]
[\gamma^m,\gamma^n]
\delta_{(d)}^{ik}
 \delta_{(d_s)}^{jn}
= 8 (d-3)(d_s-4) \delta^{lm}
- 4 ( 19 - 3 d_s - 6 d + d d_s )[\gamma^l,\gamma^m].
\end{equation}
Here we have identified terms of the type $\delta_{(d)}^{mm'}[\gamma^l,\gamma^{m'}]$ with $[\gamma^l,\gamma^{m}]$ and written both $\delta_{(d_s)}^{lm}$
and $\delta_{(d)}^{mm'}\delta_{(d_s)}^{lm'}$ simply as $\delta^{lm}$ knowing that both indices $l$ and $m$ are to be contracted with external vectors. Using this result we can write the result in terms of the symmetric and antisymmetric parts of the leading order vertex structure as
\begin{equation}
\text{num}_2^{lm}=
 4(d-3)(d_s-4) \delta^{lm}
- 4( 19 - 3 d_s - 6 d + d d_s )
\paircras_{hh'''}^{lm}.
\end{equation}
Setting $d=d_s$ here one would obtain the CDR result. It is interesting to note that in the FDH scheme $d_s=4$ the symmetric $\delta^{lm}$-term vanishes; this is an additional simplification that one gains at the expense of evaluating the algebra in $d,d_s$ dimensions. As a consistency check we can go to the limit $d=d_s=4$:
\begin{equation}
\text{num}_2^{lm} 
\underset{d=d_s \to 4 } {\to}
 -8ih\delta_{h,-h'''} \epsilon^{lm}
=
 4\paircras_{h,h'''}^{lm}.
\end{equation}
The same result can be obtained directly by taking the terms in 
\nr{eq:defnum2} in $d=d_s=4$ dimensions 
\begin{equation}
\text{num}_2^{lm} 
\underset{d=d_s = 4 } {=}
\left[-2ih\delta_{h,h'} \epsilon^{ij}\right]
\left[-2ih'\delta_{h',-h''} \epsilon^{kl} \right]
\left[2ih''\delta_{h'',h'''} \epsilon^{mn}\right]
\delta^{ik} \delta^{jn}
=-8ih\delta_{h,-h'''} \epsilon^{lm} 
=4 \paircras_{h,h'''}^{lm}.
\end{equation}
Note that as a calculational operation, the introduction and subsequent removal of the $p\!\!\!/$ happens in the same way in all combinations of emission vertices from fermions. In practice one can keep track of the terms of the calculation by writing out the vertices in terms of the $d=4$ notation involving 2-dimensional Levi-Civita tensors. Then, whenever an ambiguity arises as to the meaning of products of the Levi-Civita-tensors, one replaces $\epsilon^{ij}$ by $[\gamma^i,\gamma^j]$, orders the vertices following the fermion line, performs contractions of the $\gamma$-matrices with $d$- and $d_s$-dimensional external tensors, expresses the result in terms of $[\gamma^i,\gamma^j]$ and $\delta^{ij}$ and identifies these in terms of the vertex structure of the leading order diagram. This procedure greatly simplifies the appearence of a factorized form for the loop corrections, which appear as multiplicative corrections to the corresponding leading order wave functions.

\subsubsection*{Note on \cite{Lappi:2016oup}}

Let us briefly note the difference between the formulation introduced here and the one used in our earlier work~\cite{Lappi:2016oup}. There we first calculated the $(d-2)$-dimensional loop tensorial integrals, which result in a structure that contains $(d-2)$-dimensional Kronecker deltas. These were then contracted with the $(d_s-2=2)$-dimensional gluon polarization vectors as $\delta_{(d)}^{ij} \epst_\lambda^j \to \epst_\lambda^i$. The error in this calculation comes when the resulting polarization vectors were then treated again as $(d_s-2)$-dimensional ones in order to perform the polarization sums. In fact, contracting with a lower dimensional Kronecker delta projects the polarization vector  into a lower dimensional subspace; this was not taken into account in the calculation of \cite{Lappi:2016oup}. We have checked that with the correct treatment presented in this paper, the power  divergences in the longitudinal cutoff $\alpha$, present in the final result of \cite{Lappi:2016oup}, cancel.

Let us finally point out an essential  technical aspect that enables the correct way to calculate the polarization sums. One has to write all 3-particle vertices in a form where the only dependence on the gluon polarization is in the linear dependence on the polarization vector of each gluon, see e.g.~\eq\nr{eq:vertexqtoqgd} or \eq\nr{eq:vertexgtogg}. Then the expression for a given diagram becomes quadratic in the internal gluon polarization vectors, and the polarization sum can be evaluated using \eq\nr{eq:polsum}. The resulting $(d_s-2)$-dimensional Kronecker delta can then be correctly contracted with both $(d_s-2)$- and $(d-2)$-dimensional objects. In contrast, writing the elementary vertex \nr{eq:gluonemission} in a form like $\delta_{h,h'}\left(\delta_{\lambda,h} + (1-z) \delta_{\lambda,-h}\right)\epst_\lambda \cdot \qt$ as in Ref.~\cite{Lappi:2016oup}, while correct, has an additional dependence on the polarization $\lambda$. This results in expressions where summing over the internal polarizations correctly is difficult.

\subsection{Other vertices}
\label{sec:otherdiags}

\begin{figure}[t]
\centerline{
\includegraphics[width=6.00cm]{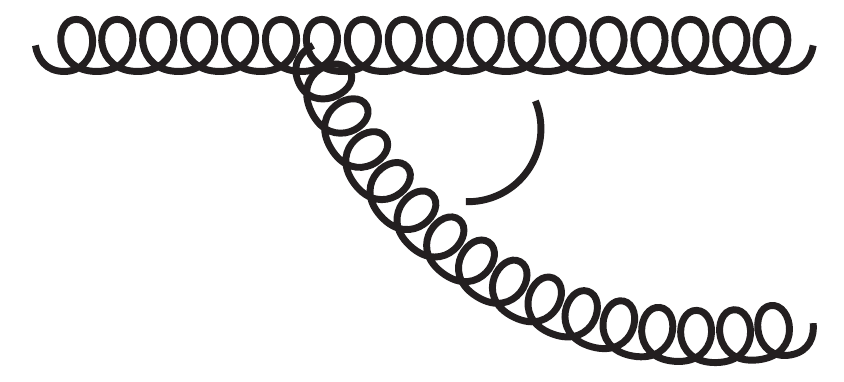}
\begin{tikzpicture}[overlay]
\node[anchor=south west] at (-6cm,2.7cm) {$\pvec,\lambda_1,a$};
\node[anchor=south west] at (-2cm,2.7cm) {$\ppvec \equiv\pvec-\kvec,\lambda_2,b$};
\node[anchor=south west] at (-2cm,0.7cm) {$\kvec,\lambda_3,c; \quad k^+ = z p^+$};
\node[anchor=north west] at (-2.3cm,2cm) {$\qt \equiv \kt - z\pt$};
\end{tikzpicture}
\rule{8em}{0pt}
\reflectbox{\includegraphics[width=6.00cm]{diags/vertexgtogg}}
\begin{tikzpicture}[overlay]
\node[anchor=south west] at (-1.5cm,2.7cm) {$\pvec,\lambda_1,a$};
\node[anchor=south west] at (-6cm,2.7cm) {$\ppvec \equiv \pvec-\kvec,\lambda_2,b$};
\node[anchor=south west] at (-7cm,0.7cm) {$\kvec,\lambda_3,c; \quad k^+ = z p^+$};
\node[anchor=north east] at (-4.3cm,2cm) {$\qt \equiv \kt - z\pt$};
\end{tikzpicture}
}
\caption{Left: Gluon splitting vertex 
$\Gamma^{a; bc}_{\lambda_1; \lambda_2,\lambda_3}(\qt,z)$ \eq\nr{eq:vertexgtogg}, where $a, b, c$ are the gluon colors and $\lambda_1, \lambda_2, \lambda_3$  gluon helicities. 
Right: Gluon merging vertex 
$\Gamma^{ bc;a}_{ \lambda_2,\lambda_3;\lambda_1}(\qt,z)$ \eq\nr{eq:vertexggtog}.
}\label{fig:vertexgtogg}
\end{figure}

In addition, we also have two different type of LC elementary vertices with 3-gluon self interaction: The elementary vertex for $1\rightarrow 2$ gluon splitting shown in \fig\ref{fig:vertexgtogg} is given by
\begin{equation}\label{eq:vertexgtogg}
\Gamma^{a; b,c}_{\lambda_1; \lambda_2,\lambda_3}
(\qt,z)
=
 -2ig f^{abc} \biggl [  \frac{\epst_{\lambda_2}^{\ast j}\epst_{\lambda_3}^{\ast k}\epst_{\lambda_1}^{l}}{1-z}       
 + \frac{\epst_{\lambda_3}^{\ast j}\epst_{\lambda_2}^{\ast k}\epst_{\lambda_1}^{l}}{z} 
- \epst_{\lambda_1}^{j}\epst_{\lambda_3}^{\ast k}\epst_{\lambda_2}^{\ast l} 
\biggr ]\delta^{ij}\delta^{kl}\qt^{i}.
\end{equation}
Similarly, the $2\rightarrow 1$ gluon merging vertex is given by
\begin{equation}\label{eq:vertexggtog}
\Gamma^{ b,c;a}_{ \lambda_2,\lambda_3;\lambda_1}
(\qt,z) =
 +2ig f^{abc} \biggl [  \frac{\epst_{\lambda_2}^{j}\epst_{\lambda_3}^{k}\epst_{\lambda_1}^{\ast l}}{1-z}       
 + \frac{\epst_{\lambda_3}^{j}\epst_{\lambda_2}^{k}\epst_{\lambda_1}^{\ast l}}{z} 
- \epst_{\lambda_1}^{\ast j}\epst_{\lambda_3}^{k}\epst_{\lambda_2}^{l} 
\biggr ]\delta^{ij}\delta^{kl}\qt^{i}.
\end{equation}

As we will discuss in more detail below, the instantaneous interaction diagrams contribute to the one-loop  wave functions and to the 3-particle final states. We will not present here the full set of instantaneous vertices (see \cite{Brodsky:1997de}) but merely the ones needed here, and for the combinations of helicities needed for our calculation. Similarly as above, one can easily derive more general expressions as discussed above, here we present only the ones in $d_s=4$ dimensions that are needed for our present calculation.

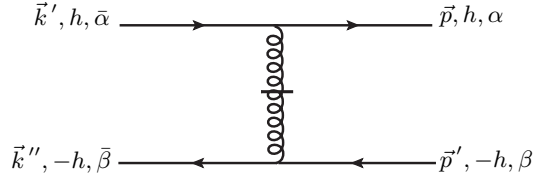
\begin{figure}[tp]
\centerline{
\includegraphics[width=4.5cm]{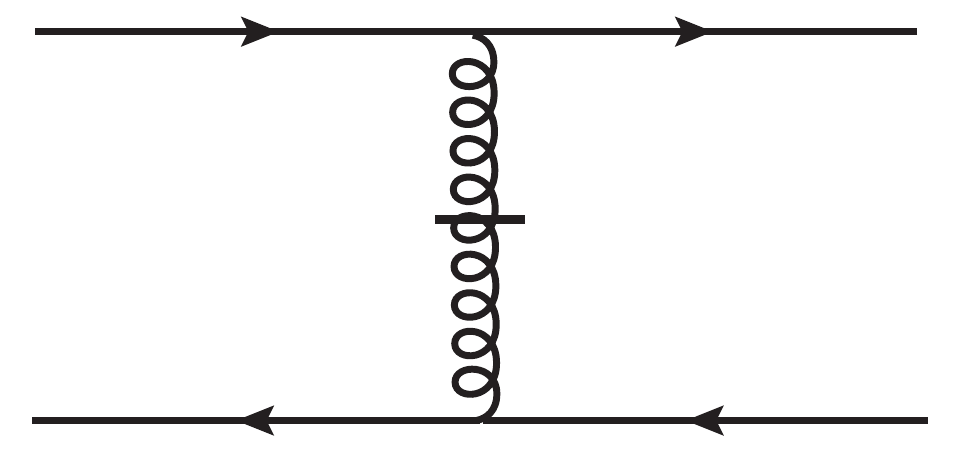}
\begin{tikzpicture}[overlay]
\node[anchor=west] at (-0.3,2.1) {$\pvec,h,\alpha$};
\node[anchor=west] at (-0.3,0.2) {$\ppvec,-h,\beta$};
\node[anchor=west] at (-6.0,0.2) {$\kppvec, -h,\bar{\beta} $};
\node[anchor=south west] at (-5.7,1.8) {$\kpvec,h,\bar{\alpha}$};
 \end{tikzpicture}
}
\rule{0pt}{1ex}
\caption{Time ordered (momenta flows from left to right) instantaneous vertex contributing to the $q\bar{q}$-component of the longitudinal virtual photon wave function at NLO.}
\label{fig:qqbarqqbarinst}
 \end{figure}

\begin{figure}[tp]
\centerline{
\includegraphics[width=5.5cm]{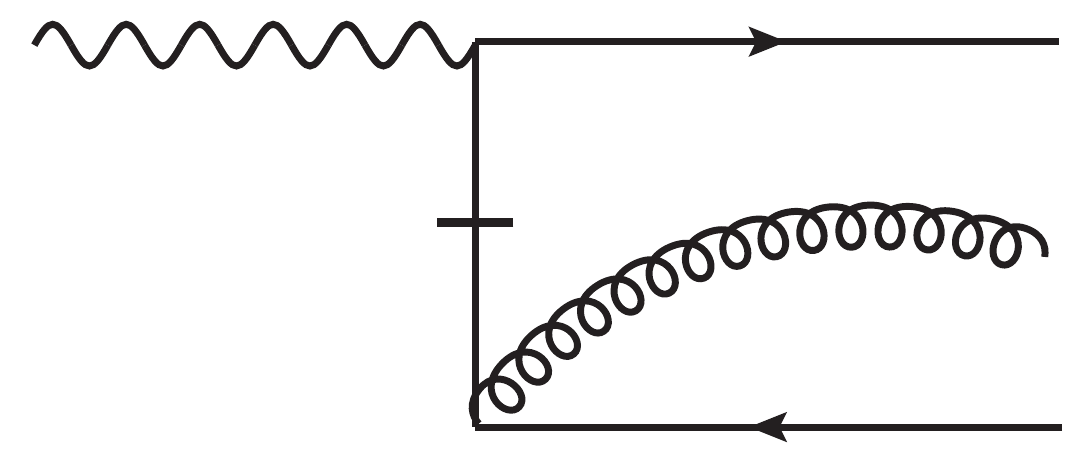}
\begin{tikzpicture}[overlay]
\node[anchor=west] at (-0.3,2.2) {$\pvec,h,\alpha$};
\node[anchor=west] at (-0.3,0.2) {$\ppvec,-h,\beta$};
\node[anchor=west] at (-0.3,1.2) {$\kvec,\sigma,a$};
\node[anchor=south west] at (-5.3,1.4) {$\qvec,\lambda$};
 \end{tikzpicture}
}
\rule{0pt}{1ex}
\caption{Time ordered (momenta flows from left to right) instantaneous diagram contributing to the $q\bar{q}g$-component of the transverse virtual photon wave function at NLO.}
\label{fig:meqqbarginst}
 \end{figure}

\begin{figure}[tp]
\centerline{
\includegraphics[width=5.5cm]{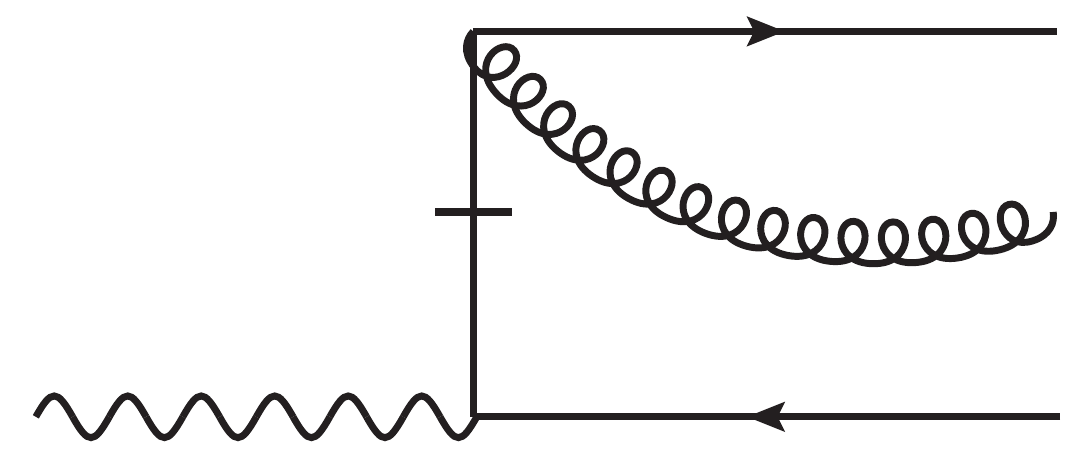}
\begin{tikzpicture}[overlay]
\node[anchor=west] at (-0.3,2.2) {$\pvec,h,\alpha$};
\node[anchor=west] at (-0.3,1.3) {$\kvec,\sigma,a$};
\node[anchor=west] at (-0.3,0.2) {$\ppvec,-h,\beta$};
\node[anchor=west] at (-5.2,0.65) {$\qvec, \lambda$};
 \end{tikzpicture}
}
\rule{0pt}{1ex}
\caption{Time ordered (momenta flows from left to right) instantaneous diagram contributing to the $q\bar{q}g$-component of the transverse virtual photon wave function at NLO. }
\label{fig:meqgqbarinst}
 \end{figure}

The instantaneous gluon exchange diagram \fig\ref{fig:qqbarqqbarinst} is given by the following matrix element 
\begin{equation}\label{eq:qqbarqqbarinst}
\mathfrak{I}^{(\ref{fig:qqbarqqbarinst})} = -g^2	t^{a}_{\alpha\bar{\alpha}}t^{a}_{\beta\bar{\beta}}\biggl [\bar{u}_h(p)\gamma^+u_h(k') \biggr ]\frac{1}{(k'^+-p^+)^2}\biggl [\bar{v}_{-h}(k'')\gamma^+v_{-h}(p') \biggr ]
\end{equation}
which simplifies in $d_s =4$ to 
\begin{equation}
\mathfrak{I}^{(\ref{fig:qqbarqqbarinst})} =-4g^2 t^{a}_{\alpha\bar{\alpha}}t^{a}_{\beta\bar{\beta}}\frac{\sqrt{p^+p'^+k'^+k''^+}}{(k'^+-p^+)^2},
\end{equation}
where the momenta are labeled as in \fig\ref{fig:qqbarqqbarinst}.

The matrix element for $\gamma\to q\bar {q}g$ via the exchange of an instantaneous quark, diagram \fig\ref{fig:meqqbarginst}, is given by
\begin{equation}
\label{eq:meqqbarginst}
\mathfrak{I}^{(\ref{fig:meqqbarginst})} = \frac{-ee_fgt^{a}_{\alpha\beta}}{2}\bar{u}_h(p)\epsl_{\lambda}(q)\frac{\gamma^+}{(p'^+ + k^+)}\epsl^{\ast}_{\sigma}(k)v_{-h}(p')
\end{equation}
which, in $d_s=4$, can be expressed in the helicity basis as 
\begin{equation}
\mathfrak{I}^{(\ref{fig:meqqbarginst})}  = -ee_fgt^{a}_{\alpha\beta} \frac{\sqrt{p^+p'^+}}{(p'^+ + k^+)}\biggl [\delta^{ij} - ih\epsilon^{ij} \biggr ]\epst^{\ast i}_{\sigma}\epst^{j}_{\lambda}.
\end{equation}

Similarly, the matrix element for the other instantaneous quark $\gamma\to q\bar {q}g$ diagram \fig\nr{fig:meqgqbarinst} is given by
\begin{equation}
\label{eq:meqgqbarinst}
\mathfrak{I}^{(\ref{fig:meqgqbarinst})} =  \frac{+ee_fgt^{a}_{\alpha\beta}}{2}\bar{u}_h(p)\epsl^{\ast}_{\sigma}(k)\frac{\gamma^+}{(p^+ + k^+)}\epsl_{\lambda}(q)v_{-h}(p'),
\end{equation}
which in the helicity basis and $d_s=4$ reduces to
\begin{equation}
\mathfrak{I}^{(\ref{fig:meqgqbarinst})} = +ee_fgt^{a}_{\alpha\beta} \frac{\sqrt{p^+p'^+}}{(p^+ + k^+)}\biggl [\delta^{ij} + ih\epsilon^{ij} \biggr ]\epst^{\ast i}_{\sigma}\epst^{j}_{\lambda}.
\end{equation}

\section{Calculating the DIS cross section from light cone wave functions}
\label{sec:sigma}

We consider a setup where a relativistic projectile moving in the  light-cone $x^+$ direction scatters on a very dense and highly boosted target moving in the  light-cone $x^-$ direction. At high energy the target consists of a gluon field, and the scattering can be evaluated using the eikonal approximation in terms of Wilson lines in this field~\cite{Bjorken:1970ah,Weigert:2005us}. The total cross section for a virtual photon scattering from a classical gluon field can be obtained by the optical theorem as twice the forward inelastic scattering amplitude. With the appropriate normalization~\cite{Bjorken:1970ah} this results in: 
\begin{equation}
\sigma^{\gamma^{\ast}}[A] = \frac{2}{2q^+(2\pi)\delta(q'^+-q^+)}\R\biggl [{}_{\text{i}}\langle \gamma^{\ast}(\qpvec,Q^2,\lambda')\vert 1- \hat{S}_E\vert \gamma^{\ast}(\qvec,Q^2,\lambda)\rangle_{\text{i}}\biggr ].
\end{equation}
The full perturbative Fock state decomposition for the virtual photon in the momentum space with momentum $\qvec$, virtuality $Q$, and helicity $\lambda$ is given by 
\begin{equation}
\label{eq:vpfockdecompostion}
\begin{split}	
\vert \gamma^{\ast}(\qvec,Q^2,\lambda)\rangle_{\rm i}  = & \sqrt{Z_{\gamma^{\ast}}(q^+)}\biggl [\vert \gamma(\qvec,\lambda)\rangle_{\rm b}  + \int \ddp\dpp(2\pi)^3\delta^{(3)}(\qvec-\pvec -\ppvec)\psi^{\gamma^{\ast}\rightarrow q\bar{q}}\vert q(\pvec,h,\alpha)\bar{q}(\ppvec,h',\beta)\rangle\\
& + \int \ddp\dpp\dk(2\pi)^3\delta^{(3)}(\qvec-\pvec -\ppvec -\kvec)\psi^{\gamma^{\ast}\rightarrow q\bar{q}g}\vert q(\pvec,h,\alpha)\bar{q}(\ppvec,h',\beta)g(\kvec,\sigma,a)\rangle + \cdots \biggr ],
\end{split}              
\end{equation}
where $\vert \gamma^{\ast}(\qvec,Q^2,\lambda)\rangle_{\rm i}$ is the physical one particle state in the interaction picture and $\vert \gamma(\qvec,\lambda)\rangle_{\rm b}$ the corresponding free bare state. Note that the free bare states are defined by creation operators, depending  only on the spatial momentum $\qvec$, operating on the vacuum.  Thus the bare state $\vert \gamma(\qvec,\lambda)\rangle_{\rm b}$ is independent of $Q^2$ and on shell, as are all LCPT free states.
The full interacting theory state $\vert \gamma^{\ast}(\qvec,Q^2,\lambda)\rangle_{\rm i}$, on the other hand, ``knows'' that it has a virtuality $-Q^2$. This is reflected in the wave functions $\psi^{\gamma^{\ast}\rightarrow q\bar{q}}$ etc. via the energy denominators that depend on the light cone energy of the initial state\footnote{We thank G. Beuf for pointing this out to us.}.  We have ignored electromagnetic  contributions (i.e. $\gamma^{\ast}\rightarrow \ell \bar{\ell}$ and $\gamma^{\ast}\rightarrow \ell \bar{\ell}\gamma$, etc) since we are only interested in the order $\mathcal{O}(\alpha_\text{e.m.}\as)$ NLO correction to the order $\mathcal{O}(\alpha_\text{e.m.})$ leading order cross section.  The Fock states are defined as 
\begin{equation}
\begin{split}
\vert q(\pvec,h,\alpha)\bar{q}(\ppvec,h',\beta)\rangle &= b^{\dagger}(\pvec,h,\alpha)d^{\dagger}(\ppvec,h',\beta)\vert 0\rangle\\ 
\vert q(\pvec,h,\alpha)\bar{q}(\ppvec,h',\beta)g(\kvec,\sigma,a)\rangle & = b^{\dagger}(\pvec,h,\alpha)d^{\dagger}(\ppvec,h',\beta)a^{\dagger}(\kvec,\sigma,a)\vert 0\rangle\\
&\cdots 
\end{split}
\end{equation}
where the operators $b^{\dagger}$ ($d^{\dagger}$) create quark $q$ (anti-quark $\bar{q}$) with momentum $\pvec$ ($\ppvec$) and helicity $h$ ($h'$) and the fundamental color index $\alpha$ ($\beta$), and similarly $a^{\dagger}$ create gluon $g$ with momentum $\kvec$, helicity $\sigma$ and adjoint color index $a$. The normalization of the operators $b, d$ and $a$ is chosen such that commutation and anti-commutation rules in momentum space satisfy
\begin{equation}
\begin{split}
\{b(\pvec,h,\alpha),b^{\dagger}(\qvec,s,\beta)\} & = \{d(\pvec,h,\alpha),d^{\dagger}(\qvec,s,\beta)\} = 2p^+(2\pi)^3\delta^{(3)}(\pvec-\qvec)\delta_{h,s}\delta_{\alpha,\beta}\\
[a(\kvec,\sigma,a),a^{\dagger}(\qvec,s,b)] & = 2k^+(2\pi)^3\delta^{(3)}(\kvec-\qvec)\delta_{\sigma,s}\delta_{a,b}. 
\end{split}
\end{equation}
The renormalization constant $\sqrt{Z_{\gamma^{\ast}}}$ can be determined from the normalization requirement 
\begin{equation}
_{\text{int}}\langle \gamma^{\ast}(\qpvec,Q^2,\lambda') \vert \gamma^{\ast}(\qvec,Q^2,\lambda)\rangle_{\text{int}} = 2q^+(2\pi)^3\delta^{(3)}(\qpvec - \qvec)\delta_{\lambda',\lambda}. 
\end{equation}
However, since all the corrections to the photon wave function are proportional to the electromagnetic coupling, $Z= 1 + \mathcal{O}(\alpha_\text{e.m.})$. Thus working at lowest order in $\alpha_\text{e.m.}$ we can drop the photon wave function renormalization.


The Fock state representation in momentum space $(k^+,\kt)$ is switched to the mixed space representation $(k^+,\xt)$ by the transverse Fourier transform of all the creation operators present in the state, with 
\begin{equation}
\label{eq:ftransformoperators}
\begin{split}
a^{\dagger}(\kvec,\sigma,a) &= \int_{\xt} e^{i\kt\cdot\xt}a^{\dagger}(k^+,\xt,\sigma,a)\\
b^{\dagger}(\pvec,h,\alpha) &= \int_{\xt} e^{i\pt\cdot\xt}b^{\dagger}(p^+,\xt,h,\alpha)\\
d^{\dagger}(\pvec,h,\alpha) &= \int_{\xt} e^{i\pt\cdot\xt}d^{\dagger}(p^+,\xt,h,\alpha),\\
\end{split}
\end{equation}
where 
\begin{equation}
\int_{\xt} = \int \ud^2\xt.
\end{equation}
The mixed space operators satisfy
\begin{equation}
\label{eq:normalizationMS}
\begin{split}
\{b(p^+,\xt,h,\alpha),b^{\dagger}(q^+,\yt,s,\beta)\} & = \{d(p^+,\xt,h,\alpha),d^{\dagger}(q^+,\yt,s,\beta)\} = 2p^+(2\pi)\delta(p^+-q^+)\delta^{(2)}(\xt-\yt)\delta_{h,s}\delta_{\alpha,\beta}\\
[a(k^+,\xt,\sigma,a),a^{\dagger}(q^+,\yt,s,b)] & = 2k^+(2\pi)\delta(k^+-q^+)\delta^{(2)}(\xt-\yt)\delta_{\sigma,s}\delta_{a,b}. 
\end{split}
\end{equation}
Since the high energy scattering of the projectile partons  off the gluon target is eikonal, the scattering operator $\hat{S}_{E}$ acts on Fock states by only color rotating each partons by a Wilson line defined along the partons trajectory through the target: 
\begin{equation}
\label{eq:eikonaloperatorqqbar}
\hat{S}_{E}b^{\dagger}(p^+,\xt,h,\alpha)d^{\dagger}(p'^+,\yt,h',\beta)\vert 0\rangle = \sum_{\bar{\alpha},\bar{\beta}}[U[A](\xt)]_{\bar{\alpha}\alpha}[U^{\dagger}[A](\yt)]_{\beta\bar{\beta}}b^{\dagger}(p^+,\xt,h,\bar\alpha)d^{\dagger}(p'^+,\yt,h',\bar\beta)\vert 0\rangle
\end{equation}
and 
\begin{equation}
\label{eq:eikonaloperatorqqbarg}
\begin{split}
\hat{S}_{E}b^{\dagger}(p^+,\xt,h,\alpha)d^{\dagger}(p'^+,\yt,h',\beta)a^\dagger(k^+,\zt,\sigma,a)\vert 0\rangle = \sum_{\bar{\alpha},\bar{\beta},b}&[U[A](\xt)]_{\bar{\alpha}\alpha}[U^{\dagger}[A](\yt)]_{\beta\bar{\beta}}[V[A](\zt)]_{ba}\\
& \times b^{\dagger}(p^+,\xt,h,\bar\alpha)d^{\dagger}(p'^+,\yt,h',\bar\beta)a^\dagger(k^+,\zt,\sigma,b)\vert 0\rangle,
\end{split}
\end{equation}
where the fundamental and adjoint Wilson line are respectively defined as the path ordered exponential for a classical gluon target $A$:
\begin{equation}
\begin{split}
U[A](\xt) & = \mathcal{P}\exp\biggl [ig\int \ud x^+ t^a A^{-}_a(x^+,0,\xt) \biggr ]\\
V[A](\xt) & = \mathcal{P}\exp\biggl [ig\int \ud x^+ T^a A^{-}_a(x^+,0,\xt) \biggr ].
\end{split}
\end{equation}

Applying \eq\nr{eq:ftransformoperators}, we can define the amplitudes corresponding to the $q\bar{q}$-component and $q\bar{q}g$-component of the Fock state decomposition in the mixed space:
\begin{equation}
\label{eq:mixedspaceqbarqfull}
\vert \gamma^{\ast}(q^+,Q^2,\lambda)\rangle_{q\bar{q}}  =  \mathcal{PS}^{+}_{(2)}\int_{\xt \yt} 
\;
\widetilde{\psi}^{\gamma^{\ast}\rightarrow q\bar{q}}
\vert q(p^+,\xt,h,\alpha)\bar{q}(p'^+,\yt,h',\beta)\rangle
\end{equation}
\begin{equation}
\label{eq:mixedspaceqbarqgfull}
\vert \gamma^{\ast}(q^+,Q^2,\lambda)\rangle_{q\bar{q}g} = \mathcal{PS}^{+}_{(3)}\int_{\xt\yt\zt }
\;
\widetilde{\psi}^{\gamma^{\ast}\rightarrow q\bar{q}g}
\vert q(p^+,\xt, h,\alpha)\bar{q}(p'^+,\yt,h',\beta)g(k^+,\zt,\sigma,a)\rangle,
\end{equation}
where the two and three particle longitudinal phase space factors  $\mathcal{PS}^{+}_{(2)}$ and $\mathcal{PS}^{+}_{(3)}$, respectively, are defined as 
\begin{equation}
\label{eq:plusmomPS}
\begin{split}
\mathcal{PS}^{+}_{(2)} &= \int_0^\infty\frac{\ud p^+}{2p^+(2\pi)}\int_0^\infty\frac{\ud p'^+}{2p'^+(2\pi)}(2\pi)\delta(q^+-p^+-p'^+)
\\
\mathcal{PS}^{+}_{(3)} &= \int_0^\infty\frac{\ud p^+}{2p^+(2\pi)}\int_0^\infty\frac{\ud p'^+}{2p'^+(2\pi)}\int_0^\infty\frac{\ud k^+}{2k^+(2\pi)}(2\pi)\delta(q^+-p^+-p'^+ - k^+)
.
\end{split}
\end{equation} 
The mixed space wave functions $\widetilde\psi$ are transverse Fourier transforms of the  LCWF's:
\begin{equation}
\label{eq:mixedspaceqbarqfullphi}
\widetilde{\psi}^{\gamma^{\ast}\rightarrow q\bar{q}}
\left(\xt,p^+,\yt,p'^{+}\right)
= \int\frac{\ud^2\pt}{(2\pi)^2}\int\frac{\ud^2\ptp}{(2\pi)^2}(2\pi)^2\delta^{(2)}(\qt-\pt -\ptp)
\psi^{\gamma^{\ast}\rightarrow q\bar{q}}\left(\pt,p^+,\ptp,p'^{+}\right)
e^{i\pt\cdot \xt}e^{i\ptp\cdot \yt}
\end{equation}
and
\begin{multline}
\label{eq:mixedspaceqbarqgfullphi}
\widetilde{\psi}^{\gamma^{\ast}\rightarrow q\bar{q}g}
\left(\xt,p^+,\yt,p'^{+},\zt,k^+\right)
= \int\frac{\ud^2\pt}{(2\pi)^2}\int\frac{\ud^2\ptp}{(2\pi)^2}\int\frac{\ud^2\kt}{(2\pi)^2}(2\pi)^2\delta^{(2)}(\qt-\pt -\ptp-\kt)
\\
\psi^{\gamma^{\ast}\rightarrow q\bar{q}g}
\left(\pt,p^+,\ptp,p'^{+},\kt,k^+\right)
e^{i\pt\cdot \xt}e^{i\ptp\cdot \yt}e^{i\kt\cdot \zt}.
\end{multline}

Using the shorthand notation for the different Fock state components the virtual photon state \nr{eq:vpfockdecompostion} is
\begin{equation}
\vert \gamma^{\ast}(q^+,Q^2,\lambda)\rangle_{\rm i} = 
\vert \gamma(q^+,\lambda)\rangle_{\rm b} +
\vert \gamma^{\ast}(q^+,Q^2,\lambda)\rangle_{q\bar{q}} + \vert \gamma^{\ast}(q^+,Q^2,\lambda)\rangle_{q\bar{q}g} + \dots,
\end{equation}
with the two last terms on the right-hand side given in mixed space by \eqs\nr{eq:mixedspaceqbarqfull} and \nr{eq:mixedspaceqbarqgfull}. In this decomposition the photon cross section at NLO accuracy can be written as 
\begin{equation}
\label{eq:CSfinal}
\begin{split}
\sigma^{\gamma^{\ast}}[A]  = \frac{2}{2q^+(2\pi)\delta(q'^+-q^+)}\biggl \{ {}_{q\bar{q}}\langle &\gamma^{\ast}(q'^+,Q^2,\lambda')\vert1-\hat{S}_{E}\vert\gamma^{\ast}(q^+,Q^2,\lambda)\rangle_{q\bar{q}}\\
& + {}_{q\bar{q}g}\langle \gamma^{\ast}(q'^+,Q^2,\lambda')\vert1-\hat{S}_{E}\vert\gamma^{\ast}(q^+,Q^2,\lambda)\rangle_{q\bar{q}g} \biggr \}.
\end{split}
\end{equation}
Here the $q\bar{q}$-component contains the leading order (LO) contribution and the NLO contribution coming from the the one-loop virtual diagrams,  and the $q\bar{q}g$-component contains the NLO contribution coming from the radiative correction diagrams.
 
We shall now set out to calculate the wave function
$\widetilde{\psi}^{\gamma^{\ast}\rightarrow q\bar{q}}$ to one-	loop accuracy
 and
$\widetilde{\psi}^{\gamma^{\ast}\rightarrow q\bar{q}g}$ at tree level, and using these results return to the cross section~\nr{eq:CSfinal} in Sec.~\ref{sec:nlodis}.

\section{Leading order wave function}
\label{sec:lo}

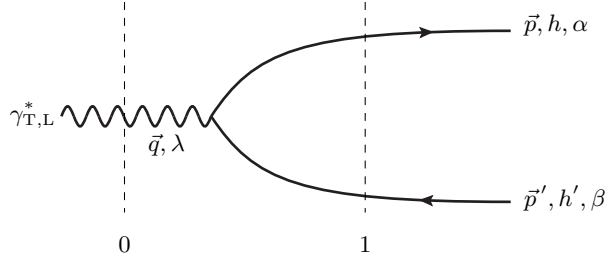
\begin{figure}[tb]
\centerline{
\includegraphics[width=6.4cm]{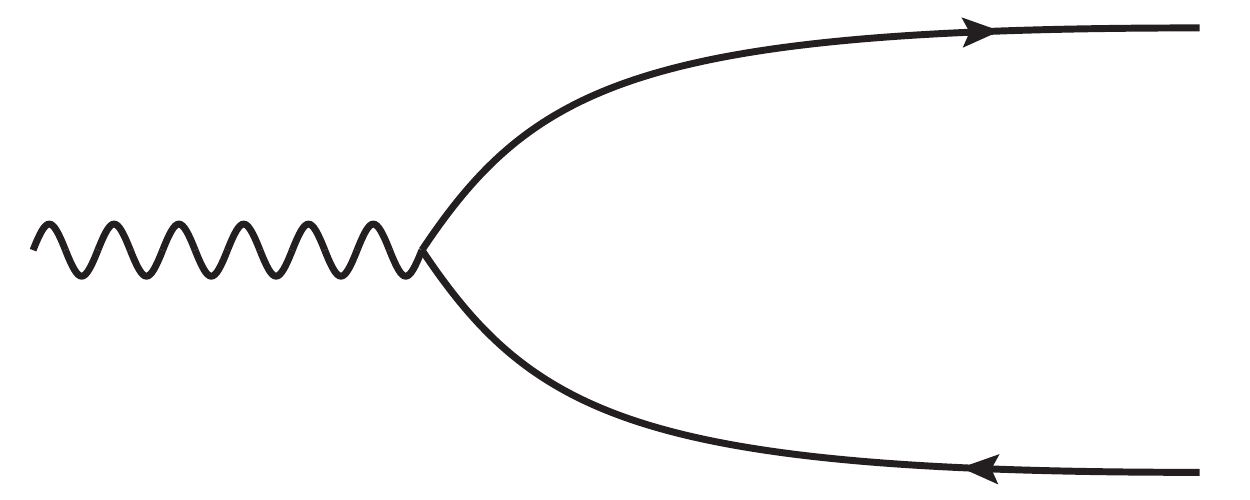}
\begin{tikzpicture}[overlay]
\draw [dashed] (-5.5,2.8) -- (-5.5,0);
\node[anchor=north] at (-5.5cm,-0.2cm) {0};
\draw [dashed] (-2.3,2.8) -- (-2.3,0);
\node[anchor=north] at (-2.3cm,-0.2cm) {1};
\node[anchor=west] at (-0.3,2.5) {$\pvec,h,\alpha$};
\node[anchor=west] at (-0.3,0.2) {$\ppvec,h',\beta$};
\node[anchor=west] at (-5.3,0.9) {$\qvec, \lambda $};
\node[anchor=east] at (-6.3,1.3) {$\gamma^{\ast}_{\rm T,L}$};
 \end{tikzpicture}
}
\rule{0pt}{1ex}
\caption{Time ordered (momenta flows from left to right) diagram contributing to the $q\bar{q}$-component of the transverse and longitudinal virtual photon wave function at leading order with energy denominators and kinematics. Momentum conservation: $\qvec=\pvec+\ppvec$. The longitudinal momentum fractions for quark and anti-quark are parametrized as $p^+ = zq^+$ and $p'^+ = (1-z)q^+$.}
\label{fig:lovertex}
 \end{figure}

The leading order $\gamma^*\to q\bar{q}$ wave functions shown in \fig\ref{fig:lovertex} are well known, but we will briefly write them down here to set the normalization in our conventions. Following the diagrammatic rules listed in \cite{Lappi:2016oup} the light cone wave function contributing to the transverse or longitudinal virtual photon splitting into a quark anti-quark dipole is given by 
\begin{equation}
\label{eq:lowf}
\psi_{\rm LO}^{\gamma^{\ast}_{\rm T/L}\rightarrow q\bar{q}} = \frac{-ee_f\delta_{\alpha\beta}}{\Delta_{01}^{-}}\biggl [\bar{u}_h(p)\epsl_{\lambda,\mathrm{T/L}}(q)v_{h'}(p') \biggr ],
\end{equation}
where the LC energy denominator can be cast in the following form
\begin{equation}
\begin{split}
\label{eq:loed}
\Delta_{01}^{-}  = q^- - (p^- + p'^-) &= -\frac{Q^2}{2q^+} - \left (\frac{\pt^2}{2p^+} + \frac{\pt^2}{2p'^+}\right )\\
& = \frac{1}{(-2q^+)z(1-z)}\biggl [\pt^2 + \overline{Q}^2 \biggr ]
\end{split}
\end{equation} 
with $\overline{Q}^2 = z(1-z)Q^2$. Note that we are working in a frame where the transverse momentum of the photon is zero and thus $\pt=-\ptp$; otherwise the transverse momentum argument $\pt$ would be replaced by the center of mass momentum $\pt-z\qt$.

\subsection{Transversely polarized virtual photon}

In the $\qt = 0$ frame the polarization vector for a transversely polarized virtual photon in the LC gauge is given by 
\begin{equation}
\label{eq:epsilont}
\varepsilon^{\mu}_{\lambda, \mathrm{T}}(q) = (0, \frac{\qt \cdot \epst_\lambda}{q^+},\epst_\lambda) = (0,0,\epst_\lambda).
\end{equation} 
Using \eq\nr{eq:epsilont} the  light cone wave function for the transversely polarized virtual photon in \eq\nr{eq:lowf} can be expressed in the explicit helicity basis as 
\begin{equation}
\label{eq:lowfT}
\psi_{\rm LO}^{\gamma^{\ast}_{\rm T}\rightarrow q\bar{q}}(\pt,z) = \frac{\delta_{\alpha\beta}}{\Delta_{01}^{-}}\paircr_{\lambda;h,h'}^{\gamma^{\ast}_{\rm T}}(\pt,z)
\end{equation} 
with $\paircr_{\lambda;h,h'}^{\gamma^{\ast}_{\rm T}}$ defined as for the gluon vertex in 
\nr{eq:vertexgtoqqb}:
\begin{equation}
\label{eq:LOTvirtualphoton}
\paircr_{\lambda;h,h'}^{\gamma^{\ast}_{\rm T}}(\pt,z) = \frac{-2ee_f}{\sqrt{z(1-z)}}\biggl [
\left (z-\frac{1}{2}\right)\delta^{ij}\delta_{h,-h'} + \frac{1}{4}\paircras^{ij}_{h,h'}
\biggr ]\pt^i \epst^j_\lambda.
\end{equation}
Here we have kept the helicity notation as general as possible. However, for a massless quarks the helicity is conserved in the light cone vertices. This implies that for the $\gamma q\bar{q}$-vertex $h = -h'$.

\subsection{Longitudinally polarized virtual photon}

Strictly speaking there is no such thing as a longitudinal photon in the spectrum of physical states in the theory. In stead, a longitudinal photon in DIS is a part of an instantaneous interaction vertex with the lepton. However, for calculational purposes  we will here leave out the lepton and simply define a longitudinal virtual photon polarization vector, treating the longitudinal photon analogously to the transverse one.
The polarization vector for a  longitudinally polarized 
virtual photon in the LC gauge can be expressed as 
\begin{equation}
\varepsilon^{\mu}_{\lambda, \mathrm{L}}(q) = (0, \frac{\sqrt{Q^2-\qt^2}}{q^+}, \frac{\qt}{\sqrt{Q^2-\qt^2}}) = (0,\frac{Q}{q^+},\mathbf{0}).
\end{equation} 
Thus the light cone wave function for the longitudinally polarized virtual photon in \eq\nr{eq:lowf} can be cast in the following form
\begin{equation}
\label{eq:lowfL}
\psi_{\rm LO}^{\gamma^{\ast}_{\rm L}\rightarrow q\bar{q}}(Q,z) = \frac{\delta_{\alpha\beta}}{\Delta_{01}^{-}}\paircr_{\lambda;h,h'}^{\gamma^{\ast}_{\rm L}}(Q,z),
\end{equation} 
where 
\begin{equation}
\label{eq:LOLvirtualphoton}
\paircr_{\lambda;h,h'}^{\gamma^{\ast}_{\rm L}}(Q,z) = -ee_f \frac{Q}{q^+}\biggr [\bar{u}_h(p)\gamma^+v_{h'}(p') \biggr ] = -2ee_f\frac{Q}{q^+}\sqrt{p^+p'^+}\delta_{h,-h'} = -2ee_fQ\sqrt{z(1-z)}\delta_{h,-h'}.
\end{equation}

\section{Virtual photon LCWF'S}
\label{sec:loop}

\subsection{Quark self-energy diagrams}

\begin{figure}[tb]
\centerline{
\includegraphics[width=6.4cm]{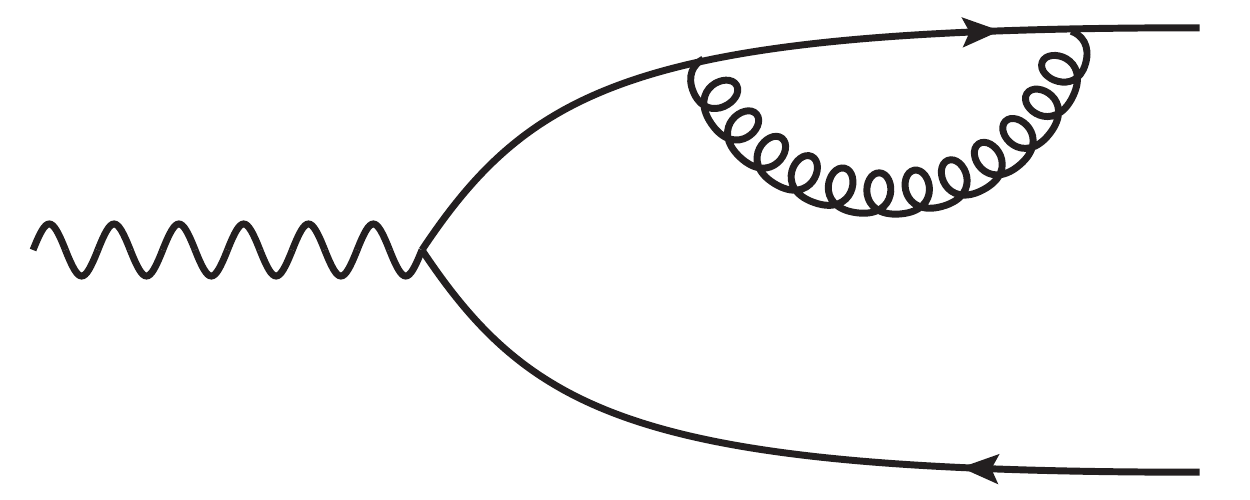}
\begin{tikzpicture}[overlay]
\draw [dashed] (-5.5,2.8) -- (-5.5,0);
\node[anchor=north] at (-5.5cm,-0.2cm) {0};
\draw [dashed] (-3.3,2.8) -- (-3.3,0);
\node[anchor=north] at (-3.3cm,-0.2cm) {1};
\draw [dashed] (-2.0,2.8) -- (-2.0,0);
\node[anchor=north] at (-2.0cm,-0.2cm) {2};
\draw [dashed] (-0.6,2.8) -- (-0.6,0);
\node[anchor=north] at (-0.6cm,-0.2cm) {3};
\node[anchor=west] at (-0.3,2.5) {$\pvec,h,\alpha$};
\node[anchor=west] at (-2.0,1.2) {$\kvec,\sigma,a$};
\node[anchor=west] at (-2.9,2.6) {$\kpvec,h,\bar\alpha$};
\node[anchor=west] at (-4.9,2.3) {$\pppvec,h,\beta$};
\node[anchor=west] at (-2.6,2.0) {$\mt$};
\node[anchor=west] at (-4.0,1.3) {$\pt$};
\node[anchor=west] at (-0.3,0.2) {$\ppvec,-h,\beta$};
\node[anchor=west] at (-5.3,0.9) {$\qvec, \lambda $};
\node[anchor=east] at (-6.3,1.3) {$\gamma^{\ast}_{\rm T/L}$};
\node[anchor=south west] at (-7cm,0cm) {\namediag{diag:oneloopSEUPT}};
 \end{tikzpicture}
}
\rule{0pt}{1ex}
\caption{Quark self-energy  diagram \ref{diag:oneloopSEUPT} contributing to the $q\bar{q}$-component of the transverse  virtual photon wave function at NLO with energy denominators and kinematics. Momentum conservation: $\qvec=\pvec+\ppvec$, $\pppvec = \kpvec + \kvec$ and $ \kpvec + \kvec = \pvec$. The longitudinal momentum fractions for quark and anti-quark are parametrized as $p^+ = zq^+$ and $p'^+ = (1-z)q^+$. The momentum fraction of the  virtual photon splitting into a $q\bar{q}$ dipole is $p^+/q^+ = z$ and the natural momentum is $\pt - z\qt = \pt$ (note $\qt = \ot$).
The momentum fraction of the gluon emission and absorption is $k^+/p^+ = z'/z$, and the natural momentum in the gluon loop is $\mt =  \kt - (z'/z)\pt$.
}
\label{fig:oneloopSEUPT}
 \end{figure}

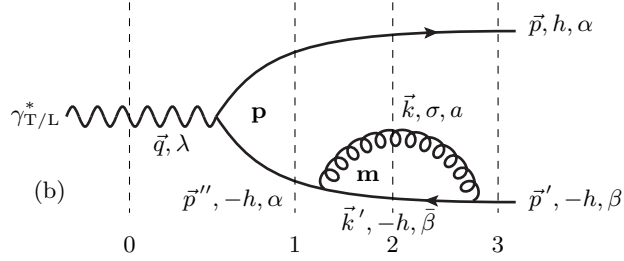
\begin{figure}[tb]
\centerline{
\includegraphics[width=6.4cm]{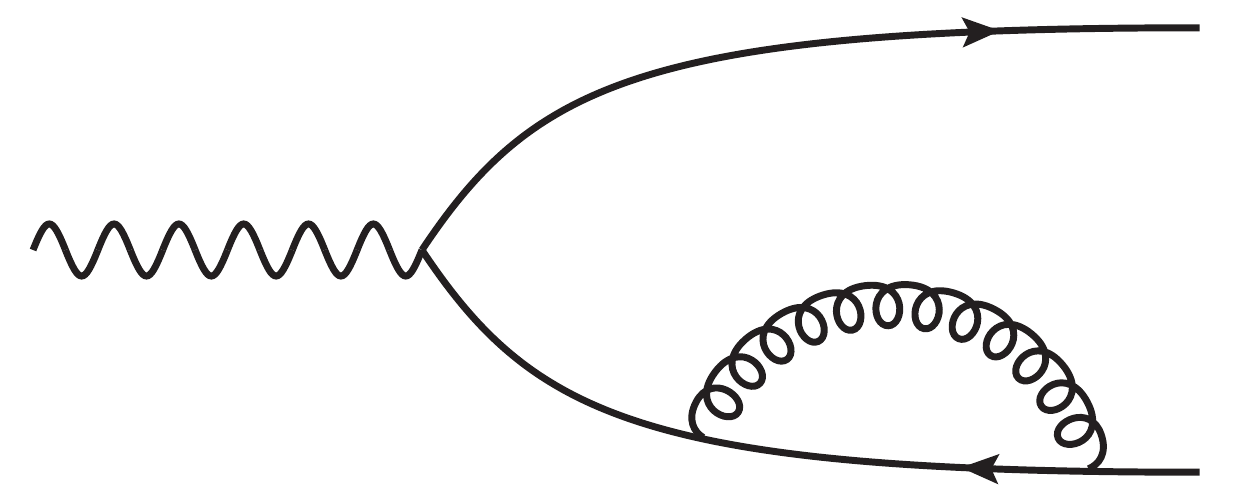}
\begin{tikzpicture}[overlay]
\draw [dashed] (-5.5,2.8) -- (-5.5,0);
\node[anchor=north] at (-5.5cm,-0.2cm) {0};
\draw [dashed] (-3.3,2.8) -- (-3.3,0);
\node[anchor=north] at (-3.3cm,-0.2cm) {1};
\draw [dashed] (-2.0,2.8) -- (-2.0,0);
\node[anchor=north] at (-2.0cm,-0.2cm) {2};
\draw [dashed] (-0.6,2.8) -- (-0.6,0);
\node[anchor=north] at (-0.6cm,-0.2cm) {3};
\node[anchor=west] at (-0.3,2.5) {$\pvec,h,\alpha$};
\node[anchor=west] at (-2.0,1.4) {$\kvec,\sigma,a$};
\node[anchor=west] at (-2.8,-0.1) {$\kpvec,-h,\bar\beta$};
\node[anchor=west] at (-2.6,0.5) {$\mt$};
\node[anchor=west] at (-4.9,0.2) {$\pppvec,-h,\alpha$};
\node[anchor=west] at (-4.0,1.3) {$\pt$};
\node[anchor=west] at (-0.3,0.2) {$\ppvec,-h,\beta$};
\node[anchor=west] at (-5.3,0.9) {$\qvec, \lambda $};
\node[anchor=east] at (-6.3,1.3) {$\gamma^{\ast}_{\rm T/L}$};
\node[anchor=south west] at (-7cm,0cm) {\namediag{diag:oneloopSEDOWNT}};
 \end{tikzpicture}
}
\rule{0pt}{1ex}
\caption{Antiquark self-energy diagram \ref{diag:oneloopSEDOWNT} contributing to the $q\bar{q}$-component of the transverse  virtual photon wave function at NLO with energy denominators and kinematics. Momentum conservation: $\qvec=\pvec+\ppvec$, $\pppvec = \kpvec + \kvec$ and $\kpvec + \kvec = \ppvec$. The longitudinal momentum fractions for quark and anti-quark are parametrized as $p^+ = zq^+$ and $p'^+ = (1-z)q^+$. The momentum fraction of the  virtual photon splitting vertex is $p^+/q^+ = z$ and the natural momentum $\pt - z\qt = \pt$ (note $\qt = \ot$). 
The momentum fraction of the gluon emission and absorption is $k^+/p'^+ = z'/(1-z)$, and the natural momentum in the gluon loop is $\mt =  \kt - (z'/(1-z))\pt$.
}
\label{fig:oneloopSEDOWNT}
\end{figure}

The quark self energy diagrams are explicitly proportional to the leading order $\gamma^*\to q\bar{q}$ vertex, so one does not need to calculate them separately for the different virtual photon polarizations. There are two diagrams that contribute, the ones shown in \figs\ref{fig:oneloopSEUPT} and~\ref{fig:oneloopSEDOWNT}.
 
The LCWF for quark self-energy diagram \ref{diag:oneloopSEUPT} shown in  \fig\ref{fig:oneloopSEUPT} is given by
\begin{equation}
\label{eq:LCWFTLa}
\begin{split}
\psi^{\gamma^{\ast}_{\rm T/L}\rightarrow q\bar{q}}_{\ref{diag:oneloopSEUPT}} = \int \dk\dkp\dppp (2\pi)^{d-1}\delta^{(d-1)}&(\pppvec-\kpvec -\kvec)(2\pi)^{d-1}\delta^{(d-1)}(\kpvec + \kvec - \pvec)
\frac{1}{\Delta_{01}^{-}\Delta_{02}^{-}\Delta_{03}^{-}}\mathrm{num}\bigg\vert_{\ref{diag:oneloopSEUPT}},
\end{split}
\end{equation}
where the Lorentz invariant measure in $d$ dimensions is defined as $\dk \equiv \frac{\ud k^+ \ud^{d-2}\kt}{2k^+(2\pi)^{d-1}}$, and the  numerator for the transversersally polarized virtual photon becomes
\begin{equation}
\label{eq:numa}
\mathrm{num}\bigg\vert_{\ref{diag:oneloopSEUPT}} = V^{\bar{\alpha},a;\alpha}_{h,\sigma;h}(\mt,z'/z)V^{\beta;\bar{\alpha},a}_{h;h,\sigma}(\mt,z'/z)\paircr^{\gamma^{\ast}_{\rm T}}_{\lambda,h,-h}(\pt,z).
\end{equation}
Correspondingly, the numerator for longitudinally polarized virtual photon in \eq\nr{eq:LCWFTLa} is obtained by the trivial replacement 
\begin{equation}
\paircr^{\gamma^{\ast}_{\rm T}}_{\lambda,h,-h}(\pt,z) \rightarrow \paircr^{\gamma^{\ast}_{\rm L}}_{\lambda,h,-h}(Q,z).
\end{equation}
In \eq\nr{eq:numa} the vertex for the  virtual photon splitting into a $q\bar{q}$ dipole is given in \eq\nr{eq:LOTvirtualphoton} or \eq\nr{eq:LOLvirtualphoton} depending on the polarization and the gluon emission and absorption vertices 
$V^{\beta;\bar{\alpha},a}_{h;h\sigma}(\mt,z'/z)$ and $V^{\bar{\alpha},a;\alpha}_{h\sigma;h}(\mt,z'/z)$ are given by \eqs\nr{eq:vertexqtoqgd} and~\nr{eq:vertexqgtoqd}. The LC energy denominators in \eq\nr{eq:LCWFTLa} are 
\begin{equation}
\Delta_{01}^{-} = \Delta_{03}^{-} = \frac{1}{(-2q^+)z(1-z)}\biggl [\pt^2 + \overline{Q}^2\biggr ]
\end{equation}
and 
\begin{equation}
\Delta_{02}^{-} =\frac{z}{(-2q^+)z'(z-z')}\biggl [\mt^2 + M^{\ref{diag:oneloopSEUPT}}\biggr ] \quad \text{with}\quad  M^{\ref{diag:oneloopSEUPT}} = \frac{z'(z-z')}{z^2(1-z)}\left (\pt^2 +\overline{Q}^2 \right ).
\end{equation}
The phase space measure simplifies to 
\begin{equation}
\int \dk\dkp\ddp (2\pi)^{d-1}\delta^{(d-1)}(\kpvec - \pvec + \kvec)(2\pi)^{d-1}\delta^{(d-1)}(\pvec - \kpvec - \kvec)
=\frac{1}{16\pi(q^+)^2}\int\frac{\ud z'}{zz'(z-z')}\int \frac{\ud^{d-2}\mt}{(2\pi)^{d-2}},
\end{equation}
where $k^+ > 0$ and $k'^+ > 0$, so that  $0< z'< z$. 
 
Performing the helicity sums as described in Sec.~\ref{sec:numerators}, and integrating over the loop transverse momentum
(with dimensionally regularized integrals given in Appendix~\ref{app:loopints})
and the longitudinal momentum, regulating the soft divergence with $\alpha < z' < z$, we obtain 
 \begin{equation}
\label{eq:oneloopSEUPT}
\psi^{\gamma^{\ast}_{\rm T}\rightarrow q\bar{q}}_{\ref{diag:oneloopSEUPT}} = \psi^{\gamma^{\ast}_{\rm T}\rightarrow q\bar{q}}_{\rm LO}(\pt,z)\left (\frac{g_r^2\cf}{8\pi^2}\right )\biggr \{\biggl [\frac{3}{2} + 2\log \left (\frac{\alpha}{z}\right ) \biggr ]C_{\ref{diag:oneloopSEUPT}} - \log^2\left (\frac{\alpha}{z}\right ) -\frac{\pi^2}{3} + 3 \biggr \} + \mathcal{O}(\varepsilon),
\end{equation}
where 
\begin{equation}
C_{\ref{diag:oneloopSEUPT}} = \frac{1}{\varepsilon_{\overline{\rm MS}}} + \log \left (\frac{\mu^2}{\overline{Q}^2}\right ) - \log\left (\frac{\pt^2 + \overline{Q}^2}{\overline{Q}^2}\right ) + \log(1-z),
\end{equation}
with $\epsmsbar= 1/\varepsilon -\gamma_\text{E}+\ln (4\pi)$.
Similarly for the longitudinally polarized virtual photon,
 \begin{equation}
\label{eq:oneloopSEUPL}
\psi^{\gamma^{\ast}_{\rm L}\rightarrow q\bar{q}}_{\ref{diag:oneloopSEUPT}} = \psi^{\gamma^{\ast}_{\rm L}\rightarrow q\bar{q}}_{\rm LO}(Q,z)\left (\frac{g_r^2\cf}{8\pi^2}\right )\biggr \{\biggl [\frac{3}{2} + 2\log \left (\frac{\alpha}{z}\right ) \biggr ]C_{\ref{diag:oneloopSEUPT}} - \log^2\left (\frac{\alpha}{z}\right ) -\frac{\pi^2}{3} + 3 \biggr \} + \mathcal{O}(\varepsilon).
\end{equation}
The LCWF for diagram \ref{diag:oneloopSEDOWNT} shown in \fig\ref{fig:oneloopSEDOWNT} can be now easily obtained by using the symmetry  between the diagrams \ref{diag:oneloopSEUPT} and \ref{diag:oneloopSEDOWNT} (i.e. by making the substitution $z \leftrightarrow 1-z$ and $\pt \rightarrow -\pt$ simultaneously) as 
\begin{equation}
\label{eq:oneloopSEDOWNT}
\psi^{\gamma^{\ast}_{\rm T}\rightarrow q\bar{q}}_{\ref{diag:oneloopSEDOWNT}} = \psi^{\gamma^{\ast}_{\rm T}\rightarrow q\bar{q}}_{\rm LO}(\pt,z)\left (\frac{g_r^2\cf}{8\pi^2}\right )\biggr \{\biggl [\frac{3}{2} + 2\log \left (\frac{\alpha}{1-z}\right ) \biggr ]C_{\ref{diag:oneloopSEDOWNT}}  - \log^2\left (\frac{\alpha}{1-z}\right )  -\frac{\pi^2}{3} + 3\biggr \}
\end{equation}
and
\begin{equation}
\label{eq:oneloopSEDOWNL}
\psi^{\gamma^{\ast}_{\rm L}\rightarrow q\bar{q}}_{\ref{diag:oneloopSEDOWNT}} = \psi^{\gamma^{\ast}_{\rm L}\rightarrow q\bar{q}}_{\rm LO}(Q,z)\left (\frac{g_r^2\cf}{8\pi^2}\right )\biggr \{\biggl [\frac{3}{2} + 2\log \left (\frac{\alpha}{1-z}\right ) \biggr ]C_{\ref{diag:oneloopSEDOWNT}}  - \log^2\left (\frac{\alpha}{1-z}\right )  -\frac{\pi^2}{3} + 3\biggr \}
\end{equation}
with
\begin{equation}
C_{\ref{diag:oneloopSEDOWNT}} = \frac{1}{\varepsilon_{\overline{\rm MS}}} + \log \left (\frac{\mu^2}{\overline{Q}^2}\right ) - \log\left (\frac{\pt^2 + \overline{Q}^2}{\overline{Q}^2}\right ) + \log(z)  .
\end{equation}

\subsection{Transverse photons}

\begin{figure}[t]
\centerline{
\includegraphics[width=6.4cm]{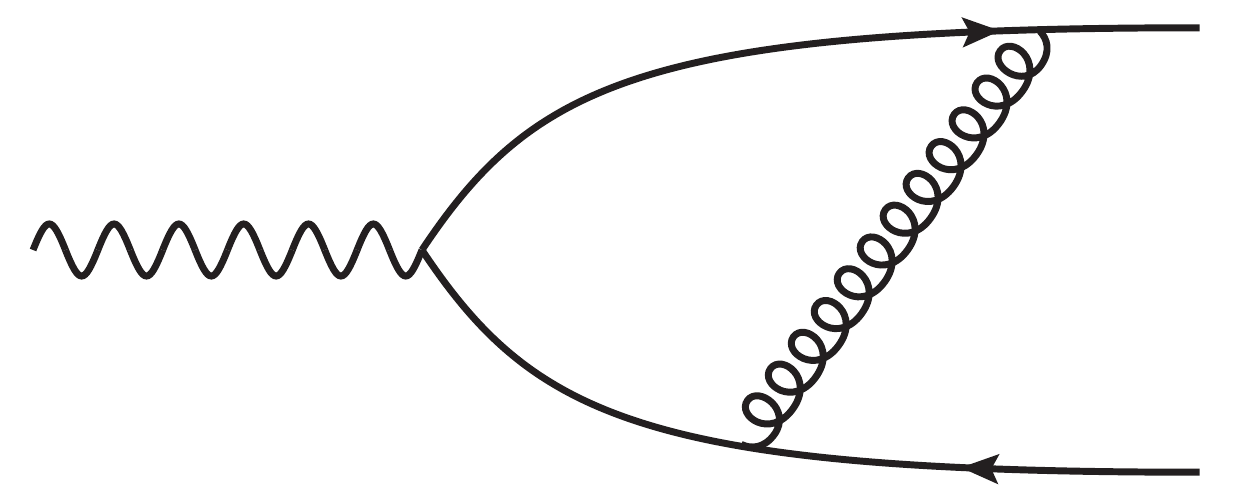}
\begin{tikzpicture}[overlay]
\draw [dashed] (-5.5,2.8) -- (-5.5,0);
\node[anchor=north] at (-5.5cm,-0.2cm) {0};
\draw [dashed] (-3.3,2.8) -- (-3.3,0);
\node[anchor=north] at (-3.3cm,-0.2cm) {1};
\draw [dashed] (-2.0,2.8) -- (-2.0,0);
\node[anchor=north] at (-2.0cm,-0.2cm) {2};
\draw [dashed] (-0.6,2.8) -- (-0.6,0);
\node[anchor=north] at (-0.6cm,-0.2cm) {3};
\node[anchor=west] at (-0.3,2.5) {$\pvec,h,\alpha$};
\node[anchor=west] at (-1.7,1.3) {$\kvec,\sigma,a$};
\node[anchor=west] at (-4.8,0.2){$\kppvec,h'',\bar\alpha$};
\node[anchor=west] at (-3.3,2.6) {$\kpvec,h',\bar\alpha$};
\node[anchor=west] at (-2.4,0.5) {$\hht$};
\node[anchor=west] at (-2.0,2.2) {$\mt$};
\node[anchor=west] at (-4.0,1.3) {$\ktp$};
\node[anchor=west] at (-0.3,0.2) {$\ppvec,-h,\beta$};
\node[anchor=west] at (-5.3,0.9) {$\qvec, \lambda $};
\node[anchor=east] at (-6.3,1.3) {$\gamma^{\ast}_{\rm T}$};
\node[anchor=south west] at (-7cm,0cm) {\namediag{diag:vertexqbaremT}};
 \end{tikzpicture}
}
\rule{0pt}{1ex}
\caption{Vertex diagram \ref{diag:vertexqbaremT} contributing to the $q\bar{q}$-component of the transverse  virtual photon wave function at NLO with energy denominators and kinematics. Momentum conservation: $\qvec=\kpvec+\kppvec=\pvec + \ppvec$, $\kpvec = \pvec- \kvec$ and $\kppvec = \kvec + \ppvec$. The longitudinal momentum fractions for quark and anti-quark are parametrized as $p^+ = zq^+$ and $p'^+ = (1-z)q^+$, and for the gluon in the loop $k^+ = z'q^+$ with $k'^+ = (z-z')q^+$ and $k''^+ = (1-z+z')q^+$. The longitudinal momentum fraction of the  virtual photon splitting into a $q\bar{q}$ dipole is $k'^+/q^+ = z-z'$ and the natural momentum is $\ktp$. The momentum fraction of the gluon emission is $k^+/k''^+ = z'/(1-z+z')$ and gluon absorption $k^+/p^+ = z'/z$. The natural momentum for the gluon emission is $\hht =  \kt - (z'/(1-z+z'))\ktpp$ and for the gluon absorption $\mt = \kt - (z'/z)\pt$. In order to use $\mt$ as the integration variable we need to know that $\ktp = -\mt + ((z-z')/z)\pt$ and $\hht = ((1-z)/(1-z+z'))(\mt + (z'/(z(1-z))\pt)$.}
\label{fig:vertexqbaremT}
\end{figure}

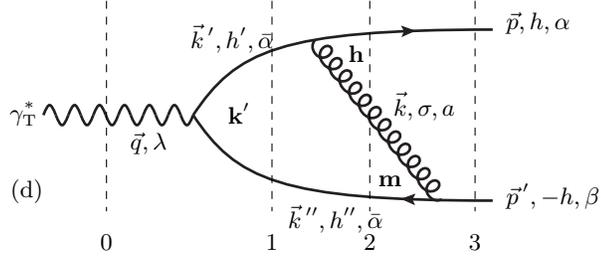
\begin{figure}[tb]
\centerline{
\includegraphics[width=6.4cm]{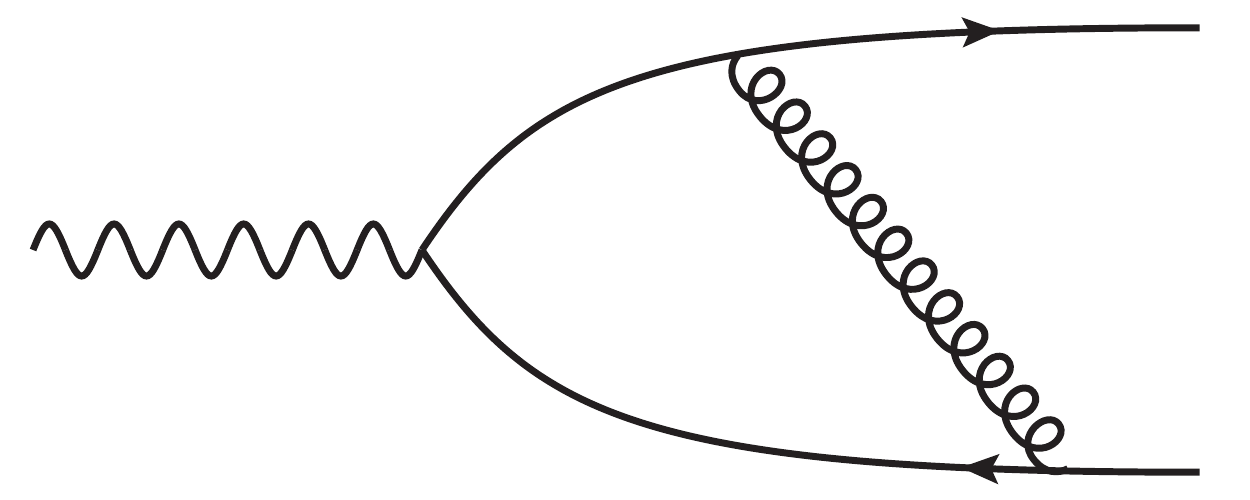}
\begin{tikzpicture}[overlay]
\draw [dashed] (-5.5,2.8) -- (-5.5,0);
\node[anchor=north] at (-5.5cm,-0.2cm) {0};
\draw [dashed] (-3.3,2.8) -- (-3.3,0);
\node[anchor=north] at (-3.3cm,-0.2cm) {1};
\draw [dashed] (-2.0,2.8) -- (-2.0,0);
\node[anchor=north] at (-2.0cm,-0.2cm) {2};
\draw [dashed] (-0.6,2.8) -- (-0.6,0);
\node[anchor=north] at (-0.6cm,-0.2cm) {3};
\node[anchor=west] at (-0.3,2.5) {$\pvec,h,\alpha$};
\node[anchor=west] at (-4.5,2.3) {$\kpvec,h',\bar\alpha$};
\node[anchor=west] at (-1.8,1.4) {$\kvec,\sigma,a$};
\node[anchor=west] at (-3.2,-0.1) {$\kppvec,h'',\bar\alpha$};
\node[anchor=west] at (-2.0,0.4) {$\mt$};
\node[anchor=west] at (-2.4,2.1) {$\hht$};
\node[anchor=west] at (-4.0,1.3) {$\ktp$};
\node[anchor=west] at (-0.3,0.2) {$\ppvec,-h,\beta$};
\node[anchor=west] at (-5.3,0.9) {$\qvec, \lambda $};
\node[anchor=east] at (-6.3,1.3) {$\gamma^{\ast}_{\rm T}$};
\node[anchor=south west] at (-7cm,0cm) {\namediag{diag:vertexqemT}};
 \end{tikzpicture}
}
\rule{0pt}{1ex}
\caption{Diagram \ref{diag:vertexqemT} contributing to the $q\bar{q}$-component of the transverse  virtual photon wave function at NLO with energy denominators and kinematics. Momentum conservation: $\qvec=\kpvec + \kppvec = \pvec+\ppvec$, $\kpvec = \kvec + \pvec$ and $\kppvec = \ppvec - \kvec$.The longitudinal momentum fractions for quark and anti-quark are parametrized as $p^+ = zq^+$ and $p'^+ = (1-z)q^+$, and for the gluon in the loop $k^+ = z'q^+$ with $k'^+ = (z+z')q^+$ and $k''^+ = (1-z-z')q^+$. The longitudinal momentum fraction of the  virtual photon splitting into a $q\bar{q}$ dipole is $k'^+/q^+ = z+z'$ and the natural momentum is $\ktp$. The momentum fraction of the gluon emission is $k^+/k'^+ = z'/(z+z')$ and gluon absorption $k^+/p'^+ = z'/(1-z)$. The natural momentum for the gluon emission is $\hht =  \kt - (z'/(z+z'))\ktp$ and for the gluon absorption $\mt = \kt + (z'/(1-z))\pt$. In order to use $\mt$ as the integration variable we need to know that $\ktp = \mt + ((1-z-z')/(1-z))\pt$ and $\hht = (z/(z+z'))(\mt - (z'/(z(1-z))\pt)$.}
\label{fig:vertexqemT}
\end{figure}

Next we calculate the LCWFs for diagrams \ref{diag:vertexqbaremT} and  \
\ref{diag:vertexqemT} shown in \figs\ref{fig:vertexqbaremT} and \ref{fig:vertexqemT}. 
Since there is a lot of symmetry between these it makes sense to present the result for the sum of the two.

For diagram \ref{diag:vertexqbaremT}, with kinematical variables as in \fig\ref{fig:vertexqbaremT}, the LCWF can be cast in the following form
\begin{equation}
\begin{split}
\psi^{\gamma^{\ast}_{\textrm{T}}\rightarrow q\bar{q}}_{\ref{diag:vertexqbaremT} } = \int \dk\dkp\dkpp (2\pi)^{d-1}\delta^{(d-1)}(\kpvec-\pvec + \kvec)&(2\pi)^{d-1}\delta^{(d-1)}(\kppvec- \kvec - \ppvec)\\
&\times 
\frac{
\qemit^{\bar\alpha,a;\alpha}_{\sigma,h';h}(\mt,\frac{z'}{z})
\paircr^{\gamma^{\ast}_{\rm T}}_{\lambda;h',h''}(\ktp,z-z')
\qbemit^{\bar\alpha;\beta,a}_{h'';-h,\sigma}(\hht, \frac{z'}{(1-z+z')})
}{\Delta^{-}_{01}\Delta^{-}_{02}\Delta^{-}_{03}},
\end{split}
\end{equation}
where the gluon emission and absorption vertices are  given by \eqs\nr{eq:vertexqgtoqd} and \nr{eq:vertexqbartoqbarg}
and the photon vertex by \eq\nr{eq:LOTvirtualphoton}.
The phase space measure  simplifies to 
\begin{equation}\label{eq:vertexqbaremTphasesp}
\int \dk\dkp\dkpp (2\pi)^{d-1}\delta^{(d-1)}(\kpvec-\pvec + \kvec)(2\pi)^{d-1}\delta^{(d-1)}(\kppvec- \kvec - \ppvec) = \frac{1}{16\pi(q^+)^2}\int_{0}^{z} \frac{\ud z'}{z'(z-z')(1-z+z')}\int \frac{\ud^{d-2}\mt}{(2\pi)^{d-2}}
\end{equation}
The LC energy denominators are given by
\begin{equation}\label{eq:vertexqbaremTlcdenom}
\begin{split}
\Delta^{-}_{01} & = \frac{1}{(-2q^+)(z-z')(1-z+z')}\biggl [\left (\mt - \frac{(z-z')}{z}\pt \right )^2 + M^{\ref{diag:vertexqbaremT}}_2 \biggr ]\\
\Delta^{-}_{02} & = \frac{z}{(-2q^+)z'(z-z')}\biggl [\mt^2 + M^{\ref{diag:vertexqbaremT}}_1 \biggr ]\\
\end{split}
\end{equation}
and 
\begin{equation}
\Delta^{-}_{03}  = \frac{1}{(-2q^+)z(1-z)}\biggl [\pt^2 + \overline{Q}^2\biggr ],
\end{equation}
where the coefficients $M^{\ref{diag:vertexqbaremT}}_1$ and $M^{\ref{diag:vertexqbaremT}}_2$ are given by
\begin{equation}\label{eq:vertexqbaremTmasses}
M^{\ref{diag:vertexqbaremT}}_1 = \frac{z'(z-z')}{z^2(1-z)}\left (\pt^2 + \overline{Q}^2\right ) \quad\text{and}\quad 
M^{\ref{diag:vertexqbaremT}}_2 = \frac{(z-z')(1-z+z')}{z(1-z)}\overline{Q}^2.
\end{equation}

For diagram \ref{diag:vertexqemT}, with kinematical variables as in \fig\ref{fig:vertexqemT}, the LCWF is 
 \begin{equation}
 \begin{split}
\psi^{\gamma^{\ast}_{\textrm{T}}\rightarrow q\bar{q}}_{ \ref{diag:vertexqemT}  } = \int \dk\dkp\dkpp (2\pi)^{d-1}\delta^{(d-1)}(\kpvec-\pvec - \kvec)&(2\pi)^{d-1}\delta^{(d-1)}(\kppvec- \ppvec + \kvec)\\
&\times 
\frac{
\qemit^{\bar\alpha;\alpha,a}_{h';\sigma,h}(\hht,\frac{z'}{z+z'})
\paircr^{\gamma^{\ast}_{\rm T}}_{\lambda;h',h''}(\ktp,z+z')
\qbemit^{\bar\alpha,a;\beta}_{h'',\sigma;-h}(\mt, \frac{z'}{(1-z)})
}{\Delta^{-}_{01}\Delta^{-}_{02}\Delta^{-}_{03}}.
\end{split}
\end{equation}
The phase space measure simplifies to 
\begin{equation}\label{eq:vertexqemTphasesp}
\int \dk\dkp\dkpp (2\pi)^{d-1}\delta^{(d-1)}(\kpvec-\pvec - \kvec)(2\pi)^{d-1}\delta^{(d-1)}(\kppvec- \ppvec + \kvec) = \frac{1}{16\pi(q^+)^2}\int_{0}^{1-z} \frac{\ud z'}{z'(z+z')(1-z-z')}\int \frac{\ud^{d-2}\mt}{(2\pi)^{d-2}}
\end{equation}
and the LC energy denominators are given by
\begin{equation}\label{eq:vertexqemTlcdenom}
\begin{split}
\Delta^{-}_{01} & = \frac{1}{(-2q^+)(z+z')(1-z-z')}\biggl [\left (\mt + \frac{(1-z-z')}{1-z}\pt \right )^2 + M^{\ref{diag:vertexqemT}}_2 \biggr ]\\
\Delta^{-}_{02} & = \frac{1-z}{(-2q^+)z'(1-z-z')}\biggl [\mt^2 + M^{\ref{diag:vertexqemT}}_1 \biggr ]\\
\end{split}
\end{equation}
and 
\begin{equation}
\Delta^{-}_{03}  = \frac{1}{(-2q^+)z(1-z)}\biggl [\pt^2 + \overline{Q}^2\biggr ],
\end{equation}
where the coefficients $M^{\ref{diag:vertexqemT}}_1$ and $M^{\ref{diag:vertexqemT}}_2$ are given by
\begin{equation}\label{eq:vertexqemTmasses}
M^{\ref{diag:vertexqemT}}_1 = \frac{z'(1-z-z')}{z(1-z)^2}\left (\pt^2 + \overline{Q}^2\right ) \quad\text{and}\quad M^{\ref{diag:vertexqemT}}_2 = \frac{(z+z')(1-z-z')}{z(1-z)}\overline{Q}^2.
\end{equation}

Now one first performs the transverse momentum integrals using the results in Appendix~\ref{app:loopints}, then performs the numerator helicity sums as described in Sec.~\ref{sec:numerators} and finally integrates over the longitudinal momentum fraction regulating the soft divergences by a cutoff $\alpha$. The result for the sum of the diagrams \ref{diag:vertexqbaremT} and  \ref{diag:vertexqemT} simlifies to
\begin{equation}
\label{eq:sumvertexWFTfinal}
\psi^{\gamma^{\ast}_{\textrm{T}}\rightarrow q\bar{q}}_{\ref{diag:vertexqbaremT} + \ref{diag:vertexqemT}} = 
\psi^{\gamma^{\ast}_{\rm T}\rightarrow q\bar{q}}_{\lo}(\pt,z)\left (\frac{g_r^2\cf}{8\pi^2}\right )\biggl \{ \biggl [-\frac{3}{2} - \log\left (\frac{\alpha}{z} \right ) - \log\left (\frac{\alpha}{1-z} \right ) \biggr ]C^{\rm T}_{\ref{diag:vertexqbaremT} + \ref{diag:vertexqemT}} + \Gamma^{\rm T} \biggr \}  + \mathcal{O}(\varepsilon),
\end{equation}
where the coefficients $C$ and $\Gamma$ are given by
\begin{equation}
C^{\rm T}_{\ref{diag:vertexqbaremT} + \ref{diag:vertexqemT}} = \frac{1}{\varepsilon_{\overline{\rm MS}}} + \log\left (\frac{\mu^2}{\overline{Q}^2}\right )  + \left (\frac{\pt^2 + \overline{Q}^2}{\pt^2}\right )\log \left (\frac{\pt^2 + \overline{Q}^2}{\overline{Q}^2} \right )
\end{equation}
and
\begin{equation}
\label{eq:gammaT}
\begin{split}
\Gamma^{\rm T} = -\frac{21}{6} + \frac{2\pi^2}{6} &- \frac{3}{2}\log(1-z) - \frac{3}{2}\log(z) + 4\log(1-z)\log(z) + \log^2\left (\frac{\alpha}{z}\right )  + \log^2\left (\frac{\alpha}{1-z}\right )\\
& -2\log(1-z)\log(\alpha)  - 2\log(z)\log(\alpha) - \mathrm{Li}_2\left (-\frac{z}{1-z} \right ) - \mathrm{Li}_2\left (-\frac{1-z}{z} \right ).
\end{split}
\end{equation}
The sum of two dilogarithm functions above can be simplified by applying the identity 
\begin{equation}
\mathrm{Li}_2\left (-x \right ) + \mathrm{Li}_2 \left (-\frac{1}{x}\right ) = -\frac{\pi^2}{6} - \frac{1}{2}\log^2(x),\quad x>0.
\end{equation}

\begin{figure}[t]
\centerline{
\includegraphics[width=0.2\textwidth]{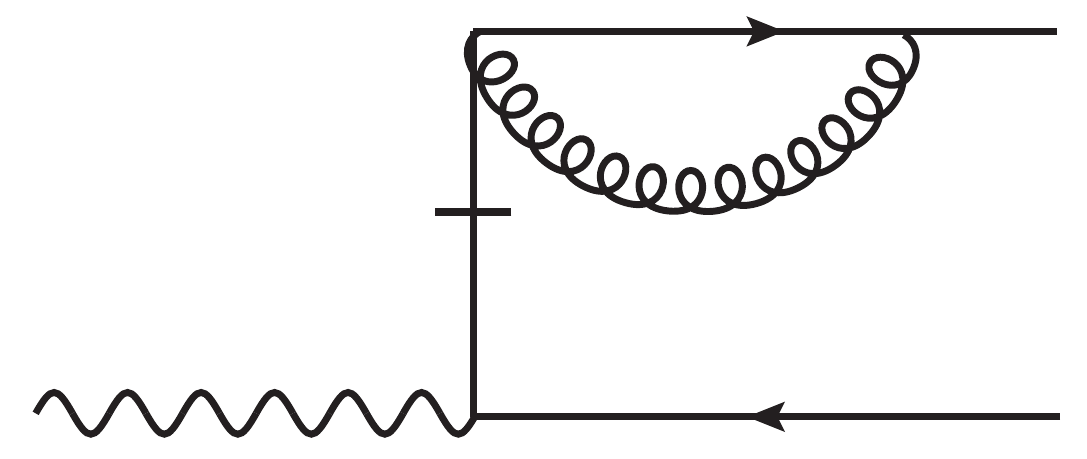}
\includegraphics[width=0.2\textwidth]{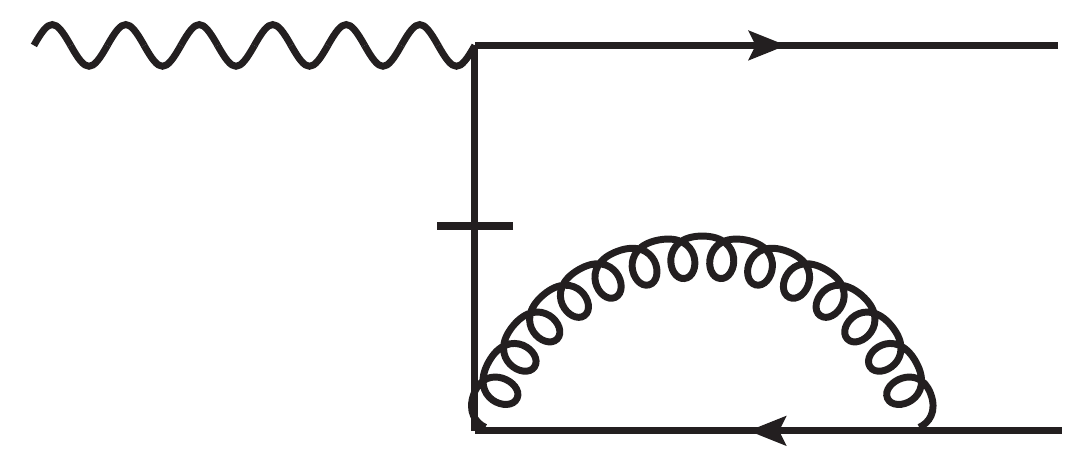}
\includegraphics[width=0.2\textwidth]{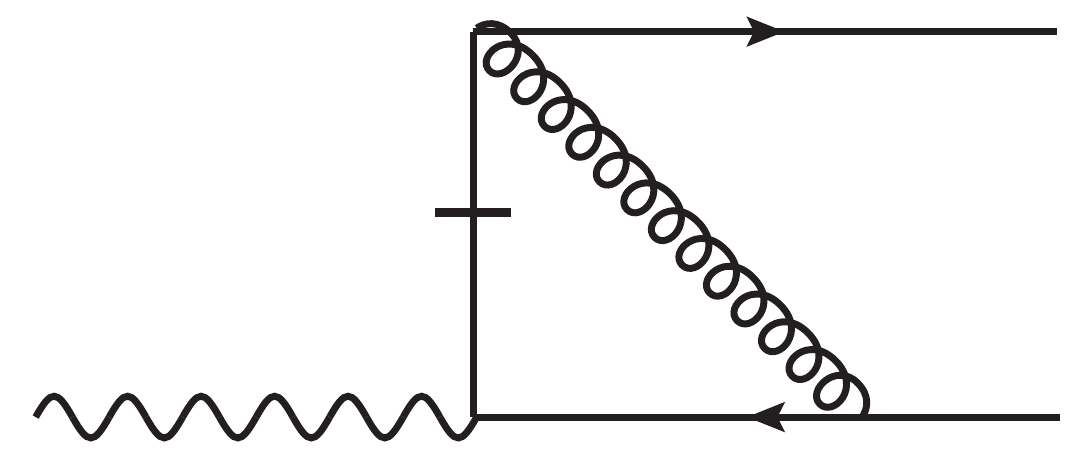}
\includegraphics[width=0.2\textwidth]{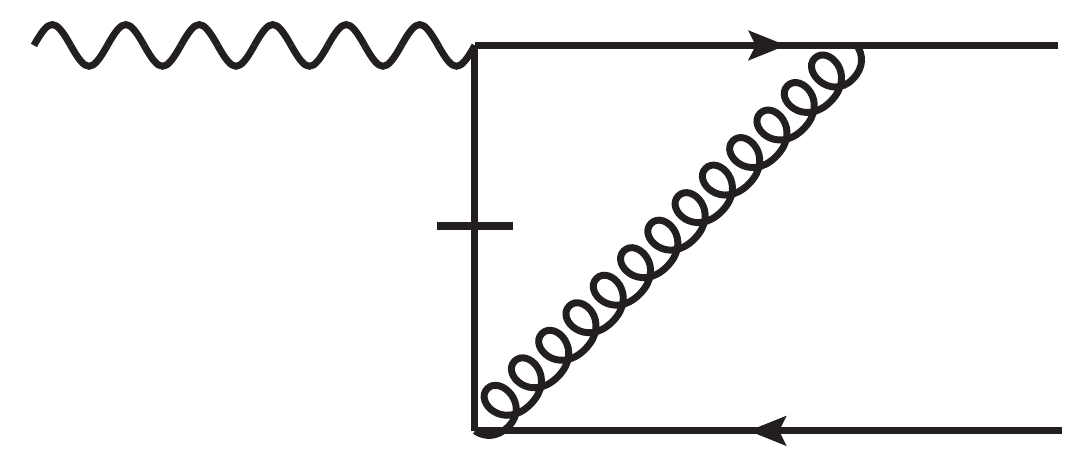}
\includegraphics[width=0.2\textwidth]{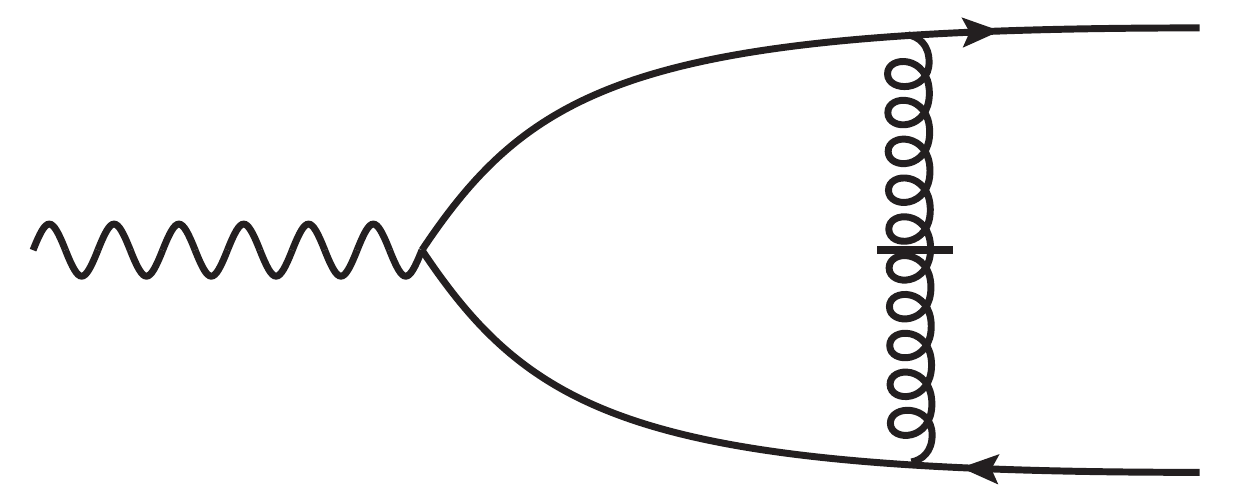}
}
\centerline{
}
\rule{0pt}{1ex}
\caption{Instantaneous diagrams for the $q\bar{q}$-component of the virtual photon wave function at NLO, yielding zero in dimensional regularization.}
\label{fig:tinst}
 \end{figure}

In principle one also has to compute the instantaneous vertex correction diagrams shown in \fig\ref{fig:tinst}. These however vanish in dimensional regularization. This is easiest to see by taking as the integration variable the natural momentum of the only non-instantantaneous vertex in the diagram, in which case this vertex and consequently the whole transvere momentum integrand are linear in the integration variable~\cite{Lappi:2016oup}.

Adding the one-loop quark self-energy corrections in \eqs\nr{eq:oneloopSEUPT} and \nr{eq:oneloopSEDOWNT} together with \eq\nr{eq:sumvertexWFTfinal}, we get the expression for the full one-loop corrected LCWF for $\gamma^{\ast}_{\rm T} \rightarrow q\bar{q}$ computed in the FDH scheme
\begin{equation}
\label{eq:fullTWFqbarq}
\begin{split}
\psi^{\gamma^{\ast}_{\textrm{T}}\rightarrow q\bar{q}}_{\nlo}\bigg\vert_{\text{FDH}} = \psi^{\gamma^{\ast}_{\textrm{T}}\rightarrow q\bar{q}}_{\lo}(\pt,z)\left (\frac{g^2_r\cf}{8\pi^2}\right )\biggl \{\biggl [\frac{3}{2} + \log\left (\frac{\alpha}{z}\right ) & + \log\left (\frac{\alpha}{1-z}\right ) \biggr ]C^{(\rm T)}_{\text{full}} + \frac{1}{2}\log^2\left (\frac{z}{1-z}\right )- \frac{\pi^2}{6}\\
& + \frac{5}{2}\biggr \}  + \mathcal{O}(\varepsilon),
\end{split}
\end{equation}
where 
\begin{equation}
C^{(\rm T)}_{\text{full}} = \frac{1}{\varepsilon_{\overline{\text{MS}}}} + \log\left (\frac{\mu^2}{\overline{Q}^2}\right ) + \left (\frac{\overline{Q}^2 - \pt^2}{\pt^2}\right )\log \left ( \frac{\pt^2 + \overline{Q}^2}{\overline{Q}^2}\right ).
\end{equation}
For comparison the full result computed in the CDR scheme~\cite{Beuf:2016wdz} and \cite{Boussarie:2016ogo} is  
\begin{equation}
\begin{split}
\psi^{\gamma^{\ast}_{\textrm{T}}\rightarrow q\bar{q}}_{\nlo}\bigg\vert_{\text{CDR}} = \psi^{\gamma^{\ast}_{\textrm{T}}\rightarrow q\bar{q}}_{\lo}(\pt,z)\left (\frac{g^2_r\cf}{8\pi^2}\right )\biggl \{\biggl [\frac{3}{2} + \log\left (\frac{\alpha}{z}\right ) & + \log\left (\frac{\alpha}{1-z}\right ) \biggr ]C^{\rm (T)}_{\text{full}} + \frac{1}{2}\log^2\left (\frac{z}{1-z}\right )- \frac{\pi^2}{6}\\
& + \frac{5}{2} + \frac{1}{2} \biggr \} + \mathcal{O}(\varepsilon),		
\end{split}
\end{equation}
where the additional factor $1/2$ is the scheme dependent part of the CDR scheme calculation.

\subsection{Longitudinal photons}

\begin{figure}[tb]
\centerline{
\includegraphics[width=6.4cm]{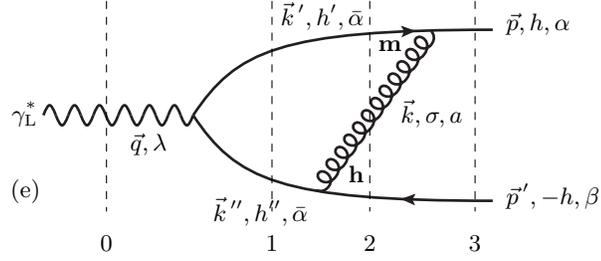}
\begin{tikzpicture}[overlay]
\draw [dashed] (-5.5,2.8) -- (-5.5,0);
\node[anchor=north] at (-5.5cm,-0.2cm) {0};
\draw [dashed] (-3.3,2.8) -- (-3.3,0);
\node[anchor=north] at (-3.3cm,-0.2cm) {1};
\draw [dashed] (-2.0,2.8) -- (-2.0,0);
\node[anchor=north] at (-2.0cm,-0.2cm) {2};
\draw [dashed] (-0.6,2.8) -- (-0.6,0);
\node[anchor=north] at (-0.6cm,-0.2cm) {3};
\node[anchor=west] at (-0.3,2.5) {$\pvec,h,\alpha$};
\node[anchor=west] at (-1.7,1.3) {$\kvec,\sigma,a$};
\node[anchor=west] at (-4.2,0.0){$\kppvec,h'',\bar\alpha$};
\node[anchor=west] at (-3.3,2.6) {$\kpvec,h',\bar\alpha$};
\node[anchor=west] at (-2.4,0.5) {$\hht$};
\node[anchor=west] at (-2.0,2.2) {$\mt$};
\node[anchor=west] at (-0.3,0.2) {$\ppvec,-h,\beta$};
\node[anchor=west] at (-5.3,0.9) {$\qvec, \lambda $};
\node[anchor=east] at (-6.3,1.3) {$\gamma^{\ast}_{\rm L}$};
\node[anchor=south west] at (-7cm,0cm) {\namediag{diag:vertexqbaremL}};
 \end{tikzpicture}
}
\rule{0pt}{1ex}
\caption{Vertex diagram \ref{diag:vertexqbaremL} contributing to the $q\bar{q}$-component of the longitudinal  virtual photon wave function. Momentum conservation: $\qvec=\kpvec+\kppvec=\pvec + \ppvec$, $\kpvec = \pvec- \kvec$ and $\kppvec = \kvec + \ppvec$. The longitudinal momentum fractions for quark and anti-quark are parametrized as $p^+ = zq^+$ and $p'^+ = (1-z)q^+$, and for the gluon in the loop $k^+ = z'q^+$ with $k'^+ = (z-z')q^+$ and $k''^+ = (1-z+z')q^+$. The longitudinal momentum fraction of the  virtual photon splitting into a $q\bar{q}$ dipole is $k'^+/q^+ = z-z'$. The momentum fraction of the gluon emission is $k^+/k''^+ = z'/(1-z+z')$ and gluon absorption $k^+/p^+ = z'/z$. The natural momentum for the gluon emission is $\hht =  \kt - (z'/(1-z+z'))\ktpp$ and for the gluon absorption $\mt = \kt - (z'/z)\pt$. In order to use $\mt$ as the integration variable we need to know that $\hht = ((1-z)/(1-z+z'))(\mt + (z'/(z(1-z))\pt)$.}
\label{fig:vertexqbaremL}
\end{figure}

\begin{figure}[t]
\centerline{
\includegraphics[width=6.4cm]{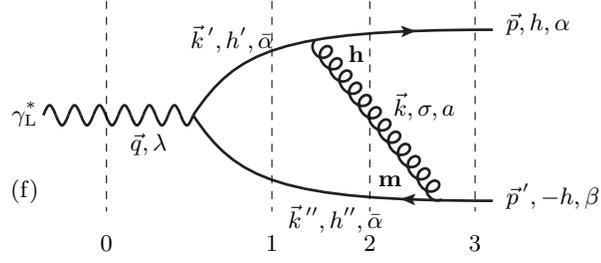}
\begin{tikzpicture}[overlay]
\draw [dashed] (-5.5,2.8) -- (-5.5,0);
\node[anchor=north] at (-5.5cm,-0.2cm) {0};
\draw [dashed] (-3.3,2.8) -- (-3.3,0);
\node[anchor=north] at (-3.3cm,-0.2cm) {1};
\draw [dashed] (-2.0,2.8) -- (-2.0,0);
\node[anchor=north] at (-2.0cm,-0.2cm) {2};
\draw [dashed] (-0.6,2.8) -- (-0.6,0);
\node[anchor=north] at (-0.6cm,-0.2cm) {3};
\node[anchor=west] at (-0.3,2.5) {$\pvec,h,\alpha$};
\node[anchor=west] at (-4.5,2.3) {$\kpvec,h',\bar\alpha$};
\node[anchor=west] at (-1.8,1.4) {$\kvec,\sigma,a$};
\node[anchor=west] at (-3.2,-0.1) {$\kppvec,h'',\bar\alpha$};
\node[anchor=west] at (-2.0,0.4) {$\mt$};
\node[anchor=west] at (-2.4,2.1) {$\hht$};
\node[anchor=west] at (-0.3,0.2) {$\ppvec,-h,\beta$};
\node[anchor=west] at (-5.3,0.9) {$\qvec, \lambda $};
\node[anchor=east] at (-6.3,1.3) {$\gamma^{\ast}_{\rm L}$};
\node[anchor=south west] at (-7cm,0cm) {\namediag{diag:vertexqemL}};
 \end{tikzpicture}
}
\rule{0pt}{1ex}
\caption{Vertex correction diagram \ref{diag:vertexqemL} contributing to the $q\bar{q}$-component of the longitudinal  virtual photon wave function at NLO with energy denominators and kinematics. Momentum conservation: $\qvec=\kpvec + \kppvec = \pvec+\ppvec$, $\kpvec = \kvec + \pvec$ and $\kppvec = \ppvec - \kvec$.The longitudinal momentum fractions for quark and anti-quark are parametrized as $p^+ = zq^+$ and $p'^+ = (1-z)q^+$, and for the gluon in the loop $k^+ = z'q^+$ with $k'^+ = (z+z')q^+$ and $k''^+ = (1-z-z')q^+$. The longitudinal momentum fraction of the  virtual photon splitting into a $q\bar{q}$ dipole is $k'^+/q^+ = z+z'$. The momentum fraction of the gluon emission is $k^+/k'^+ = z'/(z+z')$ and gluon absorption $k^+/p'^+ = z'/(1-z)$. The natural momentum for the gluon emission is $\hht =  \kt - (z'/(z+z'))\ktp$ and for the gluon absorption $\mt = \kt + (z'/(1-z))\pt$. In order to use $\mt$ as the integration variable we need to know that $\ktp = \mt + ((1-z-z')/(1-z))\pt$ and $\hht = (z/(z+z'))(\mt - (z'/(z(1-z))\pt)$.}
\label{fig:vertexqemL}
\end{figure}

For diagram \ref{diag:vertexqbaremL}, with kinematical variables as in \fig\ref{fig:vertexqbaremL}, the LCWF can be cast in the following form
\begin{equation}
\begin{split}
\psi^{\gamma^{\ast}_{\textrm{L}}\rightarrow q\bar{q}}_{\ref{diag:vertexqbaremL} } = \int \dk\dkp\dkpp (2\pi)^{d-1}\delta^{(d-1)}(\kpvec-\pvec + \kvec)&(2\pi)^{d-1}\delta^{(d-1)}(\kppvec- \kvec - \ppvec)\\
&\times
\frac{
\qemit^{\bar\alpha,a;\alpha}_{\sigma,h';h}(\mt,\frac{z'}{z})
A^{\gamma^{\ast}_{\rm L}}_{\lambda,h',h''}(Q,z-z')
\qbemit^{\bar\alpha;\beta,a}_{h'';-h,\sigma}(\hht, \frac{z'}{1-z+z'})
}{\Delta^{-}_{01}\Delta^{-}_{02}\Delta^{-}_{03}},
\end{split}
\end{equation}
where the gluon emission and absorption vertices are  given by \eqs\nr{eq:vertexqgtoqd},  \nr{eq:vertexqbartoqbarg} and the longitudinal photon splitting vertex by~\nr{eq:LOLvirtualphoton}. The LC energy denominators  are the same as in \eq\nr{eq:vertexqbaremTlcdenom} for diagram \ref{diag:vertexqbaremT}, as is the phase space measure  \nr{eq:vertexqbaremTphasesp}. Adding everything together and summing over the colors we get
\begin{equation}
\label{eq:vertexqbaremWFL}
\begin{split}
\psi^{\gamma^{\ast}_{\textrm{T}}\rightarrow q\bar{q}}_{\ref{diag:vertexqbaremL} } = \psi^{\gamma^{\ast}_{\rm L}\rightarrow q\bar{q}}_{\lo}(Q,z)\left (\frac{-g^2\cf}{\pi}\right )\int_{0}^{z} \frac{\ud z' (z-z')(1-z+z')}{(z')^2} &\int \frac{\ud^{d-2}\mt}{(2\pi)^{d-2}} \frac{m^i\left (m + \frac{z'}{z(1-z)}p\right )^n}{\biggl [\mt^2 + M_1^{\ref{diag:vertexqbaremT}} \biggr ]\biggl [\left (\mt - \frac{(z-z')}{z}\pt \right )^2 + M_2^{\ref{diag:vertexqbaremT}}  \biggr ]}\\
& \times \textrm{num}(z,z')\bigg\vert_{\ref{diag:vertexqbaremL}},
\end{split}
\end{equation}
with the mass scales $ M_1^{\ref{diag:vertexqbaremT}}$ and $ M_2^{\ref{diag:vertexqbaremT}}$ from \eq\nr{eq:vertexqbaremTmasses}.
Beause of the simple structure of the longitudinal photon splitting vertex, we can directly evaluate the numerator of \nr{eq:vertexqbaremWFL} in $d_s$ dimensions in terms of Levi-Civita tensors:
\begin{equation}
\textrm{num}(z,z')\bigg\vert_{\ref{diag:vertexqbaremL}} = \biggl [\left (1-\frac{z'}{2z}\right)\delta^{ij}_{(d_s)}- ih\left (\frac{z'}{2z}\right )\epsilon^{ij}_{(d_s)} \biggr ]\biggl [ \left ( 1-\frac{1}{2}\left (\frac{z'}{1-z+z'}\right )\right )\delta^{nm}_{(d_s)}-ih\frac{1}{2}\left (\frac{z'}{1-z+z'} \right )\epsilon^{nm}_{(d_s)} \biggr ]\delta^{jm}_{(d_s)}
.
\end{equation}

For diagram \ref{diag:vertexqemL}, with kinematical variables as in \fig\ref{fig:vertexqemL}, the LCWF is 
\begin{equation}
\begin{split}
\psi^{\gamma^{\ast}_{\textrm{L}}\rightarrow q\bar{q}}_{ \ref{diag:vertexqemL}} = \int \dk\dkp\dkpp (2\pi)^{d-1}\delta^{(d-1)}(\kpvec-\pvec - \kvec)&(2\pi)^{d-1}\delta^{(d-1)}(\kppvec- \ppvec + \kvec)\\
&\times \frac{V^{\bar\alpha;\alpha,a}_{h';\sigma,h}(\hht,\frac{z'}{z+z'})A^{\gamma^{\ast}_{\rm L}}_{\lambda;h',h''}(Q,z+z')\overline{V}^{\bar\alpha,a;\beta}_{h'',\sigma,-h}(\mt, \frac{z'}{1-z})}{\Delta^{-}_{01}\Delta^{-}_{02}\Delta^{-}_{03}},
\end{split}
\end{equation}
where the phase space measure is the same as in \eq\nr{eq:vertexqemTphasesp} for diagram \ref{diag:vertexqemT}, as are the energy denominators in \eq\nr{eq:vertexqemTlcdenom}.
Putting everything together and summing over the colors gives 
\begin{equation}
\label{eq:vertexqemWFL}
\begin{split}
\psi^{\gamma^{\ast}_{\textrm{L}}\rightarrow q\bar{q}}_{\ref{diag:vertexqemL} } = \psi^{\gamma^{\ast}_{\rm L}\rightarrow q\bar{q}}_{\lo}(Q,z)\left (\frac{-g^2\cf}{\pi}\right )\int_{0}^{1-z} \frac{\ud z'(z+z')(1-z-z')}{(z')^2} &\int \frac{\ud^{d-2}\mt}{(2\pi)^{d-2}} \frac{m^i\left (m - \frac{z'}{z(1-z)}p\right )^n}{\biggl [\mt^2 + M_1^{\ref{diag:vertexqemT}} \biggr ]\biggl [\left (\mt + \frac{(1-z-z')}{1-z}\pt \right )^2 + M_2^{\ref{diag:vertexqemT}}  \biggr ]}\\
& \times \textrm{num}(z,z')\bigg\vert_{\ref{diag:vertexqemL}}
\end{split}.
\end{equation}
with $M_1^{\ref{diag:vertexqemT}}$ and $M_2^{\ref{diag:vertexqemT}}$ from 
\eq\nr{eq:vertexqemTmasses}.
Again we can directly use the $d_s$-dimensional expression for the numerator
\begin{equation}
\begin{split}
\textrm{num}(z,z')\bigg\vert_{\ref{diag:vertexqemL}} = \biggl [\left (1-\frac{z'}{2(1-z)}\right)\delta^{ij}_{(d_s)}+ &ih\frac{1}{2}\left (\frac{z'}{1-z}\right )\epsilon^{ij}_{(d_s)} \biggr ]
\biggl [ \left ( 1-\frac{1}{2}\left (\frac{z'}{z+z'}\right )\right )\delta^{nm}_{(d_s)}+ih\frac{1}{2}\left (\frac{z'}{z+z'} \right )\epsilon^{nm}_{(d_s)} \biggr ]\delta^{jm}_{(d_s)}
\end{split}.
\end{equation}

\begin{figure}[t]
\centerline{
\includegraphics[width=6.4cm]{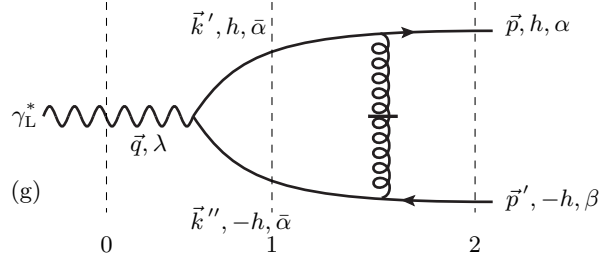}
\begin{tikzpicture}[overlay]
\draw [dashed] (-5.5,2.8) -- (-5.5,0);
\node[anchor=north] at (-5.5cm,-0.2cm) {0};
\draw [dashed] (-3.3,2.8) -- (-3.3,0);
\node[anchor=north] at (-3.3cm,-0.2cm) {1};
\draw [dashed] (-0.6,2.8) -- (-0.6,0);
\node[anchor=north] at (-0.6cm,-0.2cm) {2};
\node[anchor=west] at (-0.3,2.5) {$\pvec,h,\alpha$};
\node[anchor=west] at (-4.5,2.5) {$\kpvec,h,\bar\alpha$};
\node[anchor=west] at (-4.5,-0.1) {$\kppvec,-h,\bar\alpha$};
\node[anchor=west] at (-0.3,0.2) {$\ppvec,-h,\beta$};
\node[anchor=west] at (-5.3,0.9) {$\qvec, \lambda $};
\node[anchor=east] at (-6.3,1.3) {$\gamma^{\ast}_{\rm L}$};
\node[anchor=south west] at (-7cm,0cm) {\namediag{diag:vertexqemLinst}};
 \end{tikzpicture}
}
\rule{0pt}{1ex}
\caption{Instantaneous gluon diagram \ref{diag:vertexqemLinst} contributing to the $q\bar{q}$-component of the longitudinal virtual photon wave function at NLO with energy denominators and kinematics. Momentum conservation: $\qvec=\kpvec + \kppvec$ and $\qvec = \pvec+\ppvec$. The longitudinal momentum fractions for quark and anti-quark are parametrized as $p^+ = zq^+$ and $p'^+ = (1-z)q^+$, and  $k'^+ = z'q^+$ and $k''^+ = (1-z')q^+$. The longitudinal momentum fraction of the  virtual photon splitting into a $q\bar{q}$ dipole is $k'^+/q^+ = z'$.}
\label{fig:vertexqemLinst}
\end{figure}

For the longitudinal photon there is only one instantaneous diagram, \ref{diag:vertexqemLinst} shown in \fig\ref{fig:vertexqemLinst}, contributing to the $\gamma^{\ast}_{\rm L}\rightarrow q\bar{q}$ LCWF at one-loop level. It
is given by  
\begin{equation}
\psi^{\gamma^{\ast}_{\rm L}\rightarrow q\bar{q}}_{\ref{diag:vertexqemLinst}} = \int\dkp\dkpp (2\pi)^{d-1}\delta^{(d-1)}(\qvec - \kpvec - \kppvec)\frac{A_{\lambda;h,-h}^{\gamma^{\ast}_{\rm L}}(Q,z')
\mathfrak{I}^{(\ref{fig:qqbarqqbarinst})}
}{\Delta_{01}^{-}\Delta_{02}^{-}},
\end{equation}
where the instantaneous vertex $\mathfrak{I}^{(\ref{fig:qqbarqqbarinst})}$ is given by \eq\nr{eq:qqbarqqbarinst}.
The phase space measure simplifies to 
\begin{equation}
\int\dkp\dkpp (2\pi)^{d-1}\delta^{(d-1)}(\qvec - \kpvec - \kppvec) =  \int \frac{\ud k'^+}{2k'^+(2\pi)}\int\frac{\ud^{d-2}\ktp}{(2\pi)^{d-2}}\frac{1}{2k''^+}  = \frac{1}{8\pi q^+}\int_{0}^{1}\frac{\ud z'}{z'(1-z')}\int\frac{\ud^{d-2}\ktp}{(2\pi)^{d-2}}
\end{equation}
and the LC energy denominators are 
\begin{equation}
\begin{split}
\Delta_{01}^{-} &= \frac{1}{(-2q^+)z'(1-z')}\biggl [\ktp^2 + M \biggr ] \quad\text{with}\quad M = \frac{z'(1-z')}{z(1-z)}\overline{Q}^2\\
\Delta_{02}^{-} &= \frac{1}{(-2q^+)z(1-z)}\biggl [\pt^2 + \overline{Q}^2\biggr ].
\end{split}
\end{equation}
The instantaneous vertex $\mathfrak{I}^{(\ref{fig:qqbarqqbarinst})}$
simplifies to 
\begin{equation}
\mathfrak{I}^{(\ref{fig:qqbarqqbarinst})} =-4g^2t^a_{\alpha\bar\alpha}t^a_{\bar\alpha\beta} \frac{\sqrt{z(1-z)z'(1-z')}}{(z'-z)^2}.
\end{equation}

Now adding the results from diagrams \ref{diag:vertexqbaremL}, \ref{diag:vertexqemL} and \ref{diag:vertexqemLinst} together, we get for the full vertex correction 
\begin{equation}
\label{eq:sumvertexWFL}
\psi^{\gamma^{\ast}_{\rm L}\rightarrow q\bar{q}}_{ \ref{diag:vertexqbaremL} + \ref{diag:vertexqemL} + \ref{diag:vertexqemLinst}} = \psi^{\gamma^{\ast}_{\rm L}\rightarrow q\bar{q}}_{\lo}(Q,z)\left (\frac{g^2_r\cf}{8\pi^2}\right )\biggl \{\biggl [\frac{1}{\varepsilon_{\overline{\rm MS}}} + \log\left (\frac{\mu^2}{\overline{Q}^2}\right ) \biggr ]\left (-\frac{3}{2} - \log\left (\frac{\alpha}{z}\right ) - \log\left (\frac{\alpha}{1-z}\right )\right ) + \Gamma^{\rm L} \biggr \} + \mathcal{O}(\varepsilon),
\end{equation}
where the coefficient $\Gamma^{\rm L}$ has the same expression as in \eq\nr{eq:gammaT}. Finally, adding the one-loop quark self energy corrections \eqs\nr{eq:oneloopSEUPL} and \nr{eq:oneloopSEDOWNL} to \nr{eq:sumvertexWFL} we get the full expression for one-loop corrected LCWF for $\gamma^{\ast}_{\rm L} \rightarrow q\bar{q}$  in the FDH scheme
\begin{equation}
\label{eq:fullLWFqbarq}
\begin{split}
\psi^{\gamma^{\ast}_{\textrm{L}}\rightarrow q\bar{q}}_{\nlo}\bigg\vert_{\text{FDH}} = \psi^{\gamma^{\ast}_{\textrm{L}}\rightarrow q\bar{q}}_{\lo}(Q,z)\left (\frac{g^2_r\cf}{8\pi^2}\right )\biggl \{\biggl [\frac{3}{2} + \log\left (\frac{\alpha}{z}\right ) & + \log\left (\frac{\alpha}{1-z}\right ) \biggr ]C^{(\rm L)}_{\text{full}} + \frac{1}{2}\log^2\left (\frac{z}{1-z}\right )- \frac{\pi^2}{6}\\
& + \frac{5}{2}\biggr \} + \mathcal{O}(\varepsilon),
\end{split}
\end{equation}
where 
\begin{equation}
C^{(\rm L)}_{\text{full}} = \frac{1}{\varepsilon_{\overline{\text{MS}}}} + \log\left (\frac{\mu^2}{\overline{Q}^2}\right ) -2\log \left ( \frac{\pt^2 + \overline{Q}^2}{\overline{Q}^2}\right ).
\end{equation}
Again for comparison the full result computed in CDR scheme~\cite{Beuf:2016wdz} and \cite{Boussarie:2016ogo} is  
\begin{equation}
\begin{split}
\psi^{\gamma^{\ast}_{\textrm{L}}\rightarrow q\bar{q}}_{\nlo}\bigg\vert_{\text{CDR}} = \psi^{\gamma^{\ast}_{\textrm{L}}\rightarrow q\bar{q}}_{\lo}(Q,z)\left (\frac{g^2_r\cf}{8\pi^2}\right )\biggl \{\biggl [\frac{3}{2} + \log\left (\frac{\alpha}{z}\right ) & + \log\left (\frac{\alpha}{1-z}\right ) \biggr ]C^{(\rm L)}_{\text{full}} + \frac{1}{2}\log^2\left (\frac{z}{1-z}\right )- \frac{\pi^2}{6}\\
& + \frac{5}{2} + \frac{1}{2} \biggr \} + \mathcal{O}(\varepsilon),
\end{split}
\end{equation}
where the only difference is the term $1/2$, which was identified as a scheme dependent part in the  CDR calculation.

\section{Wave functions for gluon emission}
\label{sec:real}
 
We then move to the wave functions for quark-antiquark-gluon contributions, needed for real emission contributions to the cross section. Here all the vertices can be, in the FDH scheme,  evaluated directly in $d_s=4$ dimensions.

 \subsection{Transverse photon}

\begin{figure}[t]
\centerline{
\includegraphics[width=6.4cm]{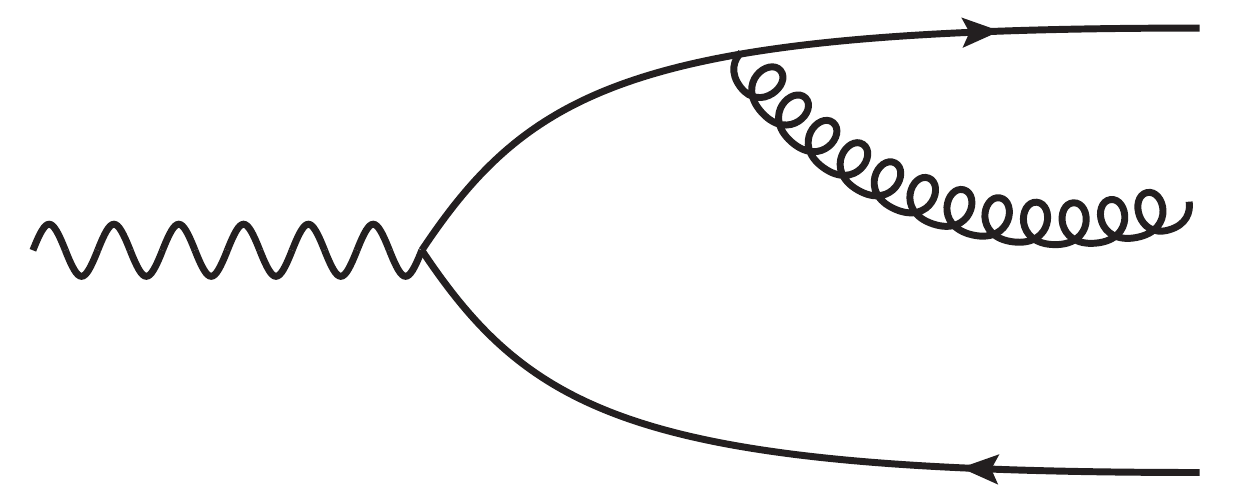}
\begin{tikzpicture}[overlay]
\draw [dashed] (-5.5,2.8) -- (-5.5,0);
\node[anchor=north] at (-5.5cm,-0.2cm) {0};
\draw [dashed] (-3.2,2.8) -- (-3.2,0);
\node[anchor=north] at (-3.2cm,-0.2cm) {1};
\draw [dashed] (-0.8,2.8) -- (-0.8,0);
\node[anchor=north] at (-0.8,-0.2) {2};
\node[anchor=west] at (-0.3,2.5) {$\pvec,h,\alpha$};
\node[anchor=west] at (-0.3,1.2) {$\kvec,\sigma,a$};
\node[anchor=west] at (-0.3,0.2) {$\ppvec,-h,\beta$};
\node[anchor=west] at (-5.3,0.9) {$\qvec, \lambda $};
\node[anchor=west] at (-1.8,1.8) {$\mt$};
\node[anchor=south west] at (-4.4,2.1) {$\kpvec,h,\beta$};
\node[anchor=east] at (-6.3,1.3) {$\gamma^{\ast}_{\rm T}$};
\node[anchor=south west] at (-7cm,0cm) {\namediag{diag:qgqbarT}};
 \end{tikzpicture}
}
\rule{0pt}{1ex}
\caption{Diagram \ref{diag:qgqbarT} contributing to the $q\bar{q}g$-component of the longitudinal virtual photon wave function at NLO with energy denominators and kinematics. Momentum conservation: $\qvec=\kpvec + \ppvec$, $\kpvec=\pvec + \kvec$ and
$\qvec=\pvec+\ppvec + \kvec$.   Momentum fractions are defined by $p^+ = z_1q^+$, $k^+ = z_2q^+$, $p'^+ = z_3q^+$  and $k'^+ = (z_1+z_2)q^+$. The momentum fraction of the transverse virtual photon splitting into a quark anti-quark dipole is $k'^+/q^+ = z_1+z_2$, and the natural momentum is $\ktp = -\ptp$. The momentum fraction of the gluon emission is $k^+/k'^+ = z_2/(z_1+z_2)$, and the natural momenta for the gluon emission vertex is $\mt \equiv \kt-(z_2/(z_1+z_2))\ktp = \kt + (z_2/(z_1+z_2))\ptp$. Note that the momentum fractions are related to each other via relation $z_1+z_2+z_3 =1$.}
\label{fig:qgqbarT}
 \end{figure}

 \begin{figure}[t]
\centerline{
\includegraphics[width=6.4cm]{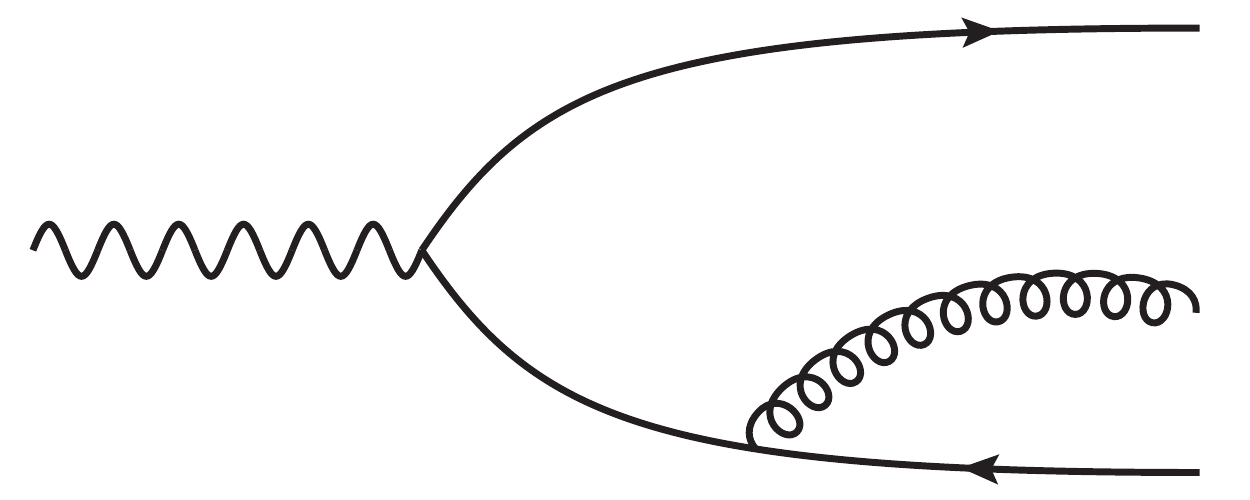}
\begin{tikzpicture}[overlay]
\draw [dashed] (-5.5,2.8) -- (-5.5,0);
\node[anchor=north] at (-5.5cm,-0.2cm) {0};
\draw [dashed] (-3.2,2.8) -- (-3.2,0);
\node[anchor=north] at (-3.2cm,-0.2cm) {1};
\draw [dashed] (-0.8,2.8) -- (-0.8,0);
\node[anchor=north] at (-0.8,-0.2) {2};
\node[anchor=west] at (-0.3,2.5) {$\pvec,h,\alpha$};
\node[anchor=west] at (-0.3,1.1) {$\kvec,\sigma,a$};
\node[anchor=west] at (-0.3,0.2) {$\ppvec,-h,\beta$};
\node[anchor=west] at (-5.3,0.9) {$\qvec, \lambda $};
\node[anchor=west] at (-1.8,0.5) {$\lt$};
\node[anchor=south west] at (-4.6,-0.1) {$\kpvec,-h,\alpha$};
\node[anchor=east] at (-6.3,1.3) {$\gamma^{\ast}_{\rm T}$};
\node[anchor=south west] at (-7cm,0cm) {\namediag{diag:qqbargT}};
 \end{tikzpicture}
}
\rule{0pt}{1ex}
\caption{Diagram \ref{diag:qqbargT} contributing to the $q\bar{q}g$-component of the longitudinal virtual photon wave function at NLO with energy denominators and kinematics. Momentum conservation: $\qvec=\pvec + \kpvec$, $\kpvec=\kvec + \ppvec$ and $\qvec=\pvec+\ppvec + \kvec$.   Momentum fractions are defined by $ p^+ = z_1q^+$, $k^+ = z_2q^+$, $p'^+ = z_3q^+$ and $k'^+ = (z_2+z_3)q^+$. The natural momentum fraction of the transverse virtual photon splitting into a quark anti-quark dipole is $p^+/q^+ = z_1$, and the momentum fraction is $\pt$. The momentum fraction of the gluon emission is $k^+/k'^+ = z_2/(z_2+z_3)$, and the natural momenta for the gluon emission vertex is $\lt \equiv \kt-(z_2/(z_2+z_3))\ktp = \kt + (z_2/(z_2+z_3))\pt$. Note that the momentum fractions are related to each other via relation $z_1+z_2+z_3 =1$.}
\label{fig:qqbargT}
 \end{figure}
 
 \begin{figure}[t]
\centerline{
\includegraphics[width=6.4cm]{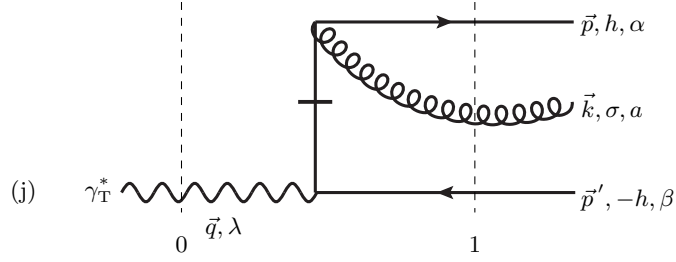}
\begin{tikzpicture}[overlay]
\draw [dashed] (-5.5,2.8) -- (-5.5,0);
\node[anchor=north] at (-5.5cm,-0.2cm) {0};
\draw [dashed] (-1.6,2.8) -- (-1.6,0);
\node[anchor=north] at (-1.6cm,-0.2cm) {1};
\node[anchor=west] at (-0.3,2.5) {$\pvec,h,\alpha$};
\node[anchor=west] at (-0.3,1.4) {$\kvec,\sigma,a$};
\node[anchor=west] at (-0.3,0.2) {$\ppvec,-h,\beta$};
\node[anchor=west] at (-5.3,-0.2) {$\qvec, \lambda $};
\node[anchor=east] at (-6.3,0.3) {$\gamma^{\ast}_{\rm T}$};
\node[anchor=south west] at (-8cm,0cm) {\namediag{diag:qgqbarinst}};
 \end{tikzpicture}
}
\rule{0pt}{1ex}
\caption{Diagram \ref{diag:qgqbarinst} contributing to the $q\bar{q}g$-component of the transverse virtual photon wave function at NLO with energy denominators and kinematics. Momentum conservation: $\qvec= \pvec + \kvec + \ppvec$. Momentum fractions are defined by $p^+ = z_1q^+$, $k^+ = z_2q^+$ and $p'^+ = z_3q^+$.}
\label{fig:qgqbarinst}
 \end{figure}
 
 \begin{figure}[t]
\centerline{
\includegraphics[width=6.4cm]{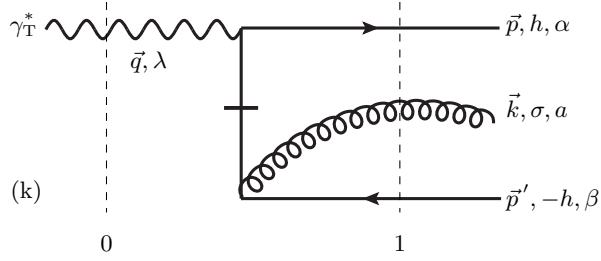}
\begin{tikzpicture}[overlay]
\draw [dashed] (-5.5,2.8) -- (-5.5,0);
\node[anchor=north] at (-5.5cm,-0.2cm) {0};
\draw [dashed] (-1.6,2.8) -- (-1.6,0);
\node[anchor=north] at (-1.6cm,-0.2cm) {1};
\node[anchor=west] at (-0.3,2.5) {$\pvec,h,\alpha$};
\node[anchor=west] at (-0.3,1.4) {$\kvec,\sigma,a$};
\node[anchor=west] at (-0.3,0.2) {$\ppvec,-h,\beta$};
\node[anchor=west] at (-5.3,2.0) {$\qvec, \lambda $};
\node[anchor=east] at (-6.3,2.5) {$\gamma^{\ast}_{\rm T}$};
\node[anchor=south west] at (-7cm,0cm) {\namediag{diag:qqbarginst}};
 \end{tikzpicture}
}
\rule{0pt}{1ex}
\caption{Diagram \ref{diag:qqbarginst} contributing to the $q\bar{q}g$-component of the transverse virtual photon wave function at NLO with energy denominators and kinematics. Momentum conservation: $\qvec=\pvec + \kvec + \ppvec$. Momentum fractions are defined by  $ p^+ = z_1q^+$, $k^+ = z_2q^+$ and $p'^+ =z_3q^+$.}
\label{fig:qqbarginst}
 \end{figure}

For transverse photons, we need to calculate the  diagrams \ref{diag:qgqbarT}-\ref{diag:qqbarginst} shown in \figs\ref{fig:qgqbarT}, \ref{fig:qqbargT}, \ref{fig:qgqbarinst} and \ref{fig:qqbarginst}. 
The LCWF for diagram \ref{diag:qgqbarT} in \fig\ref{fig:qgqbarT} can be written as 
\begin{equation}
\label{eq:TqgqbarWF}
\psi^{\gamma^{\ast}_{\rm T}\rightarrow q\bar{q}g}_{\ref{diag:qgqbarT}} = \int \dkp (2\pi)^{d-1}\delta^{(d-1)}(\kpvec -\pvec -\kvec) 
\frac{\paircr^{\gamma^{\ast}_{\rm T}}_{\lambda;h,-h}(\ktp,z_1+z_2)
\qemit^{\beta;\alpha,a}_{h;\sigma,h}(\mt,z_2/(z_1+z_2))
}{\Delta^{-}_{01}\Delta^{-}_{02}}
\end{equation}
where the vertex $\paircr^{\gamma^{\ast}_{\rm T}}_{\lambda;h,-h}(\ktp,z_1+z_2)$ for a transverse photon splitting into a quark antiquark dipole is defined in \eq\nr{eq:LOTvirtualphoton} and the gluon emission vertex $\qemit^{\beta;\alpha,a}_{h;\sigma,h}(\mt,z_2/(z_1+z_2))$ from a quark is given by \eq\nr{eq:vertexqtoqgd}. The phase space measure simplifies to 
\begin{equation}
\label{eq:TqgqbarPS}
\int \dkp (2\pi)^{d-1}\delta^{(d-1)}(\kpvec -\pvec -\kvec) = \frac{1}{2(z_1+z_2)q^+}
\end{equation}
and the LC energy denominators are given by
\begin{equation}
\label{eq:TqgqbarED}
\begin{split}
\Delta^{-}_{01} & = \frac{1}{(-2q^+)z_3(z_1+z_2)}\biggl [\ptp^2 + \overline{Q}^2_{\ref{diag:qgqbarT}} \biggr ]\\ 
\Delta^{-}_{02} & = \frac{z_1+z_2}{(-2q^+)z_1z_2}\biggl [\mt^2 + \omega_{\ref{diag:qgqbarT}}\left ( \ptp^2 + \overline{Q}^2_{\ref{diag:qgqbarT}} \right )\biggr ],\\
\end{split}
\end{equation}
where 
\begin{equation}
\overline{Q}^2_{\ref{diag:qgqbarT}} = z_3(z_1+z_2)Q^2, \quad\quad \omega_{\ref{diag:qgqbarT}} = \frac{z_1z_2}{z_3(z_1+z_2)^2}.
\end{equation}
Using \eqs\nr{eq:TqgqbarPS} and \nr{eq:TqgqbarED} as well as the expression for the vertices, we find 
\begin{equation}
\label{eq:TqgqbarWFfinal}
\begin{split}
\psi^{\gamma^{\ast}_{\rm T}\rightarrow q\bar{q}g}_{\ref{diag:qgqbarT}} &= +8q^+ee_f(gt^{a}_{\alpha\beta})(z_1z_3)^{1/2}\biggl [\left (z_1+z_2-\frac{1}{2}\right )\delta^{ij}_{(d_s)} - ih\frac{1}{2}\epsilon_{(d_s)}^{ij} \biggr ]\\
& \times \biggl [\left (1 - \frac{1}{2}\left(\frac{z_2}{z_1+z_2}\right ) \right ) \delta^{kl}_{(d_s)} + ih\frac{1}{2}\left (\frac{z_2}{z_1+z_2}\right )\epsilon_{(d_s)}^{kl} \biggr ] \frac{(-\ptp)^i\mt^{k}\epst^{j}_{\lambda}\epst^{\ast l}_{\sigma}}{\biggl [\ptp^2 + \overline{Q}^2_{\ref{diag:qgqbarT}} \biggr ]\biggl [\mt^2 + \omega_{\ref{diag:qgqbarT}} \left ( \ptp^2 + \overline{Q}^2_{\ref{diag:qgqbarT}}\right )\biggr ]}.
\end{split}
\end{equation}
 
Similarly, the LCWF for diagram \ref{diag:qqbargT} is given by
\begin{equation}
\label{eq:TqqbargWF}
\psi^{\gamma^{\ast}_{\rm T}\rightarrow q\bar{q}g}_{\ref{diag:qqbargT}} = \int \dkp (2\pi)^{d-1}\delta^{(d-1)}(\kpvec -\kvec -\ppvec) \frac{A^{\gamma^{\ast}_{\rm T}}_{\lambda;h,-h}(\pt,z_1)\overline{V}^{\alpha;\beta,a}_{-h;,\sigma,-h}(\lt,z_2/(z_2+z_3))}{\Delta^{-}_{01}\Delta^{-}_{02}}
\end{equation}
where the gluon emission vertex $\qbemit^{\alpha;\beta,a}_{-h;\sigma,-h}(\lt,z_2/(z_2+z_3))$ from an anti-quark is given by \eq\nr{eq:vertexqbartoqbarg}. The phase space measure simplifies to
\begin{equation}
\label{eq:TqqbargPS}
\int \dkp (2\pi)^{d-1}\delta^{(d-1)}(\kpvec -\kvec -\ppvec) = \frac{1}{2(z_2+z_3)q^+}
\end{equation}
and the LC energy denominators are given by 
\begin{equation}
\label{eq:TqqbargED}
\begin{split}
\Delta^{-}_{01} & = \frac{1}{(-2q^+)z_1(z_2+z_3)}\biggl [\pt^2 + \overline{Q}^2_{\ref{diag:qqbargT} } \biggr ]\\ 
\Delta^{-}_{02} & = \frac{z_2+z_3}{(-2q^+)z_2z_3}\biggl [\lt^2 + \omega_{\ref{diag:qqbargT} }\left ( \pt^2 + \overline{Q}^2_{\ref{diag:qqbargT} }\right )\biggr ]\\
\end{split}
\end{equation}
with
\begin{equation}
\overline{Q}^2_{\ref{diag:qqbargT}} = z_1(z_2+z_3)Q^2, \quad\quad \omega_{\ref{diag:qqbargT} } = \frac{z_2z_3}{z_1(z_2+z_3)^2}.
\end{equation}
Putting everything together we obtain 
\begin{equation}
\label{eq:TqqbargWFfinal}
\begin{split}
\psi^{\gamma^{\ast}_{\rm T}\rightarrow q\bar{q}g}_{\ref{diag:qqbargT}} &= -8q^+ee_f(gt^{a}_{\alpha\beta})(z_1z_3)^{1/2}\biggl [\left (z_1-\frac{1}{2}\right )\delta^{ij}_{(d_s)} - ih\frac{1}{2}\epsilon_{(d_s)}^{ij}\biggr ]\\
&\times  \biggl [\left (1 - \frac{1}{2}\left (\frac{z_2}{z_2+z_3}\right ) \right ) \delta^{kl}_{(d_s)} - ih\frac{1}{2}\left (\frac{z_2}{z_2+z_3}\right )\epsilon_{(d_s)}^{kl} \biggr ] \frac{\pt^i\lt^{k}\epst^{j}_{\lambda}\epst^{\ast l}_{\sigma}}{\biggl [\pt^2 + \overline{Q}^2_{\ref{diag:qqbargT} }  \biggr ]\biggl [\lt^2 + \omega_{\ref{diag:qqbargT} } \left ( \pt^2 + \overline{Q}^2_{\ref{diag:qqbargT} } \right )\biggr ]}.
\end{split}
\end{equation}
 
 The LCWF for the instantaneous diagram \ref{diag:qgqbarinst} shown in \fig\ref{fig:qgqbarinst} is given by 
 \begin{equation}
\label{eq:TqgqbarWFfinalinst}
\begin{split}
\psi^{\gamma^{\ast}_{\rm T}\rightarrow q\bar{q}g}_{\ref{diag:qgqbarinst}}  = -2q^+ ee_f (gt^{a}_{\alpha\beta})\frac{z_1z_2}{(z_1+z_2)^2}(z_1z_3)^{1/2}\biggl [\delta^{ij}_{(d_s)} + ih\epsilon^{ij}_{(d_s)} \biggr ]\frac{\epst^{\ast i}_{\sigma}\epst^{j}_{\lambda}}{\biggl [\mt^2 + \omega_{\ref{diag:qgqbarinst}}\left ( \ptp^2 + \overline{Q}^2_{\ref{diag:qgqbarinst}}\right ) \biggr ]}
\end{split}
\end{equation}
where the matrix element for the instantaneous interaction is defined in \eq\nr{eq:meqgqbarinst}, and the LC energy denominator is given by \eq\nr{eq:TqgqbarED} with $\mt = \kt + (z_2/(z_1+z_2))\ptp$ and 
\begin{equation}
\overline{Q}^2_{\ref{diag:qgqbarinst}} = z_3(z_1+z_2)Q^2,\quad\quad \omega_{\ref{diag:qgqbarinst}} = \frac{z_1z_2}{z_3(z_1+z_2)^2}. 
\end{equation}
Similarly, the LCWF for \ref{diag:qqbarginst} shown in \fig\nr{fig:qqbarginst} is given by 
\begin{equation}
\label{eq:qqbarginstTresult}
\begin{split}
\psi^{\gamma^{\ast}_{\rm T} \rightarrow q\bar{q}g}_{\ref{diag:qqbarginst}}  = +2q^+ ee_f (gt^{a}_{\alpha\beta})\frac{z_3z_2}{(z_2+z_3)^2}(z_1z_3)^{1/2}\biggl [\delta^{ij}_{(d_s)} - ih\epsilon^{ij}_{(d_s)} \biggr ]\frac{\epst^{\ast i}_{\sigma}\epst^{j}_{\lambda}}{\biggl [\lt^2 + \omega_{\ref{diag:qqbarginst}} \left ( \pt^2 + Q^2_{\ref{diag:qqbarginst}}\right ) \biggr ]}
\end{split}
\end{equation}
where the matrix element for the instantaneous interaction is defined in \eq\nr{eq:meqqbarginst}, and the LC energy denominator is given by \eq\nr{eq:TqqbargED} with $\lt = \kt + (z_2/(z_2+z_3))\pt$ and 
\begin{equation}
Q^2_{\ref{diag:qqbarginst}} = z_1(z_2+z_3)Q^2, \quad\quad \omega_{\ref{diag:qqbarginst}} = \frac{z_2z_3}{z_1(z_2+z_3)^2}. 
\end{equation}

Finally, adding the results in \eqs\nr{eq:TqgqbarWFfinal}, \nr{eq:TqqbargWFfinal}, \nr{eq:TqgqbarWFfinalinst} and \nr{eq:qqbarginstTresult} together, we obtain the full tree level contribution to $q\bar{q}g$-component of the transverse virtual photon wave function
\begin{equation}
\label{eq:Tradfinal}
\begin{split}
\psi^{\gamma^{\ast}_{\rm T}\rightarrow q\bar{q}g}_{}\bigg\vert_{\rm FDH} = 8q^+ee_f(gt^{a}_{\alpha\beta})(z_1z_3)^{1/2}&\biggl \{-\Sigma^{ijkl}_{\ref{diag:qgqbarT}}\frac{\ptp^i\mt^{k}\epst^{j}_{\lambda}\epst^{\ast l}_{\sigma}}{\biggl [\ptp^2 + \overline{Q}^2_{\ref{diag:qgqbarT}} \biggr ]\biggl [\mt^2 + \omega_{\ref{diag:qgqbarT}} \left ( \ptp^2 + \overline{Q}^2_{\ref{diag:qgqbarT}}\right )\biggr ]}\\
&- \Sigma^{ijkl}_{\ref{diag:qqbargT}}\frac{\pt^i\lt^{k}\epst^{j}_{\lambda}\epst^{\ast l}_{\sigma}}{\biggl [\pt^2 + \overline{Q}^2_{\ref{diag:qqbargT} }  \biggr ]\biggl [\lt^2 + \omega_{\ref{diag:qqbargT} } \left ( \pt^2 + \overline{Q}^2_{\ref{diag:qqbargT} } \right )\biggr ]}\\
& - \Sigma^{ij}_{\ref{diag:qgqbarinst}}\frac{\epst^{\ast i}_{\sigma}\epst^{j}_{\lambda}}{\biggl [\mt^2 + \omega_{\ref{diag:qgqbarinst}}\left ( \ptp^2 + \overline{Q}^2_{\ref{diag:qgqbarinst}}\right ) \biggr ]} + \Sigma^{ij}_{\ref{diag:qqbarginst}}\frac{\epst^{\ast i}_{\sigma}\epst^{j}_{\lambda}}{\biggl [\lt^2 + \omega_{\ref{diag:qqbarginst}} \left ( \pt^2 + Q^2_{\ref{diag:qqbarginst}}\right ) \biggr ]}  \biggr \}
,
\end{split}
\end{equation} 
where we have introduced the following notation:
\begin{equation}
\Sigma^{ijkl}_{\ref{diag:qgqbarT}} = \biggl [\left (z_1+z_2-\frac{1}{2}\right )\delta^{ij}_{(d_s)} - ih\frac{1}{2}\epsilon_{(d_s)}^{ij} \biggr ]\biggl [\left (1 - \frac{1}{2}\left(\frac{z_2}{z_1+z_2}\right ) \right ) \delta^{kl}_{(d_s)} + ih\frac{1}{2}\left (\frac{z_2}{z_1+z_2}\right )\epsilon_{(d_s)}^{kl} \biggr ]
\end{equation}
\begin{equation}
\Sigma^{ijkl}_{\ref{diag:qqbargT}} = \biggl [\left (z_1-\frac{1}{2}\right )\delta^{ij}_{(d_s)} - ih\frac{1}{2}\epsilon_{(d_s)}^{ij}\biggr ]\biggl [\left (1 - \frac{1}{2}\left (\frac{z_2}{z_2+z_3}\right ) \right ) \delta^{kl}_{(d_s)} - ih\frac{1}{2}\left (\frac{z_2}{z_2+z_3}\right )\epsilon_{(d_s)}^{kl} \biggr ]
\end{equation}
\begin{equation}
\Sigma^{ij}_{\ref{diag:qgqbarinst}} = \frac{1}{4}\frac{z_1z_2}{(z_1+z_2)^2}\biggl [\delta^{ij}_{(d_s)} + ih\epsilon^{ij}_{(d_s)} \biggr ]
\end{equation}
\begin{equation}
\Sigma^{ij}_{\ref{diag:qqbarginst}} = \frac{1}{4}\frac{z_3z_2}{(z_2+z_3)^2}\biggl [\delta^{ij}_{(d_s)} - ih\epsilon^{ij}_{(d_s)} \biggr ].
\end{equation}

\subsection{Longitudinal photon}
\begin{figure}[t]
\centerline{
\includegraphics[width=6.4cm]{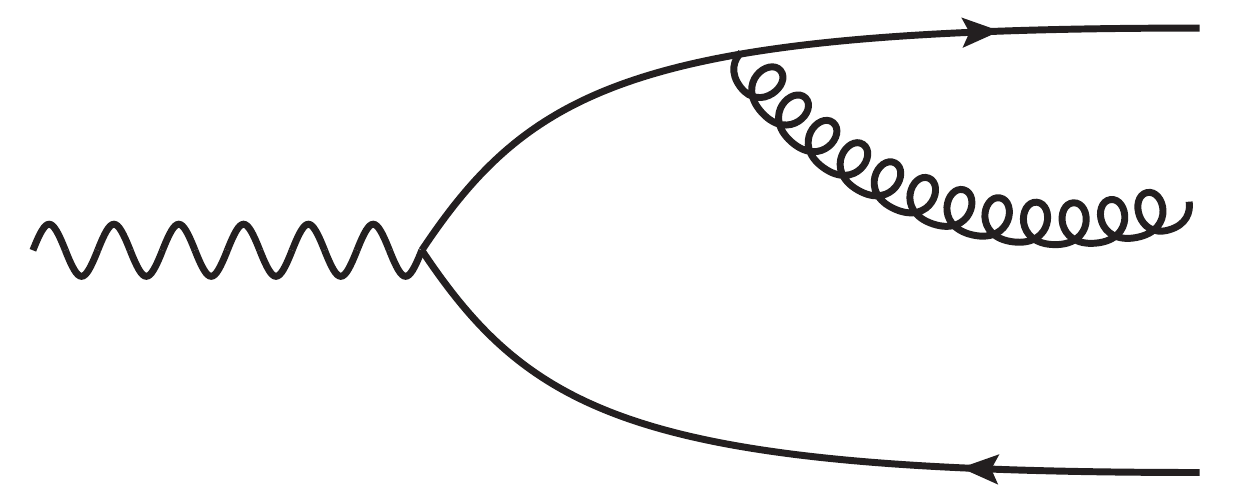}
\begin{tikzpicture}[overlay]
\draw [dashed] (-5.5,2.8) -- (-5.5,0);
\node[anchor=north] at (-5.5cm,-0.2cm) {0};
\draw [dashed] (-3.2,2.8) -- (-3.2,0);
\node[anchor=north] at (-3.2cm,-0.2cm) {1};
\draw [dashed] (-0.8,2.8) -- (-0.8,0);
\node[anchor=north] at (-0.8,-0.2) {2};
\node[anchor=west] at (-0.3,2.5) {$\pvec,h,\alpha$};
\node[anchor=west] at (-0.3,1.2) {$\kvec,\sigma,a$};
\node[anchor=west] at (-0.3,0.2) {$\ppvec,-h,\beta$};
\node[anchor=west] at (-5.3,0.9) {$\qvec, \lambda $};
\node[anchor=west] at (-1.8,1.8) {$\mt$};
\node[anchor=south west] at (-4.4,2.1) {$\kpvec,h,\beta$};
\node[anchor=east] at (-6.3,1.3) {$\gamma^{\ast}_{\rm L}$};
\node[anchor=south west] at (-7cm,0cm) {\namediag{diag:qgqbarL}};
 \end{tikzpicture}
}
\rule{0pt}{1ex}
\caption{Diagram \ref{diag:qgqbarL} contributing to the $q\bar{q}g$-component of the longitudinal virtual photon wave function at NLO with energy denominators and kinematics. Momentum conservation: $\qvec=\kpvec + \ppvec$, $\kpvec=\pvec + \kvec$ and
$\qvec=\pvec+\ppvec + \kvec$.   Momentum fractions are defined by $p^+ = z_1q^+$, $k^+ = z_2q^+$, $p'^+ = z_3q^+$  and $k'^+ = (z_1+z_2)q^+$. The momentum fraction of the virtual photon splitting into a quark anti-quark dipole is $k'^+/q^+ = z_1+z_2$ and the momentum fraction of the gluon emission is $k^+/k'^+ = z_2/(z_1+z_2)$. The natural momenta for the gluon emission vertex is $\mt \equiv \kt-(z_2/(z_1+z_2))\ktp = \kt + (z_2/(z_1+z_2))\ptp$. Note that the momentum fractions are related to each other via relation $z_1+z_2+z_3 =1$.}
\label{fig:qgqbarL}
 \end{figure}
 
\begin{figure}[t]
\centerline{
\includegraphics[width=6.4cm]{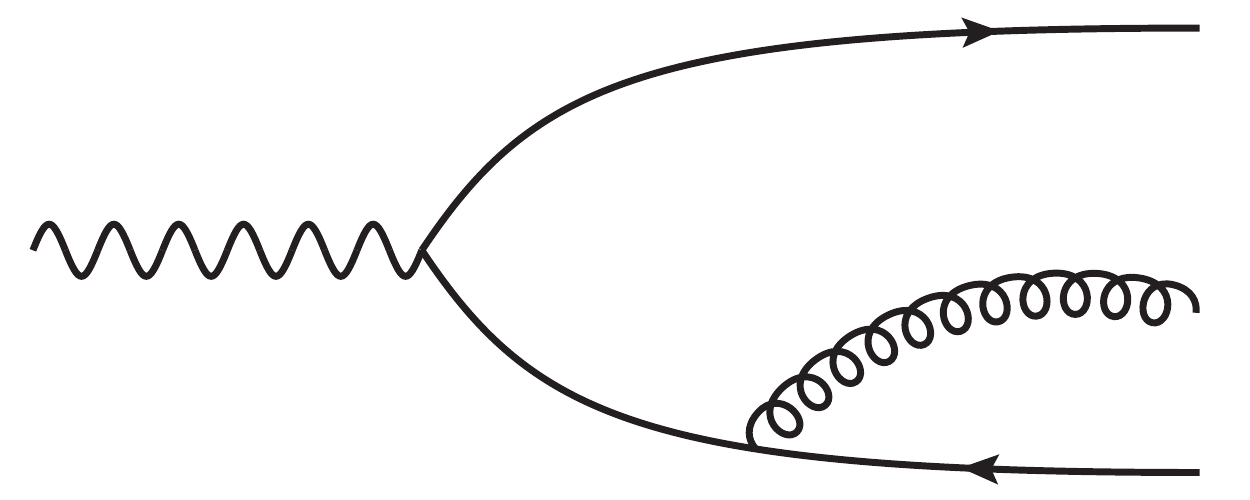}
\begin{tikzpicture}[overlay]
\draw [dashed] (-5.5,2.8) -- (-5.5,0);
\node[anchor=north] at (-5.5cm,-0.2cm) {0};
\draw [dashed] (-3.2,2.8) -- (-3.2,0);
\node[anchor=north] at (-3.2cm,-0.2cm) {1};
\draw [dashed] (-0.8,2.8) -- (-0.8,0);
\node[anchor=north] at (-0.8,-0.2) {2};
\node[anchor=west] at (-0.3,2.5) {$\pvec,h,\alpha$};
\node[anchor=west] at (-0.3,1.1) {$\kvec,\sigma,a$};
\node[anchor=west] at (-0.3,0.2) {$\ppvec,-h,\beta$};
\node[anchor=west] at (-5.3,0.9) {$\qvec, \lambda $};
\node[anchor=west] at (-1.8,0.5) {$\lt$};
\node[anchor=south west] at (-4.6,-0.1) {$\kpvec,-h,\alpha$};
\node[anchor=east] at (-6.3,1.3) {$\gamma^{\ast}_{\rm L}$};
\node[anchor=south west] at (-7cm,0cm) {\namediag{diag:qqbargL}};
 \end{tikzpicture}
}
\rule{0pt}{1ex}
\caption{Diagram \ref{diag:qqbargL} contributing to the $q\bar{q}g$-component of the longitudinal virtual photon wave function at NLO with energy denominators and kinematics. Momentum conservation: $\qvec=\pvec + \kpvec$, $\kpvec=\kvec + \ppvec$ and $\qvec=\pvec+\ppvec + \kvec$.   Momentum fractions are defined by $ p^+ = z_1q^+$, $k^+ = z_2q^+$, $p'^+ = z_3q^+$ and $k'^+ = (z_2+z_3)q^+$. The momentum fraction of the virtual photon splitting into a quark anti-quark dipole is $p^+/q^+ = z_1$ and the momentum fraction of the gluon emission is $k^+/k'^+ = z_2/(z_2+z_3)$. The natural momenta for the gluon emission vertex is $\lt \equiv \kt-(z_2/(z_2+z_3))\ktp = \kt + (z_2/(z_2+z_3))\pt$. Note that the momentum fractions are related to each other via relation $z_1+z_2+z_3 =1$.}
\label{fig:qqbargL}
\end{figure}

For gluon emission from a longitudinal photon state, we calculate the diagrams \ref{diag:qgqbarL} and \ref{diag:qqbargL} shown in \figs\nr{fig:qgqbarL} and \nr{fig:qqbargL} contributing to the $q\bar{q}g$-component of the longitudinal virtual photon wave function at NLO. There are no instantaneous diagrams to consider, because strictly speaking the longitudinal photon itself is a part of an instantaneous interaction with the emitting electron.

The LCWF for diagram \ref{diag:qgqbarL} is given by
\begin{equation}
\label{eq:LqgqbarWF}
\psi^{\gamma^{\ast}_{\rm L}\rightarrow q\bar{q}g}_{\ref{diag:qgqbarL}} = \int \dkp (2\pi)^{d-1}\delta^{(d-1)}(\kpvec -\pvec -\kvec) 
\frac{
\paircr^{\gamma^{\ast}_{\rm L}}_{\lambda;h,-h}(Q,z_1 + z_2)
\qemit^{\beta;\alpha,a}_{h;\sigma, h}(\mt,z_2/(z_1+z_2))
}{\Delta^{-}_{01}\Delta^{-}_{02}}
\end{equation}
where the vertex $A^{\gamma^{\ast}_{\rm L}}_{\lambda;h,-h}(Q,z)$ for a longitudinal photon splitting into an quark anti-quark dipole is defined in \eq\nr{eq:LOLvirtualphoton} and the gluon emission vertex $\qemit^{\beta;\alpha,a}_{h;\sigma, h}$ in \eq\nr{eq:vertexqtoqgd}. The phase space measure and LC energy denominators are the same as in \eqs\nr{eq:TqgqbarPS} and \nr{eq:TqgqbarED}, and thus we get 
\begin{equation}
\label{eq:qqbargLfinaln}
\begin{split}
\psi^{\gamma^{\ast}_{\rm L}\rightarrow q\bar{q}g}_{\ref{diag:qgqbarL}} = +8q^+ee_fQ(gt^{a}_{\alpha\beta})z_1z_3(z_1+z_2)\left (\frac{z_3}{z_1}\right )^{1/2}\biggl [&\left (1  - \frac{1}{2}\left (\frac{z_2}{z_1+z_2} \right )\right ) \delta^{ij}_{(d_s)} + ih\frac{1}{2}\left (\frac{z_2}{z_1+z_2} \right )\epsilon_{(d_s)}^{ij} \biggr ]\\
& \times \frac{\mt^{i}\epst^{\ast j}_{\sigma}}{\biggl [\ptp^2 + \overline{Q}_{\ref{diag:qgqbarL}}^2 \biggr ]\biggl [\mt^2 + \omega_{\ref{diag:qgqbarL}}\left ( \ptp^2 + \overline{Q}^2_{\ref{diag:qgqbarL}}\right )\biggr ]},
\end{split}
\end{equation}
where 
\begin{equation}
\label{eq:QW(l)}
\overline{Q}^2_{\ref{diag:qgqbarL}} = z_3(z_1+z_2)Q^2, \quad\quad \omega_{\ref{diag:qgqbarL}} = \frac{z_1z_2}{(z_1+z_2)^2z_3}.
\end{equation}
Similarly, the LCWF for diagram \ref{diag:qqbargL} can be written as 
\begin{equation}
\label{eq:LqqbargWF}
\psi^{\gamma^{\ast}_{\rm L}\rightarrow q\bar{q}g}_{\ref{diag:qqbargL}} = \int \dkp (2\pi)^{d-1}\delta^{(d-1)}(\kpvec -\kvec -\ppvec) \frac{A^{\gamma^{\ast}_{\rm L}}_{\lambda;h,-h}(Q,z_1)\overline{V}^{\alpha;\beta,a}_{-h;\sigma,-h}(\lt,z_2/(z_2+z_3))}{\Delta^{-}_{01}\Delta^{-}_{02}},
\end{equation}
where the phase space measure and LC energy denominators are given in \eqs\nr{eq:TqqbargPS} and \nr{eq:TqqbargED}. Putting everything together we obtain 
\begin{equation}
\label{eq:qqbargLfinalo}
\begin{split}
\psi^{\gamma^{\ast}_{\rm L}\rightarrow q\bar{q}g}_{\ref{diag:qqbargL}} = -8q^+ee_fQ(gt^{a}_{\alpha\beta})z_1z_3(z_2+z_3)\left (\frac{z_1}{z_3}\right )^{1/2}\biggl [&\left (1 - \frac{1}{2}\left (\frac{z_2}{z_2 + z_3}\right ) \right ) \delta^{ij}_{(d_s)} - ih\frac{1}{2}\left (\frac{z_2}{z_2 + z_3}\right )\epsilon_{(d_s)}^{ij} \biggr ]\\
& \times \frac{\lt^{i}\epst^{\ast j}_{\sigma}}{\biggl [\pt^2 + \overline{Q}^2_{\ref{diag:qqbargL}} \biggr ]\biggl [\lt^2 + \omega_{\ref{diag:qqbargL}} \left ( \pt^2 + \overline{Q}^2_{\ref{diag:qqbargL}}\right )\biggr ]},
\end{split}
\end{equation}
where 
\begin{equation}
\label{eq:QW(m)}
\overline{Q}^2_{\ref{diag:qqbargL}} = z_1(z_2+z_3)Q^2, \quad\quad \omega_{\ref{diag:qqbargL}} = \frac{z_2z_3}{(z_2+z_3)^2z_1}.
\end{equation}

Finally, summing the contributions in \eqs\nr{eq:qqbargLfinaln} and \nr{eq:qqbargLfinalo} together we get 
\begin{equation}
\label{eq:finalradLCWFlong}
\begin{split}
\psi^{\gamma^{\ast}_{\rm L}\rightarrow q\bar{q}g}_{}\bigg\vert_{\rm FDH} = 8q^+ee_fQ(gt^{a}_{\alpha\beta})z_1z_3\biggl [\Sigma^{ij}_{\ref{diag:qgqbarL}}& \frac{\mt^{i}}{\biggl [\ptp^2 + \overline{Q}^2_{\ref{diag:qgqbarL}} \biggr ]\biggl [\mt^2 + \omega_{\ref{diag:qgqbarL}}\left ( \ptp^2 + \overline{Q}^2_{\ref{diag:qgqbarL}}\right )\biggr ]}\\
& - \Sigma^{ij}_{\ref{diag:qqbargL}}\frac{\lt^{i}}{\biggl [\pt^2 + \overline{Q}^2_{\ref{diag:qqbargL}} \biggr ]\biggl [\lt^2 + \omega_{\ref{diag:qqbargL}}\left ( \pt^2 + \overline{Q}^2_{\ref{diag:qqbargL}}\right )\biggr ]} \biggr ]\epst_{\sigma}^{\ast j},
\end{split}
\end{equation}
where we have defined
\begin{equation}
\Sigma^{ij}_{\ref{diag:qgqbarL}}  = (z_1+z_2)\left (\frac{z_3}{z_1}\right )^{1/2}\biggl [\left (1  - \frac{1}{2}\left (\frac{z_2}{z_1+z_2} \right )\right ) \delta^{ij}_{(d_s)} + ih\frac{1}{2}\left (\frac{z_2}{z_1+z_2} \right )\epsilon_{(d_s)}^{ij} \biggr ]
\end{equation}
\begin{equation}
\Sigma^{ij}_{\ref{diag:qqbargL}} = (z_2+z_3)\left (\frac{z_1}{z_3}\right )^{1/2}\biggl [\left (1  - \frac{1}{2}\left (\frac{z_2}{z_2+z_3} \right )\right ) \delta^{ij}_{(d_s)} - ih\frac{1}{2}\left (\frac{z_2}{z_2+z_3} \right )\epsilon_{(d_s)}^{ij} \biggr ].
\end{equation}

\section{NLO DIS cross section}
\label{sec:nlodis}

As explained in Sec.~\ref{sec:sigma}, in order to calculate the DIS cross section we first need to  Fourier transform the final momentum space expressions of the transverse and longitudinal virtual photon LCWFs to  mixed space.  Because of the simple algebraic structure, we will first consider the longitudinal virtual photon case.

\subsection{Longitudinal photon}
\label{sec:longitsigma}
According to \eq\nr{eq:mixedspaceqbarqfull}, the mixed space expression for the $q\bar{q}$-component of the longitudinal virtual photon amplitude computed in the FDH scheme at NLO accuracy can be written as
\begin{equation}
\label{eq:longitudinalamplitudeqqbar}
\vert \gamma^{\ast}_{\rm L}(q^+,Q^2,\lambda)\rangle_{q\bar{q}} = \sum_{h}\sum_{\text{color}}
\mathcal{PS}^{+}_{(2)}\int_{ \xt  \yt }
\left (
\widetilde{\psi}^{\gamma^{\ast}_{\rm L}\rightarrow q\bar{q}}_{\lo}\bigg\vert_{\rm FDH} + \widetilde{\psi}^{\gamma^{\ast}_{\rm L}\rightarrow q\bar{q}}_{\nlo}\bigg\vert_{\rm FDH}
\right )\vert q(p^+,\xt,h,\alpha)\bar{q}(p'^+,\yt,-h,\beta)\rangle,
\end{equation}
where the two particle plus momentum phase space factor, $\mathcal{PS}^{+}_{(2)}$, defined in \eq\nr{eq:plusmomPS} is given by
\begin{equation}
\label{eq:PS2withz}
\mathcal{PS}^{+}_{(2)} = \frac{1}{8\pi q^+}\int_{0}^{1}\frac{\ud z}{z(1-z)}.
\end{equation}
The transverse Fourier transformed LO and NLO light cone wave functions for longitudinal virtual photon in the mixed space are given by
\begin{equation}
\begin{split}
\widetilde{\psi}^{\gamma^{\ast}_{\rm L}\rightarrow q\bar{q}}_{\lo/\nlo}\bigg\vert_{\rm FDH}  =
\int \frac{\ud^2 \pt}{(2\pi)^2}\left (\psi^{\gamma^{\ast}_{\rm L}\rightarrow q\bar{q}}_{\lo/\nlo}\bigg\vert_{\rm FDH}\right )e^{i\pt\cdot \rt_{xy}} 
\end{split}
\end{equation}
with $\rt_{xy} \equiv \xt-\yt$. Using \eq\nr{eq:LOLvirtualphoton} together with \eq\nr{eq:FTintegralsfinal} we get
\begin{equation}
\begin{split}
\widetilde{\psi}^{\gamma^{\ast}_{\rm L}\rightarrow q\bar{q}}_{\lo}\bigg\vert_{\rm FDH} 
& = 4q^+ee_fQ\delta_{\alpha\beta}[z(1-z)]^{3/2}\int\frac{\ud^2\pt}{(2\pi)^2}\frac{e^{i\pt\cdot \rt_{xy}}}{\biggl [\pt^2 + \overline{Q}^2\biggr ]}\\
& = \frac{4q^+ee_fQ\delta_{\alpha\beta}}{(2\pi)}[z(1-z)]^{3/2}K_{0}\left (\overline{Q}\vert \rt_{xy}\vert\right ).
\end{split}
\end{equation}
Correspondingly, using  \eqs\nr{eq:fullLWFqbarq}, \nr{eq:FTintegralsfinal} and \nr{eq:FTintegralsextra1final} gives
\begin{equation}
\widetilde{\psi}^{\gamma^{\ast}_{\rm L}\rightarrow q\bar{q}}_{\nlo}\bigg\vert_{\rm FDH} = \frac{4q^+ee_fQ\delta_{\alpha\beta}}{(2\pi)}\left (\frac{g_r^2\cf}{8\pi^2}\right )[z(1-z)]^{3/2}K_{0}\left (\overline{Q}\vert \rt_{xy}\vert\right )\mathcal{K}^{\gamma^{\ast}_{\rm L}}\bigg\vert_{\rm FDH},   
\end{equation}
where the NLO kernel for longitudinal virtual photon written in the mixed space reads  
\begin{equation}
\label{eq:NLOkernelL}
\mathcal{K}^{\gamma^{\ast}_{\rm L}}\bigg\vert_{\rm FDH}  = \biggl [\frac{3}{2} + \log\left (\frac{\alpha}{z}\right ) + \log\left (\frac{\alpha}{1-z}\right ) \biggr ]\biggl \{\frac{1}{\varepsilon_{\overline{\rm MS}}}  + \log\left (\frac{\rt_{xy}^2\mu^2}{4}\right ) - 2\Psi_{0}(1)\biggr \} + \frac{1}{2}\log^2\left (\frac{z}{1-z}\right ) - \frac{\pi^2}{6} + \frac{5}{2} + \mathcal{O}(\varepsilon).
\end{equation}
Note again that  in the CDR scheme (see \cite{Beuf:2017bpd}) there is an extra factor of $1/2$ in the expression of NLO kernel defined in \eq\nr{eq:NLOkernelL}, which is the scheme dependent part of the one-loop computation of longitudinal virtual photon LCWF. 

Next, operating on the amplitude in \eq\nr{eq:longitudinalamplitudeqqbar} with the eikonal scattering operator $(1-\hat{S}_E)$ and finally squaring the expression and simplifying the color algebra as in \eq\nr{eq:qqbarmatrixelement} we find 
\begin{equation}
\label{eq:NLOLfinsquare}
\begin{split}
{}_{q\bar{q}}\langle \gamma^{\ast}_{\rm L}(q'^+,Q^2,\lambda')&\vert 1-\hat{S}_E \vert \gamma^{\ast}_{\rm L}(q^+,Q^2,\lambda)\rangle_{q\bar{q}} = 2q^+(2\pi)\delta(q'^+-q^+)\frac{8\nc\alpha_{em}e_f^2Q^2}{(2\pi)^2}\\
&\times
\int_{\xt\yt}
\int_{0}^{1}\ud z z^2(1-z)^2[K_{0}\left (\vert \rt_{xy}\vert\overline{Q}\right )]^2\biggl [1 + \left ( \frac{\alpha_s\cf}{\pi}\right )\mathcal{K}^{\gamma^{\ast}_{\rm L}}\bigg\vert_{\rm FDH}\biggr ]\left (1-\mathcal{S}_{xy}\right ) + \mathcal{O}(\alpha_{em}\alpha_s^2),
\end{split}
\end{equation} 
where we have summed over the helicity and color, introduced the fine structure constants $\alpha_s = g_r^2/4\pi$ and $\alpha_{em} = e^2/4\pi$ and the notation
\begin{equation}
\mathcal{S}_{xy} = \frac{1}{\nc}\mathrm{Tr}\left (U[A](\xt)U^{\dagger}[A](\yt)\right ).
\end{equation}

Similarly, using \eqs\nr{eq:mixedspaceqbarqgfull}, \nr{eq:plusmomPS}, and \nr{eq:finalradLCWFlong} the mixed space expression for the $q\bar{q}g$-component of the longitudinal virtual photon amplitude computed in the FDH scheme at NLO accuracy simplifies to 
\begin{equation}
\label{eq:qbarqgamplitude}
\vert \gamma^{\ast}_{\rm L}(q^+,Q^2,\lambda)\rangle_{q\bar{q}g} = \sum_{h,\sigma}\sum_{\text{color}}\mathcal{PS}^{+}_{(3)}\int_{\xt  \yt [\zt]}
\left (\widetilde{\psi}^{\gamma^{\ast}_{\rm L}\rightarrow q\bar{q}g}_{}\bigg\vert_{\rm FDH}\right )\vert q(p^+,\xt,h,\alpha)\bar{q}(p'^+,\yt,-h,\beta)g(k^+,\zt,\sigma,a) \rangle,
\end{equation}
where we have denoted by 
\begin{equation}
\int_{[\zt]}= \int \ud^{d-2} \zt
\end{equation}
the integral over the gluon phase space, which must be done in $d$ dimensional spacetime. The quark and antiquark are ``observed'' particles in the FDH scheme, and thus the integrals over $\xt,\yt$ can be kept in 2 dimensions, simplifying the final state phase space integrations.  
The three particle plus momentum phase space factor, $\mathcal{PS}^{+}_{(3)}$, is given by
\begin{equation}
\label{eq:PS3withz}
\mathcal{PS}^{+}_{(3)} = \frac{1}{8q^+(2\pi)^2}\int_{0}^{\infty}\ud z_1\int_{0}^{\infty}\ud z_2\int_{0}^{\infty} \ud z_3\frac{1}{z_1z_2z_3}\delta(z_1+z_2+z_3-1),
\end{equation}
and
\begin{equation}
\begin{split}
\widetilde{\psi}^{\gamma^{\ast}_{\rm L}\rightarrow q\bar{q}g}_{}\bigg\vert_{\rm FDH} = 8q^+ee_fQg_rt^{a}_{\alpha\beta}z_1z_3\biggl \{ \Sigma^{ij}_{\ref{diag:qgqbarL}}\mathcal{I}^{i}(\rt_{yxz},\rt_{zx},\overline{Q}^2_{\ref{diag:qgqbarL}},\omega_{\ref{diag:qgqbarL}})- \Sigma^{ij}_{\ref{diag:qqbargL}}\mathcal{I}^{i}(\rt_{xyz},\rt_{zy},\overline{Q}^2_{\ref{diag:qqbargL}},\omega_{\ref{diag:qqbargL}})\biggr \}\epst_{\sigma}^{\ast j}.
\end{split}
\end{equation}
Here we have defined the function $\mathcal{I}^i(\bt,\rt,\overline{Q}^2,\omega)$ as  
\begin{equation}
\label{eq:Imaster}
\mathcal{I}^i(\bt,\rt,\overline{Q}^2,\omega) = \mu^{2-\frac{d}{2}}\int \frac{\ud^2\Pt}{(2\pi)^2}\int\frac{\ud^{d-2}\Kt}{(2\pi)^{d-2}}\frac{\Kt^ie^{i\Pt\cdot \bt}e^{i\Kt\cdot \rt}}{\biggl [\Pt^2 + \overline{Q}^2\biggr ]\biggl [\Kt^2 + \omega\left (\Pt^2 + \overline{Q}^2\right )\biggr ]}
\end{equation}
and introduced the notation
\begin{equation}
\rt_{yxz} = \rt_{yx} - \left (\frac{z_2}{z_1+z_2}\right )\rt_{zx}, \quad\quad \rt_{xyz} = \rt_{xy} -  \left (\frac{z_2}{z_2+z_3}\right )\rt_{zy}.
\end{equation}
The evaluation of the integral defined in \eq\nr{eq:Imaster} is outlined in Appendix~\ref{app:transversefint}, leading to
\begin{equation}
\label{eq:Ixyfinal}
\mathcal{I}^i(\bt,\rt,\overline{Q}^2,\omega) = \mu^{2-\frac{d}{2}}\frac{i}{8}\pi^{-d/2}\rt^i(\rt^2)^{1-d/2}\int_{0}^{\infty} \frac{\ud u}{u}e^{-u\overline{Q}^2}e^{-\frac{\bt^2}{4u}} \Gamma \left (\frac{d}{2}-1,\frac{\omega\rt^2}{4u} \right ).
\end{equation} 
Unfortunately, the remaining $u$-integral in \eq\nr{eq:Ixyfinal} can not be done analytically for arbitrary dimension $d$. In order to proceed further, we first square the amplitude in \eq\nr{eq:qbarqgamplitude} with the operator $(1-\hat{S}_E)$,
\begin{equation}
\begin{split}
{}_{q\bar{q}g}\langle &\gamma^{\ast}_{\rm L}(q'^+,Q^2,\lambda'\vert 1-\hat{S}_E\vert \gamma^{\ast}_{\rm L}(q^+,Q^2,\lambda)\rangle_{q\bar{q}g} 
=
\sum_{h,h',\sigma,\sigma'}\sum_{\text{color}}
\mathcal{PS}^{+}_{(3)} \mathcal{PS'}^{+}_{(3)}
\int_{\xt \yt [\zt] \xtp \yt' [\ztp]} 
\left (\widetilde{\psi}^{\gamma^{\ast}_{\rm L}\rightarrow q\bar{q}g}_{}\bigg\vert_{\rm FDH}\right )
\left (\widetilde{\psi}^{\gamma^{\ast}_{\rm L}\rightarrow q\bar{q}g}_{}\bigg\vert_{\rm FDH}\right )^{\ast}
\\
& \times \langle g(k'^+,\zt',\sigma',b)\bar{q}(\ell'^+,\yt',-h',\beta')q(p'^+,\xt',h',\alpha')\vert
1-\hat{S}_E
\vert q(p^+,\xt,h,\alpha)\bar{q}(\ell^+,\yt,-h,\beta)g(k^+,\zt,\sigma,a)\rangle.
\end{split}
\end{equation}
The color algebra is written out in detail in Appendix~\ref{app:color}.
Using the result \eq\nr{eq:qqbargmatrixelement} from the appendix we obtain
\begin{equation}
\label{eq:qqbargamplitudesquarefinal}
\begin{split}
{}_{q\bar{q}g}\langle \gamma^{\ast}_{\rm L}(q'^+,Q^2,\lambda')&\vert 1-\hat{S}_E \vert \gamma^{\ast}_{\rm L}(\qvec,Q^2,\lambda)\rangle_{q\bar{q}g} = 2q^+(2\pi)\delta(q'^+-q^+)\frac{16(2\pi)^3\nc\alpha_{em}e_f^2Q^2}{(2\pi)^2}\left (\frac{\alpha_s\cf}{\pi}\right )\int_{\xt\yt[\zt]}\\
&\times \int_{0}^{\infty}\ud z_1\int_{0}^{\infty}\ud z_3\int_{0}^{\infty} \ud z_2 \delta(z_1+z_2+z_3-1)\frac{z_1z_3}{z_2}\Theta(z_1,z_2,z_3)\left (1-\mathcal{S}_{xyz}\right ),
\end{split}
\end{equation} 
where the function $\Theta$ is given by 
\begin{equation}
\label{eq:thetadef}
\begin{split}
\Theta = \biggl [\Sigma^{ij}_{\ref{diag:qgqbarL}}\mathcal{I}^{i}(\rt_{yxz},\rt_{zx},\overline{Q}^2_{\ref{diag:qgqbarL}},\omega_{\ref{diag:qgqbarL}}) & - \Sigma^{ij}_{\ref{diag:qqbargL}}\mathcal{I}^{i}(\rt_{xyz},\rt_{zy},\overline{Q}^2_{\ref{diag:qqbargL}},\omega_{\ref{diag:qqbargL}})\biggr ]\\
&\times \biggl [\Sigma^{kl}_{\ref{diag:qgqbarL}}\mathcal{I}^{k}(\rt_{yxz},\rt_{zx},\overline{Q}^2_{\ref{diag:qgqbarL}},\omega_{\ref{diag:qgqbarL}})  - \Sigma^{kl}_{\ref{diag:qqbargL}}\mathcal{I}^{k}(\rt_{xyz},\rt_{zy},\overline{Q}^2_{\ref{diag:qqbargL}},\omega_{\ref{diag:qqbargL}})\biggr ]^{\ast}\sum_{\sigma}\epst_{\sigma}^{\ast j}\epst_{\sigma}^{l}
\end{split}
\end{equation}
and $\mathcal{S}_{xyz}$ is defined as 
\begin{equation}
\mathcal{S}_{xyz} = \frac{\nc^2}{2\cf\nc}\biggl [\mathcal{S}_{xz}\mathcal{S}_{zy} - \frac{1}{\nc^2}\mathcal{S}_{xy}\biggr ].
\end{equation}
It is important that  $\mathcal{S}_{xyz}$ must satisfy the condition $\mathcal{S}_{xyz} \rightarrow \mathcal{S}_{xy}$, when $\zt \rightarrow \xt$ or $\zt \rightarrow \yt$. This guarantees that the UV divergence in the real and virtual corrections has the same color structure from the target side, and can thus cancel between the contributions. Performing the sum over the $d_s$-dimensional gluon transverse polarization vectors and making some algebra \eq\nr{eq:thetadef} simplifies to 
\begin{equation}
\label{eq:thetafinal}
\begin{split}
\Theta = &\frac{(\mu^2)^{2-d/2}}{8^2\pi^d}\biggl \{\left (\frac{z_3}{z_1}\right )\left (z_1(z_1+z_2) + \frac{z_2^2}{2}\right )(\rt_{zx}^2)^{3-d}\mathcal{J}^2(\rt_{yxz}^2,\rt_{zx}^2,\overline{Q}^2_{\ref{diag:qgqbarL}},\omega_{\ref{diag:qgqbarL}})\\
& + \left (\frac{z_1}{z_3}\right )\left (z_3(z_2+z_3) + \frac{z_2^2}{2}\right )(\rt_{zy}^2)^{3-d}\mathcal{J}^2(\rt_{xyz}^2,\rt_{zy}^2,\overline{Q}^2_{\ref{diag:qqbargL}},\omega_{\ref{diag:qqbargL}})\\
& - \left ((z_1+z_2)z_3 + (z_2+z_3)z_1 \right )\frac{(\rt_{zx}\cdot \rt_{zy})}{(\rt_{zx}^2)^{\frac{d}{2}-1}(\rt_{zy}^2)^{\frac{d}{2}-1}}\mathcal{J}(\rt_{yxz}^2,\rt_{zx}^2,\overline{Q}^2_{\ref{diag:qgqbarL}},\omega_{\ref{diag:qgqbarL}})\mathcal{J}(\rt_{xyz}^2,\rt_{zy}^2,\overline{Q}^2_{\ref{diag:qqbargL}},\omega_{\ref{diag:qqbargL}}) \biggr \},\\
\end{split}
\end{equation}
where we have taken the limit $d_s\rightarrow 4$ and defined the function $\mathcal{J}$ as 
\begin{equation}
\label{eq:Jintegral}
\mathcal{J}(\bt^2,\rt^2,\overline{Q}^2,\omega) = \int_{0}^{\infty} \frac{\ud u}{u}e^{-u\overline{Q}^2}e^{-\frac{\bt^2}{4u}}\Gamma\left (\frac{d}{2}-1, \frac{\omega \rt^2}{4u}\right ).
\end{equation}
Now equation \nr{eq:qqbargamplitudesquarefinal} can be expressed in a more compact form by introducing the notation 
\begin{equation}
\overline{\Theta} = \overline{\Theta}_{\ref{diag:qgqbarL}} + \overline{\Theta}_{\ref{diag:qqbargL}} + \overline{\Theta}_{\ref{diag:qgqbarL}\ref{diag:qqbargL}}, 
\end{equation}
where 
\begin{equation}
\begin{split}
\overline{\Theta}_{\ref{diag:qgqbarL}} & =  \frac{(\mu^2)^{2-d/2}}{8^2\pi^d}\left (\frac{z_3}{z_1}\right )\left (z_1(z_1+z_2) + \frac{z_2^2}{2}\right )
\int_{[\zt]}(\rt_{zx}^2)^{3-d}\mathcal{J}^2(\rt_{yxz}^2,\rt_{zx}^2,\overline{Q}^2_{\ref{diag:qgqbarL}},\omega_{\ref{diag:qgqbarL}})\left (1-\mathcal{S}_{xyz}\right )\\
\overline{\Theta}_{\ref{diag:qqbargL}} & = \frac{(\mu^2)^{2-d/2}}{8^2\pi^d}\left (\frac{z_1}{z_3}\right )\left (z_3(z_2+z_3) + \frac{z_2^2}{2}\right )
\int_{[\zt]} (\rt_{zy}^2)^{3-d}\mathcal{J}^2(\rt_{xyz}^2,\rt_{zy}^2,\overline{Q}^2_{\ref{diag:qqbargL}},\omega_{\ref{diag:qqbargL}})\left (1-\mathcal{S}_{xyz}\right )
\end{split}
\end{equation}
and
\begin{equation}
\begin{split}
\overline{\Theta}_{\ref{diag:qgqbarL}\ref{diag:qqbargL}}  = - \frac{(\mu^2)^{2-d/2}}{8^2\pi^d}\left ((z_1+z_2)z_3 + (z_2+z_3)z_1\right )&
\int_{[\zt]}\frac{(\rt_{zx}\cdot \rt_{zy})}{(\rt_{zx}^2)^{\frac{d}{2}-1}(\rt_{zy}^2)^{\frac{d}{2}-1}}\\
&\times \mathcal{J}(\rt_{yxz}^2,\rt_{zx}^2,\overline{Q}^2_{\ref{diag:qgqbarL}}\omega_{\ref{diag:qgqbarL}})\mathcal{J}(\rt_{xyz}^2,\rt_{zy}^2,\overline{Q}^2_{\ref{diag:qqbargL}},\omega_{\ref{diag:qqbargL}})\left (1-\mathcal{S}_{xyz}\right ).
\end{split}
\end{equation}
Thus we obtain the following expression 
\begin{equation}
\label{eq:qqbargamplitudesquarefinalv2}
\begin{split}
{}_{q\bar{q}g}\langle \gamma^{\ast}_{\rm L}(q'^+,Q^2,\lambda')&\vert \hat{S}_E \vert \gamma^{\ast}_{\rm L}(q^+,Q^2,\lambda)\rangle_{q\bar{q}g} = 2q^+(2\pi)\delta(q'^+-q^+)\frac{16(2\pi)^3\nc\alpha_{em}e_f^2Q^2}{(2\pi)^2}\left (\frac{\alpha_s\cf}{\pi}\right ) 
 \\ \times &
\int_{ \xt \yt}
 \int_{0}^{\infty}\ud z_1\int_{0}^{\infty}\ud z_2\int_{0}^{\infty} \ud z_3 \delta(z_1+z_2+z_3-1)\frac{z_1z_3}{z_2}\biggl [ \overline{\Theta}_{\ref{diag:qgqbarL}} + \overline{\Theta}_{\ref{diag:qqbargL}} + \overline{\Theta}_{\ref{diag:qgqbarL}\ref{diag:qqbargL}}  \biggr ].
\end{split}
\end{equation} 
In \eq\nr{eq:qqbargamplitudesquarefinalv2}, the first and second term are UV-divergent when $\zt \rightarrow \xt$ and $\zt \rightarrow \yt$, respectively. The third (cross) term is UV-finite and thus one can immediately take the limit $d \rightarrow 4$. In order to make the UV subtraction between the real $q\bar{q}g$-component and the virtual $q\bar{q}$-term (which has an explicit $1/\varepsilon$) we must add and subtract a term to make this cancellation manifest. Ideally we would subtract from \eq\nr{eq:qqbargamplitudesquarefinalv2} an the same expression with the Wilson line structure 
$\left(1-\mathcal{S}_{xyz}\right )$ replaced by its UV limit 
$\left(1-\mathcal{S}_{xy}\right )$. This is, however not possible analytically since we have not been able to find an analytical expression for the required integral~\nr{eq:Jintegral}. However, there is no unique choice for the subtraction term. Indeed, since the only requirement for the subtraction is that the UV divergence needs to cancel, it is sufficient for the subtraction to approximate $\mathcal{J}(\bt^2,\rt^2,\overline{Q}^2,\omega)$ by any function that has the same value in the UV limit $\rt^2\to 0$ (for any $d$). Here we find it convenient to use the UV approximation 
\begin{equation}
\label{eq:xidef}
\mathcal{J}(\bt^2,\rt^2,\overline{Q}^2,\omega)\bigg\vert_{\text{UV}} = \int_{0}^{\infty} \frac{\ud u}{u}e^{-u\overline{Q}^2}e^{-\frac{\bt^2}{4u}}\Gamma\left (\frac{d}{2}-1\right )e^{-\frac{\rt^2}{2\bt^2\xi}} = 2K_{0}\left (\overline{Q}\vert \bt\vert\right )\Gamma\left (\frac{d}{2}-1\right )e^{-\frac{\rt^2}{2\bt^2\xi}}, \quad\quad \xi \in \Re,
\end{equation}
for which 
\begin{equation}\label{eq:Japprox}
\mathcal{J}(\bt^2,\rt^2,\overline{Q}^2,\omega)\bigg\vert_{\text{UV},\rt\to \ot}
= \mathcal{J}(\bt^2,\rt^2,\overline{Q}^2,\omega)\bigg\vert_{\rt\to \ot}.
\end{equation}
A natural way to think of this expression is that in the $u$-integral
\nr{eq:Jintegral} the exponential sets $u\sim \bt^2$. Our approximation replaces 
$\Gamma\left (\frac{d}{2}-1, \frac{\omega \rt^2}{4u}\right )$ by 
$\Gamma\left (\frac{d}{2}-1\right )e^{-\frac{\rt^2}{2\bt^2\xi}}$ which (a) is independent of $u$, allowing for an analytical calculation of the $u$-integral, (b) has the same value in the UV limit $\rt\to\ot$ and (c) is also  good and smooth approximation for large $\rt^2$. The choice of the constant  $\xi$  is somewhat arbitrary, here we adopt the value  $\xi  = e^{\gamma_\text{E}}$ that leads to simpler expressions in the following. Our choice is slightly different than  that of \cite{Beuf:2017bpd} concerning point (c) above; this difference is discussed in Appendix~\ref{app:sub}.

 Using our choice of $\mathcal{J}|_{\text{UV}}$ we now define the  UV-subtraction terms as
\begin{equation}
\begin{split}
\overline{\Theta}_{\ref{diag:qgqbarL}}\bigg\vert_{\textrm{UV};\zt\rightarrow \xt} & 
=  \frac{(\mu^2)^{2-d/2}}{8^2\pi^d}\left (\frac{z_3}{z_1}\right )\left (z_1(z_1+z_2) + \frac{z_2^2}{2}\right )
\int_{[\zt]} (\rt_{zx}^2)^{3-d}\mathcal{J}^2(\rt_{xy}^2,\rt_{zx}^2,\overline{Q}^2_{\ref{diag:qgqbarL}},\omega_{\ref{diag:qgqbarL}})\bigg\vert_{\rm UV}\left (1-\mathcal{S}_{xy}\right )\\
\overline{\Theta}_{\ref{diag:qqbargL}}\bigg\vert_{\textrm{UV};\zt\rightarrow \yt} & 
= \frac{(\mu^2)^{2-d/2}}{8^2\pi^d}\left (\frac{z_1}{z_3}\right )\left (z_3(z_2+z_3) + \frac{z_2^2}{2}\right )
\int_{[\zt]}  (\rt_{zy}^2)^{3-d}\mathcal{J}^2(\rt_{xy}^2,\rt_{zy}^2,\overline{Q}^2_{\ref{diag:qqbargL}},\omega_{\ref{diag:qqbargL}})\bigg\vert_{\rm UV}\left (1-\mathcal{S}_{xy}\right ).
\end{split}
\end{equation}
Performing the subtraction 
\begin{equation}
\label{eq:qqbargamplitudesquarefinalv2subtracted}
\begin{split}
{}_{q\bar{q}g}\langle \gamma^{\ast}_{\rm L}&(q'^+,Q^2,\lambda')\vert 1-\hat{S}_E \vert \gamma^{\ast}_{\rm L}(q^+,Q^2,\lambda)\rangle_{q\bar{q}g} = 2q^+(2\pi)\delta(q'^+-q^+)\frac{16(2\pi)^3\nc\alpha_{em}e_f^2Q^2}{(2\pi)^2}\left (\frac{\alpha_s\cf}{\pi}\right ) \\
&\times \int_{\xt \yt }\int_{0}^{\infty}\ud z_1\int_{0}^{\infty}\ud z_2\int_{0}^{\infty} \ud z_3 \delta(z_1+z_2+z_3-1)\frac{z_1z_3}{z_2}\biggl [ \left (\overline{\Theta}_{\ref{diag:qgqbarL}}- \overline{\Theta}_{\ref{diag:qgqbarL}}\bigg\vert_{\textrm{UV};\zt\rightarrow \xt} \right ) + \left (\overline{\Theta}_{\ref{diag:qqbargL}} - \overline{\Theta}_{\ref{diag:qqbargL}}\bigg\vert_{\textrm{UV};\zt\rightarrow \yt} \right )  \\
& \quad\quad\quad\quad\quad + \overline{\Theta}_{\ref{diag:qgqbarL}\ref{diag:qqbargL}} +  \overline{\Theta}_{\ref{diag:qgqbarL}}\bigg\vert_{\textrm{UV};\zt\rightarrow \xt} + \overline{\Theta}_{\ref{diag:qqbargL}}\bigg\vert_{\textrm{UV};\zt\rightarrow \yt}  \biggr ],
\end{split}
\end{equation} 
we can split the result into UV-finite terms and a divergent one as
\begin{equation}
\label{eq:qqbargamplitudesquarefinalv2subtractedfinal}
\begin{split}
{}_{q\bar{q}g}\langle \gamma^{\ast}_{\rm L}(q'^+,Q^2,\lambda')\vert 1-\hat{S}_E \vert \gamma^{\ast}_{\rm L}(q^+,Q^2,\lambda)\rangle_{q\bar{q}g} & = {}_{q\bar{q}g}\langle \gamma^{\ast}_{\rm L}(q'^+,Q^2,\lambda')\vert 1-\hat{S}_E \vert \gamma^{\ast}_{\rm L}(q^+,Q^2,\lambda)\rangle_{q\bar{q}g}\bigg\vert_{\rm UV-fin}\\
&  + {}_{q\bar{q}g}\langle \gamma^{\ast}_{\rm L}(q'^+,Q^2,\lambda')\vert 1-\hat{S}_E \vert \gamma^{\ast}_{\rm L}(q^+,Q^2,\lambda)\rangle_{q\bar{q}g}\bigg\vert_{\rm UV-div }.
\end{split}
\end{equation} 
The UV-finite part simplifies to 
\begin{equation}
\label{eq:Lfin1}
\begin{split}
&{}_{q\bar{q}g}\langle \gamma^{\ast}_{\rm L}(q'^+,Q^2,\lambda')\vert 1-\hat{S}_E \vert \gamma^{\ast}_{\rm L}(q^+,Q^2,\lambda)\rangle_{q\bar{q}g}\bigg\vert_{\rm UV-fin} = 2q^+(2\pi)\delta(q'^+-q^+)\frac{8\nc\alpha_{em}e_f^2Q^2}{(2\pi)^3}\left (\frac{\alpha_s\cf}{\pi}\right )\int_{\xt \yt \zt}\\
&\times \int_{0}^{1}\ud z_1\int_{0}^{1-z_1}\frac{\ud z_2}{z_2}\biggl \{\\
&
+z_3^2\left (2z_1(z_1+z_2) + z_2^2\right )\frac{1}{\rt_{zx}^2}\left ([K_0(\overline{Q}_{\ref{diag:qgqbarL}}\vert \Rt_{\ref{diag:qgqbarL}}\vert)]^2(1-\mathcal{S}_{xyz}) - [K_0(\overline{Q}_{\ref{diag:qgqbarL}}\vert \rt_{xy}\vert)]^2e^{-\rt_{zx}^2/(\rt_{xy}^2e^{\gamma_E})}(1-\mathcal{S}_{xy}) \right )\\
& + z_1^2\left (2z_3(z_2+z_3) + z_2^2\right )\frac{1}{\rt_{zy}^2}\left ([K_0(\overline{Q}_{\ref{diag:qqbargL}}\vert \Rt_{\ref{diag:qqbargL}}\vert)]^2(1-\mathcal{S}_{xyz}) - [K_0(\overline{Q}_{\ref{diag:qqbargL}}\vert \rt_{xy}\vert)]^2e^{-\rt_{zy}^2/(\rt_{xy}^2e^{\gamma_E})}(1-\mathcal{S}_{xy}) \right )\\
& -2\left ((z_1+z_2)z_1z_3^2 + (z_2+z_3)z_3z_1^2\right ) \frac{\rt_{zx}\cdot \rt_{zy}}{(\rt_{zx}^2)(\rt_{zy}^2)}K_0(\overline{Q}_{\ref{diag:qgqbarL}}\vert \Rt_{\ref{diag:qgqbarL}}\vert)K_0(\overline{Q}_{\ref{diag:qqbargL}}\vert \Rt_{\ref{diag:qqbargL}}\vert)(1-\mathcal{S}_{xyz})\biggr \}
\end{split}
\end{equation}
with $z_3 = 1-z_1-z_2$ and 
\begin{equation}
\Rt_{\ref{diag:qgqbarL}}^2 = \rt_{yxz}^2 + \omega_{\ref{diag:qgqbarL}}\rt_{zx}^2, \quad\quad \Rt_{\ref{diag:qqbargL}}^2 = \rt_{xyz}^2 + \omega_{\ref{diag:qqbargL}}\rt_{zy}^2. 
\end{equation}
Here the coefficients for the $\overline{Q}$'s and $\omega$'s are given by \eqs\nr{eq:QW(l)} and \nr{eq:QW(m)}. The arguments of the Bessel functions in \eq\nr{eq:Lfin1} can be expressed in a more compact form by noting that 
\begin{equation}
\label{eq:defforR}
\overline{Q}^2_{\ref{diag:qgqbarL}}\Rt_{\ref{diag:qgqbarL}}^2 = \overline{Q}^2_{\ref{diag:qqbargL}}\Rt_{\ref{diag:qqbargL}}^2 = Q^2\Rt^2,
\end{equation}
where $\Rt^2 = z_1z_3\rt_{xy}^2 + z_1z_2\rt_{zx}^2 + z_2z_3\rt_{zy}^2$, leading to 
\begin{equation}
\label{eq:Lfin2}
\begin{split}
{}_{q\bar{q}g}&\langle \gamma^{\ast}_{\rm L}(q'^+,Q^2,\lambda')\vert 1-\hat{S}_E \vert \gamma^{\ast}_{\rm L}(\qvec,Q^2,\lambda)\rangle_{q\bar{q}g}\bigg\vert_{\rm UV-fin} = 2q^+(2\pi)\delta(q'^+-q^+)\frac{8\nc\alpha_{em}e_f^2Q^2}{(2\pi)^3}\left (\frac{\alpha_s\cf}{\pi}\right )\int_{\xt \yt \zt}\\
&\times  \int_{0}^{1}\ud z_1\int_{0}^{1-z_1}\frac{\ud z_2}{z_2}\biggl \{\\
&+z_3^2\left (2z_1(z_1+z_2) + z_2^2\right )\frac{1}{\rt_{zx}^2}\left ([K_0(Q\vert \Rt\vert)]^2(1-\mathcal{S}_{xyz}) - [K_0(\overline{Q}_{\ref{diag:qgqbarL}}\vert \rt_{xy}\vert)]^2e^{-\rt_{zx}^2/(\rt_{xy}^2e^{\gamma_E})}(1-\mathcal{S}_{xy}) \right )\\
& + z_1^2\left (2z_3(z_2+z_3) + z_2^2\right )\frac{1}{\rt_{zy}^2}\left ([K_0(Q\vert \Rt\vert)]^2(1-\mathcal{S}_{xyz}) - [K_0(\overline{Q}_{\ref{diag:qqbargL}}\vert \rt_{xy}\vert)]^2e^{-\rt_{zy}^2/(\rt_{xy}^2e^{\gamma_E})}(1-\mathcal{S}_{xy}) \right )\\
& -2\left ((z_1+z_2)z_1z_3^2 + (z_2+z_3)z_3z_1^2\right ) \frac{\rt_{zx}\cdot \rt_{zy}}{(\rt_{zx}^2)(\rt_{zy}^2)}[K_0(Q\vert \Rt\vert)]^2(1-\mathcal{S}_{xyz})\biggr \}.
\end{split}
\end{equation}
In the UV-divergent term one can now analytically perform the  $\zt$-integral and the $z_2$-integral, which results in 
\begin{equation}
\label{eq:uvschemepartL}
\begin{split}
{}_{q\bar{q}g}&\langle \gamma^{\ast}_{\rm L}(q'^+,Q^2,\lambda')\vert 1-\hat{S}_E \vert \gamma^{\ast}_{\rm L}(\qvec,Q^2,\lambda)\rangle_{q\bar{q}g}\bigg\vert_{\rm UV-div} = -2q^+(2\pi)\delta(q'^+-q^+)\frac{8\nc\alpha_{em}e_f^2Q^2}{(2\pi)^2}\left (\frac{\alpha_s\cf}{\pi}\right )\int_{\xt \yt} \\
& \times  \int_{0}^{1} \ud z z^2(1-z)^2[K_0(\overline{Q}\vert \rt_{xy}\vert) ]^2\biggl [\frac{3}{2} + \log\left (\frac{\alpha}{z}\right ) + \log\left (\frac{\alpha}{1-z}\right )\biggr ]\biggl \{\frac{1}{\varepsilon_{\overline{\rm MS}}} + \log\left (\frac{\rt_{xy}^2\mu^2}{4}\right ) -2\Psi_0(1)\biggr \}(1-\mathcal{S}_{xy}).
\end{split}
\end{equation}
This expression precisely cancels the term in square brackets in \eq\nr{eq:NLOkernelL}.
After this cancellation we can write the  total cross section for longitudinal virtual photon at NLO accuracy as a sum of two finite terms
\begin{equation}
\label{eq:Lfinalresult}
\sigma^{\gamma^{\ast}_{\rm L}}[A] = \sigma^{\gamma^{\ast}_{\rm L}}\bigg\vert_{q\bar{q}} +  \sigma^{\gamma^{\ast}_{\rm L}}\bigg\vert_{q\bar{q}g},
\end{equation}
where the finite contribution to the cross section coming from the $q\bar{q}$-component is 
\begin{equation}
\label{eq:Lfinalresulta}
\begin{split}
\sigma^{\gamma^{\ast}_{\rm L}}\bigg\vert_{q\bar{q}}  = 4\nc \frac{4\alpha_{em}e_f^2Q^2}{(2\pi)^2}\int_{ \xt  \yt} &\int_{0}^{1}\ud z z^2(1-z)^2[K_{0}\left (\overline{Q}\vert \rt_{xy}\vert\right )]^2\\
&\times \biggl \{1 + \left ( \frac{\alpha_s\cf}{\pi}\right )\biggl [\frac{1}{2}\log^2\left ( \frac{z}{1-z}\right )-\frac{\pi^2}{6} + \frac{5}{2} \biggr ]\biggr \}\left (1-\mathcal{S}_{xy}\right )
\end{split}
\end{equation}
and the subtracted $q\bar{q}g$-component 
\begin{equation}
\label{eq:Lfinalresultb}
\begin{split}
\sigma^{\gamma^{\ast}_{\rm L}}\bigg\vert_{q\bar{q}g}  =  &4\nc \frac{4\alpha_{em}e_f^2Q^2}{(2\pi)^3}\left ( \frac{\alpha_s\cf}{\pi}\right )
\int_{ \xt \yt \zt} \int_{0}^{1}\ud z_1\int_{0}^{1-z_1}\frac{\ud z_2}{z_2}\biggl \{\\
&+z_3^2\left (2z_1(z_1+z_2) + z_2^2\right )\frac{1}{\rt_{zx}^2}\left ([K_0(Q\vert \Rt\vert)]^2(1-\mathcal{S}_{xyz}) - [K_0(\overline{Q}_{\ref{diag:qgqbarL}}\vert \rt_{xy}\vert)]^2e^{-\rt_{zx}^2/(\rt_{xy}^2e^{\gamma_E})}(1-\mathcal{S}_{xy}) \right )\\
& + z_1^2\left (2z_3(z_2+z_3) + z_2^2\right )\frac{1}{\rt_{zy}^2}\left ([K_0(Q\vert \Rt\vert)]^2(1-\mathcal{S}_{xyz}) - [K_0(\overline{Q}_{\ref{diag:qqbargL}}\vert \rt_{xy}\vert)]^2e^{-\rt_{zy}^2/(\rt_{xy}^2e^{\gamma_E})}(1-\mathcal{S}_{xy}) \right )\\
& -2\left ((z_1+z_2)z_1z_3^2 + (z_2+z_3)z_3z_1^2\right ) \frac{\rt_{zx}\cdot \rt_{zy}}{(\rt_{zx}^2)(\rt_{zy}^2)}[K_0(Q\vert \Rt\vert)]^2(1-\mathcal{S}_{xyz})\biggr \}.
\end{split}
\end{equation}
Now that these expressions are UV finite, all the coordinate integrals can be performed in 2 transverse dimensions.

Here we should emphasize that the scheme dependent UV contribution in \eq\nr{eq:uvschemepartL} precisely cancels
the scheme dependent UV part obtained in \eq\nr{eq:NLOLfinsquare}, and the remaining finite contribution in \eq\nr{eq:NLOLfinsquare} leads to the scheme independent final result for $q\bar{q}$-part in \eq\nr{eq:Lfinalresulta}. We have confirmed both analytically and also numerically that our final results for the cross section in \eq\nr{eq:Lfinalresulta} and \eq\nr{eq:Lfinalresultb} agree with those of G. Beuf~\cite{Beuf:2017bpd}.

In addition, we should note that in the $z_2 \rightarrow 0$ limit the part inside the curly brackets in \eq\nr{eq:Lfinalresultb} reduces to
\begin{equation}\label{eq:hassubk}
2 z_1^2 z_3^2[K_0(Q\vert \rt_{xy}\vert)]^2 \left[
\frac{\rt_{xy}^2}{\rt_{zx}^2\rt_{zy}^2} (1-\mathcal{S}_{xyz} ) 
-\frac{1}{\rt_{zx}^2} e^{-\rt_{zx}^2/(\rt_{xy}^2e^{\gamma_E})}(1-\mathcal{S}_{xy}) 
-\frac{1}{\rt_{zy}^2} e^{-\rt_{zy}^2/(\rt_{xy}^2e^{\gamma_E})}(1-\mathcal{S}_{xy}) 
\right].
\end{equation}
Noting that $(1-\mathcal{S}_{xy})$ does not depend on $\zt$ and using the integral (this is the same integral that is studied in Appendix~\ref{app:sub})
\begin{equation}\label{eq:nollaintegraali}
\int_\zt \left[
\frac{\rt_{xy}^2}{\rt_{zx}^2\rt_{zy}^2}
-\frac{1}{\rt_{zx}^2} e^{-\rt_{zx}^2/(\rt_{xy}^2e^{\gamma_E})}
-\frac{1}{\rt_{zy}^2} e^{-\rt_{zy}^2/(\rt_{xy}^2e^{\gamma_E})}
\right]=0
\end{equation}
the form \eq\nr{eq:hassubk} can, under the integral over $\zt$ in \eq\nr{eq:Lfinalresultb}, be replaced by 
\begin{equation}\label{eq:norskibk}
2z_1^2 z_3^2  [K_0(Q\vert \rt_{xy}\vert)]^2
\frac{\rt_{xy}^2}{\rt_{zx}^2\rt_{zy}^2}  \left[
 (1-\mathcal{S}_{xyz} ) -(1-\mathcal{S}_{xy}) 
\right],
\end{equation}
which is recognized as the leading order wave function times  the r.h.s. of the BK equation (or the first equation in the Balitsky hierarchy). Note that it is precisely to achieve the cancellation \eq\nr{eq:nollaintegraali} and thus to obtain the conventional BK equation that we chose the constant $\xi$ (see \eq\nr{eq:xidef}) to have the value  $e^{\gamma_E}$. Thus we see that the $z_2$-integral exhibits a small-$x$ divergence that must be absorbed into a renormalization group evolution of the target, and that this can be done using the BK equation e.g. similarly as is done in \cite{Ducloue:2017ftk}.

\subsection{Transverse photon}

Let us then consider the case of transverse virtual photon. Similarly as in the longitudinal photon case, the mixed space expression for the $q\bar{q}$-component of the transverse virtual photon amplitude computed in the FHD scheme at NLO accuracy is given by 
\begin{equation}
\label{eq:longitudinalamplitudeqqbarT}
\vert \gamma^{\ast}_{\rm T}(q^+,Q^2,\lambda)\rangle_{q\bar{q}} = \sum_{h,\lambda}\sum_{\text{color}}\mathcal{PS}^{+}_{(2)}\int \ud^2 \xt \int \ud^2 \yt 
\left (
\widetilde{\psi}^{\gamma^{\ast}_{\rm T}\rightarrow q\bar{q}}_{\lo}\bigg\vert_{\rm FDH} + \widetilde{\psi}^{\gamma^{\ast}_{\rm T}\rightarrow q\bar{q}}_{\nlo}\bigg\vert_{\rm FDH}
\right )\vert q(p^+,\xt,h,\alpha)\bar{q}(p'^+,\yt,-h,\beta)\rangle.
\end{equation}
Here the factor $\mathcal{PS}^{+}_{(2)}$ is given in \eq\nr{eq:PS2withz}, and the transverse Fourier transformed LO and NLO LCWF's for transverse virtual photon in the mixed space can be written as  
\begin{equation}
\begin{split}
\widetilde{\psi}^{\gamma^{\ast}_{\rm T}\rightarrow q\bar{q}}_{\lo/\nlo}\bigg\vert_{\rm FDH}  =
\int \frac{\ud^2 \pt}{(2\pi)^2}\left (\psi^{\gamma^{\ast}_{\rm T}\rightarrow q\bar{q}}_{\lo/\nlo}\bigg\vert_{\rm FDH}\right )e^{i\pt\cdot \rt_{xy}}. 
\end{split}
\end{equation}
The momentum space expressions for the LO and NLO wave functions are given in \eqs\nr{eq:LOTvirtualphoton} and \nr{eq:fullTWFqbarq}, respectively. Using the result given in \eqs\nr{eq:FTintegralsfinal}, \nr{eq:FTintegralsextra2final} and \nr{eq:FTintegralsextra3final} we obtain for the LO part
\begin{equation}
\widetilde{\psi}^{\gamma^{\ast}_{\rm T}\rightarrow q\bar{q}}_{\lo}\bigg\vert_{\rm FDH} = \frac{4iq^+ee_fQ\delta_{\alpha\beta}}{(2\pi)}z(1-z) \biggl [\left (z-\frac{1}{2}\right )\delta^{ij}-ih\frac{1}{2}\epsilon^{ij}\biggr ]\frac{(\rt_{xy})^i\epst^{j}_{\lambda}}{\vert \rt_{xy}\vert}K_{1}\left (\overline{Q}\vert \rt_{xy}\vert\right )
\end{equation}
and similarly for the NLO part
\begin{equation}
\begin{split}
\widetilde{\psi}^{\gamma^{\ast}_{\rm T}\rightarrow q\bar{q}}_{\nlo}\bigg\vert_{\rm FDH} = \frac{4iq^+ee_fQ\delta_{\alpha\beta}}{(2\pi)}\left (\frac{\alpha_s\cf}{2\pi}\right )\biggl \{z(1-z)\biggl [\left (z-\frac{1}{2}\right )\delta^{ij} -ih\frac{1}{2}\epsilon^{ij}\biggr ]\mathcal{K}^{\gamma^{\ast}_{\rm T}}\bigg\vert_{\rm FDH}\biggr \}\frac{(\rt_{xy})^i\epst^{j}_{\lambda}}{\vert \rt_{xy}\vert}K_{1}\left (\overline{Q}\vert \rt_{xy}\vert\right ),
\end{split}
\end{equation}
where the NLO kernel in the FDH scheme simplifies to 
\begin{equation}
\label{eq:NLOkernelTF}
\mathcal{K}^{\gamma^{\ast}_{\rm T}}\bigg\vert_{\rm FDH}  = \biggl [\frac{3}{2} + \log\left (\frac{\alpha}{z}\right ) + \log\left (\frac{\alpha}{1-z}\right ) \biggr ]\biggl \{\frac{1}{\varepsilon_{\overline{\rm MS}}}  + \log\left (\frac{\rt_{xy}^2\mu^2}{4}\right ) - 2\Psi_{0}(1)\biggr \} + \frac{1}{2}\log^2\left (\frac{z}{1-z}\right ) - \frac{\pi^2}{6} + \frac{5}{2}  +  \mathcal{O}(\varepsilon).
\end{equation}
Note that after the Fourier transform to mixed space (but not before), the NLO correction $\mathcal{K}^{\gamma^{\ast}_{\rm T}}$ for transverse photons is the same one as for the longitudinal ones in \eq\nr{eq:NLOkernelL}.
Squaring the LCWF in \eq\nr{eq:longitudinalamplitudeqqbarT} (summed over the helicity) together with the eikonal scattering operator $(1-\hat{S}_E)$ we find the final (unsubtracted) result for the $q\bar{q}$ part of the cross section as
\begin{equation}
\label{eq:LOTunsub}
\begin{split}
{}_{q\bar{q}}\langle &\gamma^{\ast}_{\rm T}(q'^+,Q^2,\lambda')\vert 1-\hat{S}_E \vert \gamma^{\ast}_{\rm T}(q^+,Q^2,\lambda)\rangle_{q\bar{q}} = 2q^+(2\pi)\delta(q'^+-q^+)\frac{4\nc\alpha_{em}e_f^2Q^2}{(2\pi)^2}\\
&\times \int_{\xt \yt}\int_{0}^{1}\ud z [K_{1}\left (\overline{Q}\vert \rt_{xy}\vert\right )]^2z(1-z)\biggl \{1-2z(1-z)\biggr \}\biggl [1 + \left ( \frac{\alpha_s\cf}{\pi}\right )\mathcal{K}^{\gamma^{\ast}_{\rm T}}\bigg\vert_{\rm FDH}\biggr ] \left (1-\mathcal{S}_{xy}\right ) + \mathcal{O}(\alpha_{em}\alpha_s^2).
\end{split}
\end{equation}

The mixed space expression for the $q\bar{q}g$-component of the transverse virtual photon amplitude computed in the FDH scheme at NLO accuracy is given by 
\begin{equation}
\label{eq:qbarqgamplitudeT}
\vert \gamma^{\ast}_{\rm T}(q^+,Q^2,\lambda)\rangle_{q\bar{q}g} = \sum_{h,\sigma,\lambda}\sum_{\text{color}}\mathcal{PS}^{+}_{(3)}
\int_{\xt \yt [\zt]}
\left (\tilde{\psi}^{\gamma^{\ast}_{\rm T}\rightarrow q\bar{q}g}_{}\bigg\vert_{\rm FDH}\right )\vert q(p^+,\xt,h,\alpha)\bar{q}(p'^+,\yt,-h,\beta)g(k^+,\zt,\sigma,a) \rangle,
\end{equation}
where $\mathcal{PS}^{+}_{(3)}$ is given in \eq\nr{eq:PS3withz}, and from \eq\nr{eq:Tradfinal} the NLO expression of transverse virtual photon LCWF in the mixed space simplifies to   
\begin{equation}
\label{eq:Tradmixed}
\begin{split}
\widetilde{\psi}^{\gamma^{\ast}_{\rm T}\rightarrow q\bar{q}g}_{}\bigg\vert_{\rm FDH} = -8ee_fg_rt^{a}_{\alpha\beta}(z_1z_3)^{1/2}
 &\biggl \{ \Sigma^{ijkl}_{\ref{diag:qgqbarT}}\mathcal{I}^{ik}(\rt_{yxz},\rt_{zx},\overline{Q}^2_{\ref{diag:qgqbarT}},\omega_{\ref{diag:qgqbarT}})\epst^{\ast l}_{\sigma} + \Sigma^{ijkl}_{\ref{diag:qqbargT}
}\mathcal{I}^{ik}(\rt_{xyz},\rt_{zy},\overline{Q}^2_{\ref{diag:qqbargT}},\omega_{\ref{diag:qqbargT}})\epst^{\ast l}_{\sigma}\\
& + \Sigma^{ij}_{\ref{diag:qgqbarinst}}\mathcal{I}(\rt_{yxz},\rt_{zx},\overline{Q}^2_{\ref{diag:qgqbarinst}},\omega_{\ref{diag:qgqbarinst}})\epst^{\ast i}_{\sigma} - \Sigma^{ij}_{\ref{diag:qqbarginst}}\mathcal{I}(\rt_{xyz},\rt_{zy},\overline{Q}^2_{\ref{diag:qqbarginst}},\omega_{\ref{diag:qqbarginst}})\epst^{\ast i}_{\sigma}\biggr \}\epst^{j}_{\lambda}.
\end{split}
\end{equation}
Similarly as in the case of longitudinal photon, we have introduced the notation  
\begin{equation}
\label{eq:Imasterrank2}
\mathcal{I}^{ik}(\bt,\rt,\overline{Q}^2,\omega) = \mu^{2-\frac{d}{2}}\int \frac{\ud^2\Pt}{(2\pi)^2}\int\frac{\ud^{d-2}\Kt}{(2\pi)^{d-2}}\frac{\Pt^i\Kt^ke^{i\Pt\cdot \bt}e^{i\Kt\cdot \rt}}{\biggl [\Pt^2 + \overline{Q}^2\biggr ]\biggl [\Kt^2 + \omega\left (\Pt^2 + \overline{Q}^2\right )\biggr ]}
\end{equation}
and 
\begin{equation}
\label{eq:Imasterrank0}
\mathcal{I}(\bt,\rt,\overline{Q}^2,\omega) = \mu^{2-\frac{d}{2}}\int \frac{\ud^2\Pt}{(2\pi)^2}\int\frac{\ud^{d-2}\Kt}{(2\pi)^{d-2}}\frac{e^{i\Pt\cdot \bt}e^{i\Kt\cdot \rt}}{\biggl [\Kt^2 + \omega\left (\Pt^2 + \overline{Q}^2\right )\biggr ]}.
\end{equation}
Squaring the amplitude in \eq\nr{eq:qbarqgamplitudeT} with the eikonal operator $(1-\hat{S}_E)$, and using the result \eq\nr{eq:qqbargmatrixelement} we get
\begin{equation}
\label{eq:qqbargamplitudesquarefinalT}
\begin{split}
{}_{q\bar{q}g}\langle \gamma^{\ast}_{\rm T}(q'^+,Q^2,\lambda')&\vert 1-\hat{S}_E \vert \gamma^{\ast}_{\rm T}(q^+,Q^2,\lambda)\rangle_{q\bar{q}g} = 2q^+(2\pi)\delta(q'^+-q^+)\frac{16(2\pi)^3\nc\alpha_{em}e_f^2}{(2\pi)^2}\left (\frac{\alpha_s\cf}{\pi}\right )\\
&
\int_{\xt\yt[\zt]}
\int_{0}^{\infty}\ud z_1\int_{0}^{\infty}\ud z_3\int_{0}^{\infty} \frac{\ud z_2}{z_2} \delta(z_1+z_2+z_3-1)\Theta(z_1,z_2,z_3)\left (1-\mathcal{S}_{xyz}\right ).
\end{split}
\end{equation} 
Here, following the same notation as in section \ref{sec:longitsigma}, we have defined the function $\Theta$ as 
\begin{equation}
\label{eq:thetafull}
\begin{split}
\Theta = \sum_{\sigma,\lambda}&\biggl [\Sigma^{ijkl}_{\ref{diag:qgqbarT}}\mathcal{I}^{ik}(\rt_{yxz},\rt_{zx},\overline{Q}^2_{\ref{diag:qgqbarT}},\omega_{\ref{diag:qgqbarT}})\epst^{\ast l}_{\sigma} + \Sigma^{ijkl}_{\ref{diag:qqbargT}
}\mathcal{I}^{ik}(\rt_{xyz},\rt_{zy},\overline{Q}^2_{\ref{diag:qqbargT}},\omega_{\ref{diag:qqbargT}})\epst^{\ast l}_{\sigma}\\
& + \Sigma^{ij}_{\ref{diag:qgqbarinst}}\mathcal{I}(\rt_{yxz},\rt_{zx},\overline{Q}^2_{\ref{diag:qgqbarinst}},\omega_{\ref{diag:qgqbarinst}})\epst^{\ast i}_{\sigma} - \Sigma^{ij}_{\ref{diag:qqbarginst}}\mathcal{I}(\rt_{xyz},\rt_{zy},\overline{Q}^2_{\ref{diag:qqbarginst}},\omega_{\ref{diag:qqbarginst}})\epst^{\ast i}_{\sigma}\biggr ]\\
&\times \biggl [\Sigma^{mnrs}_{\ref{diag:qgqbarT}}\mathcal{I}^{mr}(\rt_{yxz},\rt_{zx},\overline{Q}^2_{\ref{diag:qgqbarT}},\omega_{\ref{diag:qgqbarT}})\epst^{\ast s}_{\sigma} + \Sigma^{mnrs}_{\ref{diag:qqbargT}
}\mathcal{I}^{mr}(\rt_{xyz},\rt_{zy},\overline{Q}^2_{\ref{diag:qqbargT}},\omega_{\ref{diag:qqbargT}})\epst^{\ast s}_{\sigma}\\
& + \Sigma^{mn}_{\ref{diag:qgqbarinst}}\mathcal{I}(\rt_{yxz},\rt_{zx},\overline{Q}^2_{\ref{diag:qgqbarinst}},\omega_{\ref{diag:qgqbarinst}})\epst^{\ast m}_{\sigma} - \Sigma^{mn}_{\ref{diag:qqbarginst}}\mathcal{I}(\rt_{xyz},\rt_{zy},\overline{Q}^2_{\ref{diag:qqbarginst}},\omega_{\ref{diag:qqbarginst}})\epst^{\ast m}_{\sigma}\biggr ]^{\ast}\epst_{\lambda}^{j}\epst_{\lambda}^{\ast n}
\end{split}
\end{equation}
which corresponds to the  full $q\bar{q}g$-sector wave function (see \eq\nr{eq:Tradmixed}) squared and summed over the gluon and photon polarization vectors. The expression in \nr{eq:thetafull} can be simplified further by introducing the notation 
\begin{equation}
\label{eq:thetadefTT}
\overline{\Theta} = \overline{\Theta}_{\ref{diag:qgqbarT}} + \overline{\Theta}_{\ref{diag:qqbargT}} + \overline{\Theta}_{\ref{diag:qgqbarinst}} + \overline{\Theta}_{\ref{diag:qqbarginst}}  + \overline{\Theta}_{\ref{diag:qgqbarT}\ref{diag:qqbargT}\ref{diag:qgqbarinst}\ref{diag:qqbarginst}}
\end{equation}
with
\begin{equation}
\label{eq:qqbargamplitudesquarefinalTv2}
\begin{split}
{}_{q\bar{q}g}\langle \gamma^{\ast}_{\rm T}(q'^+,Q^2,\lambda')&\vert 1-\hat{S}_E \vert \gamma^{\ast}_{\rm T}(q^+,Q^2,\lambda)\rangle_{q\bar{q}g} = 2q^+(2\pi)\delta(q'^+-q^+)\frac{16(2\pi)^3\nc\alpha_{em}e_f^2}{(2\pi)^2}\left (\frac{\alpha_s\cf}{\pi}\right )\\
&
\int_{\xt\yt}
\int_{0}^{\infty}\ud z_1\int_{0}^{\infty}\ud z_3\int_{0}^{\infty} \frac{\ud z_2}{z_2} \delta(z_1+z_2+z_3-1)\overline{\Theta}(z_1,z_2,z_3),
\end{split}
\end{equation} 
where the individual terms coming from the wave function squared are given by
\begin{equation}
\label{eq:diag:qgqbarT:sqr}
\overline{\Theta}_{\ref{diag:qgqbarT}} = \int_{[\zt]} \biggl \{\Sigma^{ijkl}_{\ref{diag:qgqbarT}}\mathcal{I}^{ik}(\rt_{yxz},\rt_{zx},\overline{Q}^2_{\ref{diag:qgqbarT}},\omega_{\ref{diag:qgqbarT}})\left (\Sigma^{mjrl}_{\ref{diag:qgqbarT}}\mathcal{I}^{mr}(\rt_{yxz},\rt_{zx},\overline{Q}^2_{\ref{diag:qgqbarT}},\omega_{\ref{diag:qgqbarT}}) \right )^{\ast}\biggr \}\left (1-\mathcal{S}_{xyz}\right )
\end{equation}
\begin{equation}
\label{eq:diag:qqbargT:sqr}
\overline{\Theta}_{\ref{diag:qqbargT}} = \int_{[\zt]} \biggl \{\Sigma^{ijkl}_{\ref{diag:qqbargT}}\mathcal{I}^{ik}(\rt_{xyz},\rt_{zy},\overline{Q}^2_{\ref{diag:qqbargT}},\omega_{\ref{diag:qqbargT}})\left (\Sigma^{mjrl}_{\ref{diag:qqbargT}}\mathcal{I}^{mr}(\rt_{xyz},\rt_{zy},\overline{Q}^2_{\ref{diag:qqbargT}},\omega_{\ref{diag:qqbargT}}) \right )^{\ast}\biggr \}\left (1-\mathcal{S}_{xyz}\right )
\end{equation}
\begin{equation}
\label{eq:diag:qgqbarinst:sqr}
\overline{\Theta}_{\ref{diag:qgqbarinst}} = \int_{[\zt]} \biggl \{\Sigma^{ij}_{\ref{diag:qgqbarinst}}\mathcal{I}(\rt_{yxz},\rt_{zx},\overline{Q}^2_{\ref{diag:qgqbarinst}},\omega_{\ref{diag:qgqbarinst}})\left (\Sigma^{ij}_{\ref{diag:qgqbarinst}}\mathcal{I}(\rt_{yxz},\rt_{zx},\overline{Q}^2_{\ref{diag:qgqbarinst}},\omega_{\ref{diag:qgqbarinst}}) \right )^{\ast}\biggr \}\left (1-\mathcal{S}_{xyz}\right )
\end{equation}
\begin{equation}
\label{eq:diag:qqbarginst:sqr}
\overline{\Theta}_{\ref{diag:qqbarginst}} = \int_{[\zt]} \biggl \{\Sigma^{ij}_{\ref{diag:qqbarginst}}\mathcal{I}(\rt_{xyz},\rt_{zy},\overline{Q}^2_{\ref{diag:qqbarginst}},\omega_{\ref{diag:qqbarginst}})\left (\Sigma^{ij}_{\ref{diag:qqbarginst}}\mathcal{I}(\rt_{xyz},\rt_{zy},\overline{Q}^2_{\ref{diag:qqbarginst}},\omega_{\ref{diag:qqbarginst}}) \right )^{\ast}\biggr \}\left (1-\mathcal{S}_{xyz}\right )
\end{equation}
and the possible cross terms:
\begin{equation}
\label{eq:thetaT}
\begin{split}
\overline{\Theta}_{\ref{diag:qgqbarT}\ref{diag:qqbargT}\ref{diag:qgqbarinst}\ref{diag:qqbarginst}} = & 2 \int_{[\zt]}\Re e\biggl [\Sigma^{ijkl}_{\ref{diag:qgqbarT}}\mathcal{I}^{ik}(\rt_{yxz},\rt_{zx},\overline{Q}^2_{\ref{diag:qgqbarT}},\omega_{\ref{diag:qgqbarT}})\left (\Sigma^{mjrl}_{\ref{diag:qqbargT}}\mathcal{I}^{mr}(\rt_{xyz},\rt_{zy},\overline{Q}^2_{\ref{diag:qqbargT}},\omega_{\ref{diag:qqbargT}})\right )^{\ast}\\
& + \left (\Sigma^{ijkl}_{\ref{diag:qgqbarT}}\mathcal{I}^{ik}(\rt_{yxz},\rt_{zx},\overline{Q}^2_{\ref{diag:qgqbarT}},\omega_{\ref{diag:qgqbarT}}) + \Sigma^{ijkl}_{\ref{diag:qqbargT}}\mathcal{I}^{ik}(\rt_{xyz},\rt_{zy},\overline{Q}^2_{\ref{diag:qqbargT}},\omega_{\ref{diag:qqbargT}}) \right )\\
&\times \left (\Sigma^{lj}_{\ref{diag:qgqbarinst}}\mathcal{I}(\rt_{yxz},\rt_{zx},\overline{Q}^2_{\ref{diag:qgqbarinst}},\omega_{\ref{diag:qgqbarinst}}) - \Sigma^{lj}_{\ref{diag:qqbarginst}}\mathcal{I}(\rt_{xyz},\rt_{zy},\overline{Q}^2_{\ref{diag:qqbarginst}},\omega_{\ref{diag:qqbarginst}})  \right )^{\ast}\\
& - \Sigma^{ij}_{\ref{diag:qgqbarinst}}\mathcal{I}(\rt_{yxz},\rt_{zx},\overline{Q}^2_{\ref{diag:qgqbarinst}},\omega_{\ref{diag:qgqbarinst}})\left (\Sigma^{ij}_{\ref{diag:qqbarginst}}\mathcal{I}(\rt_{xyz},\rt_{zy},\overline{Q}^2_{\ref{diag:qqbarginst}},\omega_{\ref{diag:qqbarginst}})  \right )^{\ast} \biggr ]\left (1-\mathcal{S}_{xyz}\right ).
\end{split}
\end{equation}
The contributions in \eqs\nr{eq:diag:qgqbarT:sqr}, \nr{eq:diag:qqbargT:sqr}, \nr{eq:diag:qgqbarinst:sqr} and \nr{eq:diag:qqbarginst:sqr} correspond to the squared amplitudes from diagrams \ref{diag:qgqbarT}, \ref{diag:qqbargT}, \ref{diag:qgqbarinst} and \ref{diag:qqbarginst} respectively, and the contribution in \eq\nr{eq:thetaT} contains the cross terms of these diagrams.  

In order to simplify the individual contributions above we note that
\begin{equation}
\overline{Q}^2_{\ref{diag:qgqbarT}} = \overline{Q}^2_{\ref{diag:qgqbarinst}} = \overline{Q}^2_{\ref{diag:qgqbarL}},\quad \overline{Q}^2_{\ref{diag:qqbargT}} = \overline{Q}^2_{\ref{diag:qqbarginst}} = \overline{Q}^2_{\ref{diag:qqbargL}} 
,
\end{equation}
and 
\begin{equation}
\omega_{\ref{diag:qgqbarT}} = \omega_{\ref{diag:qgqbarinst}} = \omega_{\ref{diag:qgqbarL}}, \quad \omega_{\ref{diag:qqbargT}} = \omega_{\ref{diag:qqbarginst}} = \omega_{\ref{diag:qqbargL}}
.
\end{equation}
This implies that 
\begin{equation}
\overline{Q}^2_{\ref{diag:qgqbarT}}\Rt_{\ref{diag:qgqbarT}}^2 = \overline{Q}^2_{\ref{diag:qqbargT}}\Rt_{\ref{diag:qqbargT}}^2 = Q^2\Rt^2, \quad \overline{Q}^2_{\ref{diag:qgqbarinst}}\Rt_{\ref{diag:qgqbarinst}}^2 = \overline{Q}^2_{\ref{diag:qqbarginst}}\Rt_{\ref{diag:qqbarginst}}^2 = Q^2\Rt^2,
\end{equation} 
where $\Rt^2$ is defined in \eq\nr{eq:defforR}. Using the definitions in \eq\nr{eq:Tradfinal} and result derived in \eq\nr{eq:Imasterrank2res} we obtain 
\begin{equation}
\label{eq:qgqbarT}
\Theta_{\ref{diag:qgqbarT}} = \frac{F(z_1,z_2,z_3)(\mu^2)^{2-d/2}}{16^2\pi^d}\int_{[\zt]}  (\rt_{yxz}\cdot \rt_{yxz})(\rt_{zx}^2)^{3-d}\mathcal{L}^2(\rt_{yxz}^2,\rt_{zx}^2,\overline{Q}^2_{\ref{diag:qgqbarT}},\omega_{\ref{diag:qgqbarT}})\left (1-\mathcal{S}_{xyz}\right )
\end{equation}
and 
\begin{equation}
\label{eq:qqbargT}
\Theta_{\ref{diag:qqbargT}} = \frac{G(z_1,z_2,z_3)(\mu^2)^{2-d/2}}{16^2\pi^d}\int_{[\zt]}  (\rt_{xyz}\cdot \rt_{xyz})(\rt_{zy}^2)^{3-d}\mathcal{L}^2(\rt_{xyz}^2,\rt_{zy}^2,\overline{Q}^2_{\ref{diag:qqbargT}},\omega_{\ref{diag:qqbargT}}) \left (1-\mathcal{S}_{xyz}\right ),
\end{equation}
where 
\begin{equation}
\begin{split}
(\rt_{yxz}\cdot \rt_{yxz}) &= \frac{\Rt^2}{z_3(z_1 + z_2)} - \frac{z_1z_2}{z_3(z_1+z_2)^2}\rt_{zx}^2 \\
(\rt_{xyz}\cdot \rt_{xyz}) &= \frac{\Rt^2}{z_1(z_2 + z_3)} - \frac{z_3z_2}{z_1(z_2+z_3)^2}\rt_{zy}^2. \\
\end{split}
\end{equation} 
The two functions $F$ and $G$ above are given by    
\begin{equation}
\begin{split}
F(z_2,z_3) & = \biggl [\left (z_1 + z_2 - \frac{1}{2}\right )^2 + \frac{1}{4} \biggr ]\biggl [\left (1 - \frac{1}{2}\left (\frac{z_2}{z_1+z_2}\right )\right )^2 + \frac{1}{4}\left (\frac{z_2}{z_1+z_2}\right )^2 \biggr ]\\
& = \frac{1}{4(z_1+z_2)^2}\biggl [1 - 2z_3(1-z_3) \biggr ]\biggl [2z_1(z_1+z_2) + z_2^2 \biggr ]
\end{split}
\end{equation}
and
\begin{equation}
\begin{split}
G(z_2,z_1) & =  \biggl [\left (z_1 - \frac{1}{2}\right )^2 + \frac{1}{4} \biggr ]\biggl [\left (1 - \frac{1}{2}\left (\frac{z_2}{z_2+z_3}\right )\right )^2 + \frac{1}{4}\left (\frac{z_2}{z_2+z_3}\right )^2 \biggr ]\\
& = \frac{1}{4(z_2+z_3)^2}\biggl [1 - 2z_1(1-z_1) \biggr ]\biggl [2z_3(z_2+z_3) + z_2^2 \biggr ],
\end{split}
\end{equation}
and, similarly as in the longitudinal case, we have defined the function
\begin{equation}
\mathcal{L}(\bt^2,\rt^2,\overline{Q}^2,\omega) = \int_{0}^{\infty} \frac{\ud u}{u^2}e^{-u\overline{Q}^2}e^{-\frac{\bt^2}{4u}}\Gamma\left (\frac{d}{2}-1, \frac{\omega \rt^2}{4u}\right ).
\end{equation}

Now the UV divergences in the $\Theta_{\ref{diag:qgqbarT}}$ and $\Theta_{\ref{diag:qqbargT}}$ terms can be subtracted in the same way as in the longitudinal photon case. Introducing the subtraction terms
\begin{equation}
\label{eq:subt1}
\overline{\Theta}_{\ref{diag:qgqbarT}}\bigg\vert_{\textrm{UV};\zt\rightarrow \xt} = \frac{F(z_1,z_2,z_3)(\mu^2)^{2-d/2}}{16^2\pi^d}\int_{[\zt]}  \biggl [\rt_{xy}^2(\rt_{zx}^2)^{3-d}\biggr ]\mathcal{L}^2(\rt_{xy}^2,\rt_{zx}^2,\overline{Q}^2_{\ref{diag:qgqbarT}},\omega_{\ref{diag:qgqbarT}})\bigg\vert_{\rm{UV}} \left (1-\mathcal{S}_{xy}\right )
\end{equation}
and 
\begin{equation}
\label{eq:subt2}
\overline{\Theta}_{\ref{diag:qqbargT}}\bigg\vert_{\textrm{UV};\zt\rightarrow \yt} = \frac{G(z_1,z_2,z_3)(\mu^2)^{2-d/2}}{16^2\pi^d}\int_{[\zt]} \biggl [\rt_{xy}^2(\rt_{zy}^2)^{3-d}\biggr ]\mathcal{L}^2(\rt_{xy}^2,\rt_{zy}^2,\overline{Q}^2_{\ref{diag:qqbargT}},\omega_{\ref{diag:qqbargT}})\bigg\vert_{\rm{UV}} \left (1-\mathcal{S}_{xy}\right ),
\end{equation}
where
\begin{equation}
\begin{split}
\mathcal{L}(\bt^2,\rt^2,\overline{Q}^2,\omega)\bigg\vert_{\rm{UV}}  & =  \int_{0}^{\infty} \frac{\ud u}{u^2}e^{-u\overline{Q}^2}e^{-\frac{\bt^2}{4u}}\Gamma\left (\frac{d}{2}-1\right )e^{-\frac{\rt^2}{2\bt^2\xi}} = \frac{4\overline{Q}}{\vert\bt\vert} K_{1}\left (\overline{Q}\vert \bt\vert\right )\Gamma\left (\frac{d}{2}-1\right )e^{-\frac{\rt^2}{2\bt^2\xi}}, \quad \xi\in \Re\\
\end{split}
\end{equation}
we can write down the UV subtracted contributions for \nr{eq:qgqbarT} and \nr{eq:qqbargT} 
\begin{equation}
\begin{split}
\overline{\Theta}_{\ref{diag:qgqbarT}} -\overline{\Theta}_{\ref{diag:qgqbarT}}\bigg\vert_{\textrm{UV};\zt\rightarrow \xt}  = &\frac{Q^2}{4(2\pi)^4}f(z_1,z_2,z_3)\int_{\zt} \biggl [\left (\frac{1}{\rt_{zx}^2} - \frac{z_1z_2}{(z_1+z_2)}\frac{1}{\Rt^2}  \right )[K_1(Q\vert \Rt\vert)]^2\left (1-\mathcal{S}_{xyz}\right )\\
& - \frac{1}{\rt_{zx}^2}[K_1(\overline{Q}_{\ref{diag:qgqbarT}}\vert \rt_{xy}\vert)]^2e^{-\rt_{zx}^2/(\rt_{xy}^2\xi)} \left (1-\mathcal{S}_{xy}\right )  \biggr ]
\end{split}
\end{equation}
and 
\begin{equation}
\begin{split}
\overline{\Theta}_{\ref{diag:qqbargT}} -\overline{\Theta}_{\ref{diag:qqbargT}}\bigg\vert_{\textrm{UV};\zt\rightarrow \yt}  = &\frac{Q^2}{4(2\pi)^4}g(z_1,z_2,z_3)\int_{\zt} \biggl [\left (\frac{1}{\rt_{zy}^2} - \frac{z_3z_2}{(z_2+z_3)}\frac{1}{\Rt^2}  \right )[K_1(Q\vert \Rt\vert)]^2\left (1-\mathcal{S}_{xyz}\right )\\
& - \frac{1}{\rt_{zy}^2}[K_1(\overline{Q}_{\ref{diag:qqbargT}}\vert \rt_{xy}\vert)]^2e^{-\rt_{zy}^2/(\rt_{xy}^2\xi)} \left (1-\mathcal{S}_{xy}\right )  \biggr ]
\end{split}
\end{equation}
with 
\begin{equation}
\begin{split}
f(z_1,z_2) &= \frac{z_3}{(z_1+z_2)}\biggl [1 - 2z_3(1-z_3) \biggr ]\biggl [2z_1(z_1+z_2) + z_2^2 \biggr ]\\
g(z_1,z_2) &= \frac{z_1}{(z_2+z_3)}\biggl [1 - 2z_1(1-z_1) \biggr ]\biggl [2z_3(z_2+z_3) + z_2^2 \biggr ].
\end{split}
\end{equation}

The contributions in \eqs\nr{eq:diag:qgqbarinst:sqr}, \nr{eq:diag:qqbarginst:sqr} and \nr{eq:thetaT} are all UV-finite and hence one can perform the computation in four dimension. The calculation of these terms is lengthy but straightforward. Thus in here we only show the final result and include the detailed derivation in Appendix \ref{app:finite}: 
\begin{equation}\label{eq:Tfinitefromapp}
\begin{split}
\overline{\Theta}\bigg\vert_{\text{UV-finite}} =  \frac{Q^2}{4(2\pi)^4}& \int_{\zt}\frac{[K_1(Q\vert \Rt\vert)]^2}{\Rt^2}\biggl [-2z_1z_3\biggl \{ \biggl [(1-z_1)^2 + z_1^2 \biggr ] + \biggl [(1-z_3)^2 + z_3^2 \biggr ]  \biggr \}\frac{\Rt^2(\rt_{zx}\cdot \rt_{zy})}{\rt_{zx}^2\rt_{zy}^2}\\
& + \frac{2z_1(z_2z_3)^2}{(z_1+z_2)}\frac{(\rt_{zx}\cdot \rt_{zy})}{\rt_{zx}^2}  + \frac{2z_3(z_2z_1)^2}{(z_2+z_3)}\frac{(\rt_{zx}\cdot \rt_{zy})}{\rt_{zy}^2} + z_1z_2z_3\biggl \{(z_1+z_2)^2 + (z_2+z_3)^2\\
& + 2z_1^2 + 2z_3^2 + \frac{(z_1z_3)^2}{(z_1+z_2)^2} +  \frac{(z_1z_3)^2}{(z_2+z_3)^2} - z_2\biggl [\frac{(z_1+z_2)}{(z_2+z_3)} + \frac{(z_2 + z_3)}{(z_1+z_2)}  \biggr ]\biggr \} \biggr ]\left (1-\mathcal{S}_{xyz}\right ).
\end{split}
\end{equation}
Adding all the pieces together gives the result
\begin{equation}
\label{eq:qqbargamplitudesquarefinalTv3}
\begin{split}
{}_{q\bar{q}g}\langle \gamma^{\ast}_{\rm T}(q'^+,Q^2,\lambda')&\vert 1-\hat{S}_E \vert \gamma^{\ast}_{\rm T}(q^+,Q^2,\lambda)\rangle_{q\bar{q}g} = 2q^+(2\pi)\delta(q'^+-q^+)\frac{16(2\pi)^3\nc\alpha_{em}e_f^2}{(2\pi)^2}\left (\frac{\alpha_s\cf}{\pi}\right )\int_{\xt\yt}\\
&\int_{0}^{\infty}\ud z_1\int_{0}^{\infty}\ud z_3\int_{0}^{\infty} \frac{\ud z_2}{z_2} \delta(z_1+z_2+z_3-1)\biggl \{\biggl [ \overline{\Theta}_{\ref{diag:qgqbarT}} -\overline{\Theta}_{\ref{diag:qgqbarT}}\bigg\vert_{\textrm{UV};\zt\rightarrow \xt} \biggr ] + \biggl [ \overline{\Theta}_{\ref{diag:qqbargT}} -\overline{\Theta}_{\ref{diag:qqbargT}}\bigg\vert_{\textrm{UV};\zt\rightarrow \yt} \biggr ]\\
& +  \overline{\Theta}\bigg\vert_{\text{UV-finite}} + \overline{\Theta}_{\ref{diag:qgqbarT}}\bigg\vert_{\textrm{UV};\zt\rightarrow \xt} + \overline{\Theta}_{\ref{diag:qqbargT}}\bigg\vert_{\textrm{UV};\zt\rightarrow \yt}\biggr \}.
\end{split} 
\end{equation}
Dividing this expression into finite and UV-divergent parts as 
\begin{equation}
\begin{split}
{}_{q\bar{q}g}\langle \gamma^{\ast}_{\rm T}(q'^+,Q^2,\lambda')\vert 1-\hat{S}_E \vert \gamma^{\ast}_{\rm T}(q^+,Q^2,\lambda)\rangle_{q\bar{q}g} & = {}_{q\bar{q}g}\langle \gamma^{\ast}_{\rm T}(q'^+,Q^2,\lambda')\vert 1-\hat{S}_E \vert \gamma^{\ast}_{\rm T}(q^+,Q^2,\lambda)\rangle_{q\bar{q}g}\bigg\vert_{\rm UV-fin}\\
&  + {}_{q\bar{q}g}\langle \gamma^{\ast}_{\rm T}(q'^+,Q^2,\lambda')\vert 1-\hat{S}_E \vert \gamma^{\ast}_{\rm T}(q^+,Q^2,\lambda)\rangle_{q\bar{q}g}\bigg\vert_{\rm UV-div }
\end{split}
\end{equation} 
and carrying out some algebra we otain
\begin{equation}
\begin{split}
&{}_{q\bar{q}g}\langle \gamma^{\ast}_{\rm T}(q'^+,Q^2,\lambda')\vert 1-\hat{S}_E \vert \gamma^{\ast}_{\rm T}(q^+,Q^2,\lambda)\rangle_{q\bar{q}g}\bigg\vert_{\rm UV-fin} = 2q^+(2\pi)\delta(q'^+-q^+)\frac{4\nc\alpha_{em}Q^2e_f^2}{(2\pi)^3}\left (\frac{\alpha_s\cf}{\pi}\right )\int_{\xt\yt\zt}\\
&\times \int_{0}^{1}\ud z_1\int_{0}^{1-z_1}\frac{\ud z_2}{z_2}\biggl \{\frac{f}{\rt_{zx}^2}\biggl [[K_1(Q\vert \Rt\vert)]^2\left (1-\mathcal{S}_{xyz}\right )-[K_1(\overline{Q}_{\ref{diag:qgqbarT}}\vert \rt_{xy}\vert)]^2e^{-\rt_{zx}^2/(\rt_{xy}^2\xi)} \left (1-\mathcal{S}_{xy}\right )  \biggr ]\\
& +\frac{g}{\rt_{zy}^2}\biggl [[K_1(Q\vert \Rt\vert)]^2\left (1-\mathcal{S}_{xyz}\right ) -[K_1(\overline{Q}_{\ref{diag:qqbargT}}\vert \rt_{xy}\vert)]^2e^{-\rt_{zy}^2/(\rt_{xy}^2\xi)} \left (1-\mathcal{S}_{xy}\right )  \biggr ] +\frac{[K_1(Q\vert \Rt\vert)]^2}{\Rt^2}\Pi\left (1- \mathcal{S}_{xyz}\right )\biggr \},
\end{split}
\end{equation}
where
\begin{equation}
\begin{split}
\Pi = &-2z_1z_3\biggl \{ \biggl [(1-z_1)^2 + z_1^2 \biggr ] + \biggl [(1-z_3)^2 + z_3^2 \biggr ]  \biggr \}\frac{\Rt^2(\rt_{zx}\cdot \rt_{zy})}{\rt_{zx}^2\rt_{zy}^2} + \frac{2z_1(z_2z_3)^2}{(z_1+z_2)}\frac{(\rt_{zx}\cdot \rt_{zy})}{\rt_{zx}^2}  + \frac{2z_3(z_2z_1)^2}{(z_2+z_3)}\frac{(\rt_{zx}\cdot \rt_{zy})}{\rt_{zy}^2}\\
& + z_1z_2z_3\biggl \{(z_1+z_2)^2 + (z_2+z_3)^2 + 2z_1^2 + 2z_3^2 + \frac{(z_1z_3)^2}{(z_1+z_2)^2} +  \frac{(z_1z_3)^2}{(z_2+z_3)^2} - z_2\biggl [\frac{(z_1+z_2)}{(z_2+z_3)} + \frac{(z_2 + z_3)}{(z_1+z_2)}  \biggr ]\\
& -\frac{\biggl [1 - 2z_3(1-z_3) \biggr ]\biggl [2z_1(z_1+z_2) + z_2^2 \biggr ]}{(z_1+z_2)^2} -\frac{\biggl [1 - 2z_1(1-z_1) \biggr ]\biggl [2z_3(z_2+z_3) + z_2^2 \biggr ]}{(z_2+z_3)^2} \biggr \} 
\end{split}
\end{equation}
and
\begin{equation}
\label{eq:uvschemepartT}
\begin{split}
&{}_{q\bar{q}g}\langle \gamma^{\ast}_{\rm T}(q'^+,Q^2,\lambda')\vert 1-\hat{S}_E \vert \gamma^{\ast}_{\rm T}(q^+,Q^2,\lambda)\rangle_{q\bar{q}g}\bigg\vert_{\rm UV-div} =  -2q^+(2\pi)\delta(q'^+-q^+)\frac{4\nc\alpha_{em}e_f^2Q^2}{(2\pi)^2}\left (\frac{\alpha_s\cf}{\pi}\right )\int_{\xt\yt}\int_{0}^{1}\ud z\\
& \times  [K_1(\overline{Q}\vert\rt_{xy}\vert)]^2 z(1-z)\biggl \{1-2z(1-z)\biggr \}\biggl [\frac{3}{2} + \log\left (\frac{\alpha}{z}\right ) + \log\left (\frac{\alpha}{1-z}\right )\biggr ]\biggl \{\frac{1}{\varepsilon_{\overline{\rm MS}}} + \log\left (\frac{\rt_{xy}^2\mu^2}{4}\right ) -2\Psi_0(1)\biggr \} \left (1-\mathcal{S}_{xy}\right )
.
\end{split}
\end{equation}
Like in the longitudinal photon case, the last expresion above cancels the UV-divergent term in square brackets in \eq\nr{eq:LOTunsub}.    
Finally thanks to \eq\nr{eq:CSfinal}, the total cross section for transverse virtual photon (averaged over the two incoming transverse virtual photon polarization states) at NLO accuracy is given by 
\begin{equation}
\label{eq:Tfinalresult}
\sigma^{\gamma^{\ast}_{\rm T}}[A] = \sigma^{\gamma^{\ast}_{\rm T}}\bigg\vert_{q\bar{q}} +  \sigma^{\gamma^{\ast}_{\rm T}}\bigg\vert_{q\bar{q}g},
\end{equation}
where the $q\bar{q}$-term is   
\begin{equation}
\label{eq:Tfinalresulta}
\begin{split}
\sigma^{\gamma^{\ast}_{\rm T}}\bigg\vert_{q\bar{q}}  = 4\nc \frac{\alpha_{em}e_f^2Q^2}{(2\pi)^2}\int_{\xt\yt}&\int_{0}^{1}\ud z [K_{1}\left (\overline{Q}\vert \rt_{xy}\vert\right )]^2z(1-z)\biggl \{1-2z(1-z)\biggr \}\\
&\times \biggl \{1 + \left ( \frac{\alpha_s\cf}{\pi}\right )\biggl [\frac{1}{2}\log^2\left ( \frac{z}{1-z}\right )-\frac{\pi^2}{6} + \frac{5}{2} \biggr ]\biggr \}\left (1-\mathcal{S}_{xy}\right )
\end{split}
\end{equation}
and for the $q\bar{q}g$-component we find
\begin{equation}
\label{eq:Tfinalresultb}
\begin{split}
\sigma^{\gamma^{\ast}_{\rm T}}\bigg\vert_{q\bar{q}g}  &=  4\nc \frac{\alpha_{em}e_f^2Q^2}{(2\pi)^3}\left ( \frac{\alpha_s\cf}{\pi}\right )
\int_{  \xt  \yt  \zt} \int_{0}^{1}\ud z_1\int_{0}^{1-z_1}\frac{\ud z_2}{z_2}\\
&\times \biggl \{\frac{f}{\rt_{zx}^2}\biggl [[K_1(Q\vert \Rt\vert)]^2\left (1-\mathcal{S}_{xyz}\right )-[K_1(\overline{Q}_{\ref{diag:qgqbarT}}\vert \rt_{xy}\vert)]^2e^{-\rt_{zx}^2/(\rt_{xy}^2\xi)} \left (1-\mathcal{S}_{xy}\right )  \biggr ]\\
& +\frac{g}{\rt_{zy}^2}\biggl [[K_1(Q\vert \Rt\vert)]^2\left (1-\mathcal{S}_{xyz}\right ) -[K_1(\overline{Q}_{\ref{diag:qqbargT}}\vert \rt_{xy}\vert)]^2e^{-\rt_{zy}^2/(\rt_{xy}^2\xi)} \left (1-\mathcal{S}_{xy}\right )  \biggr ]
 +\frac{[K_1(Q\vert \Rt\vert)]^2}{\Rt^2}\Pi\left (1- \mathcal{S}_{xyz}\right )\biggr \}.
\end{split}
\end{equation}

Again, as for the longitudinal case the scheme dependent UV contribution in \eq\nr{eq:uvschemepartT} cancels
the scheme dependent UV part obtained in \eq\nr{eq:LOTunsub}, and the remaining finite contribution in \eq\nr{eq:LOTunsub} leads to the scheme independent final result for $q\bar{q}$-part in \eq\nr{eq:Tfinalresulta}. In addition, like in the longitudinal case, we have confirmed both analytically and also numerically that our final results for the cross section in \eq\nr{eq:Tfinalresulta} and \eq\nr{eq:Tfinalresultb} agree with \cite{Beuf:2017bpd}, and we have checked that the part inside the curly brackets in \eq\nr{eq:Tfinalresultb} reduces to the r.h.s. of the BK equation.

\section{Conclusions and outlook}
\label{sec:conc}

As a concrete result, we have in this paper derived the NLO cross section for deep inelastic scattering in the dipole picture, with the final results given in 
\eq\nr{eq:Lfinalresult} (with \eqs\nr{eq:Lfinalresulta} and~\nr{eq:Lfinalresultb}) for the longitudinal and in \eq\nr{eq:Tfinalresult} (with \eqs\nr{eq:Tfinalresulta} and~\nr{eq:Tfinalresultb}) for the transverse virtual photon polarization. We have confirmed both analytically and numerically that our results agree with those of G. Beuf in~\cite{Beuf:2017bpd}. Being derived in a different regularization scheme, they are an indication of the scheme-independence of this result. As a small difference, we believe that our choice of the subtraction term to cancel the UV divergence is, while equivalent, somewhat more benign numerically.

Nevertheless, the most important purpose of this paper has been to develop calculational techniques that should enable further NLO calculations to be more efficiently performed in LCPT. We have demonstrated how to express the elementary vertices of the theory systematically in terms of their natural variables, the center-of-mass splitting momentum, splitting momentum fraction and the helicities of the particles involved. Using our expressions the evaluation of the scheme-independent parts of the cross section reduces to multiplications of 2-dimensional vectors and tensors, and simple scalar integrations over longitudinal momentum fractions. They can be easily automated by symbolic manipulation prograns such as \textsc{form}~\cite{Vermaseren:2000nd} or \textsc{FeynCalc}~\cite{Shtabovenko:2016sxi}.
The scheme dependent parts require some more work, where at one point one must reduce expressions of Dirac matrices contracted with $(d_s-2)$- and $(d-2)$-dimensional Kronecker deltas. However, this procedure also is readily automated.  We hope that the method developed here can be useful in future work. As an immediate future application with clear phenomenological relevance, the next step is  to include quark masses in the DIS cross section calculation.

\section*{Acknowledgments} 
We thank G. Beuf for numerous discussions and providing his results in~\cite{Beuf:2017bpd} to us already prior to publication. This work has been supported by the Academy of Finland, projects 273464 and 303756, and  by the European Research Council, grants
ERC-2015-CoG-681707 and ERC-2016-CoG-725369.

\appendix

\section{Decomposition of LC vertices }
\label{appendix:decomplcvertices}

In this section we show how to decompose the general LC vertex to the symmetric and antisymmetric parts as discussed in section \ref{sec:vertices}. The most general form for the LC vertex (without the coupling and color structure) in the LC gauge is given by 
\begin{equation}
\label{eq:genralLCvertex}
\bar{\chi}_{h}(p')\epsl_\lambda(q)\omega_{s}(p) = \frac{\qt^i \epst_{\lambda}^j}{q^+}\delta^{ij}  \bar{\chi}_{h}(p')\gamma^{+}\omega_{s}(p) - \epst_{\lambda}^i\bar{\chi}_{h}(p')\gamma^{i}\omega_{s}(p)  ,
\end{equation}
where $\chi$ and $\omega$ can be either positive or negative energy massless spinors, i.e.  $u$ or $v$. For massless quarks the spinors  $\chi$ and $\omega$ satisfy the Dirac equations:
\begin{equation}
\label{eq:diraceq}
\begin{split}
\psl \omega_{s}(p) &= \left (\gamma^{+}p^- + \gamma^{-}p^+ - \gamma^j \pt^j \right )\omega_{s}(p) = 0,\\
\bar{\chi}_{h}(p')\ppsl & = \bar{\chi}_{h}(p')\left (\gamma^{+}p'^{-} + \gamma^{-}p'^{+} - \gamma^{j}\ptp^j \right )= 0.
\end{split}
\end{equation}
Applying the Clifford algebra one can write 
\begin{equation}
\bar{\chi}_{h}(p')\gamma^{+}\gamma^i\gamma^-\omega_{s}(p) = - \bar{\chi}_{h}(p')\gamma^{+}\gamma^-\gamma^i\omega_{s}(p) = -2\bar{\chi}_{h}(p')\gamma^i\omega_{s}(p)  + \bar{\chi}_{h}(p')\gamma^{-}\gamma^+\gamma^i\omega_{s}(p)
\end{equation}
which gives 
\begin{equation}
\label{eq:form1}
-2\bar{\chi}_{h}(p')\gamma^i\omega_{s}(p) = \bar{\chi}_{h}(p')\gamma^{+}\gamma^i\gamma^-\omega_{s}(p) -  \bar{\chi}_{h}(p')\gamma^{-}\gamma^+\gamma^i\omega_{s}(p).
\end{equation}
Furthemore, using the Dirac equation \nr{eq:diraceq} we find  
\begin{equation}
\bar{\chi}_{h}(p')\gamma^{+}\gamma^i\gamma^-\omega_{s}(p) = \frac{1}{p^+}\bar{\chi}_{h}(p')\gamma^{+}\gamma^i\gamma^-p^+\omega_{s}(p) = -\frac{1}{p^+}\bar{\chi}_{h}(p')\gamma^{+}\gamma^i\left (\gamma^+p^- - \gamma^j\pt^j \right )\omega_{s}(p)
\end{equation}
and
\begin{equation}
\bar{\chi}_{h}(p')\gamma^{-}\gamma^+\gamma^i\omega_{s}(p) = \frac{1}{p'^+}\bar{\chi}_{h}(p')\gamma^{-}p'^+\gamma^+\gamma^i\omega_{s}(p) = -\frac{1}{p'^+}\bar{\chi}_{h}(p')\left (\gamma^+p'^- -\gamma^j\ptp^j\right )\gamma^{+}\gamma^i \omega_{s}(p).
\end{equation}
Since $\gamma^+\gamma^+ = 0$, these simplify to 
\begin{equation}
\begin{split}
\label{eq:form2}
\bar{\chi}_{h}(p')\gamma^{+}\gamma^i\gamma^-\omega_{s}(p) & = \frac{\pt^j}{p^+}\bar{\chi}_{h}(p')\gamma^{+}\gamma^i\gamma^j\omega_{s}(p)\\
\bar{\chi}_{h}(p')\gamma^{-}\gamma^+\gamma^i\omega_{s}(p) & = -\frac{\ptp^j}{p'^+}\bar{\chi}_{h}(p')\gamma^{+}\gamma^j\gamma^i\omega_{s}(p).
\end{split}
\end{equation}
Combining \eqs\nr{eq:form1} and \nr{eq:form2} we obtain
\begin{equation}
\label{eq:form3}
-2\bar{\chi}_{h}(p')\gamma^i\omega_{s}(p) = \frac{\pt^j}{p^+}\bar{\chi}_{h}(p')\gamma^{+}\gamma^i\gamma^j\omega_{s}(p) + \frac{\ptp^j}{p'^+}\bar{\chi}_{h}(p')\gamma^{+}\gamma^j\gamma^i\omega_{s}(p).
\end{equation}
In order to separate the symmetric and anti-symmetric parts in \eq\nr{eq:form3} we use the identity 
\begin{equation}
\gamma^i\gamma^j = -\delta^{ij} + \frac{1}{2}[\gamma^i,\gamma^j],
\end{equation}
which gives 
\begin{equation}
\bar{\chi}_{h}(p')\gamma^i\omega_{s}(p) = \left [ \frac{\pt^j}{2p^+} + \frac{\ptp^j}{2p'^+} \right ]\delta^{ij}\bar{\chi}_{h}(p')\gamma^{+}\omega_{s}(p) - \left [ \frac{\pt^j}{4p^+} - \frac{\ptp^j}{4p'^+} \right ]\bar{\chi}_{h}(p')\gamma^{+}[\gamma^i,\gamma^j]\omega_{s}(p). 
\end{equation}
Inserting the above expression into \eq\nr{eq:genralLCvertex} gives 
\begin{equation}
\label{eq:finalLCvertex}
\bar{\chi}_{h}(p')\epsl_\lambda(q)\omega_{s}(p) = 
\left [\frac{\qt^i}{q^+} - \frac{\pt^i}{2p^+} - \frac{\ptp^i}{2p'^+} \right ] \epst_{\lambda}^j\delta^{ij}  \bar{\chi}_{h}(p')\gamma^{+}\omega_{s}(p) -  \left [ \frac{\pt^i}{4p^+} - \frac{\ptp^i}{4p'^+} \right ]\epst_{\lambda}^j\bar{\chi}_{h}(p')\gamma^{+}[\gamma^i,\gamma^j]\omega_{s}(p). 
\end{equation}
This equation is valid in arbitrary spacetime dimensions and automatically includes the plus and transverse momentum conservation.  

In the particular case $d=4$, this expression can be very compactly expresed in the helicity basis by first nothing that the commutator of Dirac transverse gamma matrices can be expressed as  
\begin{equation}
[\gamma^i,\gamma^j] = -4i\epsilon^{ij}S^3,
\end{equation}
where $\epsilon^{ij}$ is the anti-symmetric rank-two levi-civita tensor, and $S^3$ is the light cone helicity operator acting on the good component of the spinors \footnote{The projections to the good (G) and bad (B) components of a complete spinor field $\Psi$ are defined as $P_{G/B}\Psi = \Psi_{G/B}$, where $P_G = \gamma^-\gamma^+/2$ and $P_B = \gamma^+\gamma^-/2$ (see e.g. \cite{Brodsky:1997de}).}
\begin{equation}
\begin{split}
S^3u^{(G)}_{h}(p^+) & = \frac{h}{2}u^{(G)}_{h}(p^+)\\
S^3v^{(G)}_{h}(p^+) & = -\frac{h}{2}v^{(G)}_{h}(p^+),\\
\end{split}
\end{equation}
where  we denote the two fermion spin states $\pm 1/2$ by $h=\pm$ for notational simplicity. In addition, it is easy to show that the following relation between the complete spinors and good component of the spinors is satisfied  
\begin{equation}
\bar{\chi}_h(p')\gamma^+\omega_{s}(p) = \bar{\chi}^{(G)}_{h}(p'^+)\gamma^+\omega^{(G)}_{s}(p^+).
\end{equation}
Therefore, in four dimensions we find a very useful simplification of the Dirac algebra in \eq\nr{eq:vertexqtoqgd}
\begin{equation}
\bar{u}_{h'}(p')\gamma^{+}[\gamma^i,\gamma^j]u_h(p)  = -4i\epsilon^{ij}\bar{u}_{h'}(p')\gamma^{+}S^3u_h(p) = -2ih\epsilon^{ij}\bar{u}_{h'}(p')\gamma^{+}u_h(p).
\end{equation}

\section{Transverse integrals}
\label{app:loopints}

\subsection{Transversely polarized photon}

In order to compute the vertex corrections for the transversely polarized virtual photon one must evaluate the following rank-3 (r3) tensor integral
\begin{equation}
I^{(r3)}(\rt,\hat\pt,\hat\qt;M_1,M_2) = (4\pi)\int_{\mt}\frac{m^i(m-\hat{p})^j(m-\hat{q})^k}{D_0D_1},
\end{equation}
where the denominators $D_0$ and $D_1$ are defined as 
\begin{equation}
\label{eq:D0D1}
D_0 = \mt^2 + M_1, \quad D_1 = (\mt - \rt)^2 + M_2
\end{equation}
and the integral measure in $d_{\perp} = 2 - 2\varepsilon$ dimensions is
\begin{equation}
\int_{\mt} = (\mu^2)^{1-d_{\perp}/2}\int \frac{d^{d_{\perp}}\mt}{(2\pi)^{d_{\perp}}} = \mu^{2\varepsilon}\int\frac{d^{2-2\varepsilon}\mt}{(2\pi)^{2-2\varepsilon}}.
\end{equation}
Using the standard Feynman parametrization with  
\begin{equation}
(1-x)D_0 + xD_1 = (\mt - x\rt)^2 + x(1-x)\rt^2 + (1-x)M_1 + xM_2,
\end{equation}
and performing the change of variables $\nt = \mt - x\rt$ gives 
\begin{equation}
I^{(r3)}(\rt,\hat\pt,\hat\qt;M_1,M_2) = (4\pi)\int_{0}^{1} \mathrm{d}x \int_{\nt} \frac{(n + xr)^i(n+xr-\hat p)^i(n+xr-\hat q)^k}{(\nt^2 + M)^2},
\end{equation}
where
\begin{equation}
\label{eq:Mfunction}
M = x(1-x)\rt^2 + (1-x)M_1 + xM_2.
\end{equation}
Upon the integration over transverse momentum $\nt$  the numerator simplifies to 
\begin{equation}
\begin{split}
(n + xr)^i(n+xr-\hat p)^j(n+xr-\hat q)^k = \frac{\nt^2}{d_{\perp}}\biggl \{ xr^i\delta^{jk} & + (xr-\tilde p)^j\delta^{ik} + (xr-\tilde q)^k\delta^{ij} \biggr \} + xr^{i}(xr-\hat p)^j(xr-\hat q)^k\\
& + \mathcal{O}(\text{$\nt$ and $\nt^3$}),
\end{split}
\end{equation}
where the linear and cubic terms in $\nt$ goes to zero in dimensional regularization framework. Performing the transverse integrals over $\nt$ with standard momentum integrals that we have listed in \cite{Lappi:2016oup}, and expanding in power of $\varepsilon$ we obtain
\begin{equation}
I^{(r3)}(\rt,\hat\pt;M_1,M_2) = I^{(r3)}\bigg\vert_{\rm UV} + I^{(r3)}\bigg\vert_{\rm f} + I^{(r3)}\bigg\vert_{\rm F} + \mathcal{O}(\varepsilon),
\end{equation}
where the UV-divergent part of the integrals becomes
\begin{equation}
\begin{split}
I^{(r3)}\bigg\vert_{\rm UV}  &= \frac{1}{2}\biggl [\frac{1}{\varepsilon_{\overline{\rm MS}}} + \log \left (\frac{\mu^2}{\overline{Q}^2}\right ) \biggr ]\int_{0}^{1}\mathrm{d}x \left (xr^i\delta^{jk} + (xr-\tilde p)^j\delta^{ik} + (xr-\tilde q)^k\delta^{ij} \right )\\
& = \frac{1}{4}\biggl [\frac{1}{\varepsilon_{\overline{\rm MS}}} + \log \left (\frac{\mu^2}{\overline{Q}^2}\right ) \biggr ]\left ( r^i\delta^{jk} + (r - 2\tilde p)^j\delta^{ik} + (r - 2\tilde q)^k\delta^{ij}\right )
\end{split}
\end{equation}
and the UV-finite parts 
\begin{equation}
I^{(r3)}\bigg\vert_{\rm f}  = \frac{1}{2}\int_{0}^{1}\mathrm{d}x \left (x\Delta^{(r3)}_{\textrm{f1}} + \Delta^{(r3)}_{\textrm{f2}} \right )\log \left (\frac{\overline{Q}^2}{M} \right ),
\end{equation}
\begin{equation}
I^{(r3)}\bigg\vert_{\rm F} = \int_{0}^{1}\mathrm{d}x \frac{x^3\Delta^{(r3)}_{\textrm{F1}} + x^2\Delta^{(r3)}_{\textrm{F2}} + x\Delta^{(r3)}_{\textrm{F3}}}{M}. 
\end{equation}
Here the coefficients $\Delta^{(r3)}_{\textrm{Fi}}$ and $\Delta^{(r3)}_{\textrm{fj}}$ are given by
\begin{equation}
\begin{split}
\Delta^{(r3)}_{\textrm{F1}} &= r^ir^jr^k\\
\Delta^{(r3)}_{\textrm{F2}} &= -r^i(r^j\hat q^k + r^k\hat p^j)\\
\Delta^{(r3)}_{\textrm{F3}} &= r^i\hat p^j\hat q^k\\
\Delta^{(r3)}_{\textrm{f1}} & = r^i\delta^{jk} + r^j\delta^{ik} + r^k\delta^{ij}\\
\Delta^{(r3)}_{\textrm{f2}} & = -\hat p^j\delta^{ik} -\hat q^k\delta^{ij}.
\end{split}
\end{equation}

\subsection{Longitudinally polarized photon}

In order to compute the vertex corrections for the longitudinally polarized virtual photon one must evaluate the following rank-2 (r2) tensor integral
\begin{equation}
I^{(r2)}(\rt,\hat\pt;M_1,M_2) = (4\pi)\int_{\mt}\frac{m^i(m-\hat p)^j}{D_0D_1}
\end{equation}
where the denominators $D_0$ and $D_1$ are given by \eq\nr{eq:D0D1}. Performing the Feynman parametrization and the transverse integrals as in the transverse photon case we obtain 
\begin{equation}
I^{(r2)}(\rt,\hat\pt;M_1,M_2) = I^{(r2)}\bigg\vert_{\rm UV} + I^{(r2)}\bigg\vert_{\rm f} + I^{(r2)}\bigg\vert_{\rm F} + \mathcal{O}(\varepsilon)
\end{equation}
where the UV and finite parts simplify to
\begin{equation}
I^{(r2)}\bigg\vert_{\rm UV} = \frac{\delta^{ij}_{(d_{\perp})}}{2}\biggl [\frac{1}{\varepsilon_{\overline{\rm MS}}} + \log \left (\frac{\mu^2}{\overline{Q}^2}\right ) \biggr ]
\end{equation}
\begin{equation}
I^{(r2)}\bigg\vert_{\rm f}  = \frac{\delta^{ij}_{(d_{\perp})}}{2}\int_{0}^{1}\mathrm{d}x \log \left (\frac{\overline{Q}^2}{M} \right ) 
\end{equation}
and
\begin{equation}
I^{(r2)}\bigg\vert_{\rm F} = \int_{0}^{1}\mathrm{d}x \frac{x^2\Delta^{(r2)}_{\rm F1} + x\Delta^{\rm (r2)}_{\textrm{F}2}}{M}
\end{equation}
with the coefficients $\Delta^{(r2)}_{\rm F1} = r^ir^j$ and $\Delta^{(r2)}_{\rm F2} = -r^i\hat{p}^j $.
The remaining integrals over $x$ are straighforward to perform, but yield complicated expressions that we will not write out here.

\section{Transverse Fourier Integrals}
\label{app:transversefint}

In this appendix, we present the integrals that are needed to calculate the Fourier transformed LCWF's for transverse and longitudinal virtual photon in the mixed space up to NLO. The Fourier transform momentum integrals obtained in this paper can be computed by applying the Schwinger parametrization 
\begin{equation}
\label{eq:SCHPARAM}
\frac{1}{A^{\beta}} = \frac{1}{\Gamma(\beta)}\int_{0}^{\infty}\ud t t^{\beta -1}e^{-tA}, \quad A, \beta >0. 
\end{equation}
For the $q\bar{q}$-component of the longitudinal and transverse virtual photon, the two basic momentum integrals expressed in the mixed space (see section \ref{sec:nlodis}) can be written as
\begin{equation}
\label{eq:FTintegrals}
\begin{split}
\int \frac{\ud^{d-2}\Pt}{(2\pi)^{d-2}}\frac{e^{i\Pt\cdot \xt}}{\biggl [\Pt^2 + \overline{Q}^2\biggr ]} & = (4\pi)^{1-d/2}\int_{0}^{\infty}\ud t t^{1-d/2}e^{-t\overline{Q}^2}e^{-\frac{\xt^2}{4t}} \\
\int \frac{\ud^{d-2}\Pt}{(2\pi)^{d-2}}\frac{ e^{i\Pt\cdot \xt}}{\biggl [\Pt^2 + \overline{Q}^2\biggr ]}\Pt^i & = \frac{i}{2}\xt^i(4\pi)^{1-d/2}\int_{0}^{\infty}\ud t t^{-d/2}e^{-t\overline{Q}^2}e^{-\frac{\xt^2}{4t}},
\end{split}
\end{equation}
where the $(d-2)$-dimensional Gaussian integrals are performed over $\Pt$. Using the formula 
\begin{equation}
\int_{0}^{\infty} \ud t t^{\beta-1}e^{-tA}e^{-\frac{B}{t}} = 2 \left (\frac{B}{A}\right )^{\beta/2}K_{-\beta}\left (2\sqrt{AB}\right ), \quad A,B > 0
\end{equation}
where $K_{\alpha}(z)$ is the modified Bessel function of the second kind, the integrals in \eq\nr{eq:FTintegrals} simplify to 
\begin{equation}
\label{eq:FTintegralsfinal}
\begin{split}
\int \frac{\ud^{d-2}\Pt}{(2\pi)^{d-2}}\frac{e^{i\Pt\cdot \xt}}{\biggl [\Pt^2 + \overline{Q}^2\biggr ]} & = \frac{1}{2\pi}\left (\frac{\overline{Q}}{2\pi\vert\xt\vert}\right )^{d/2-2}K_{\frac{d}{2}-2}\left (\vert\xt\vert\overline{Q}\right ) \\
\int \frac{\ud^{d-2}\Pt}{(2\pi)^{d-2}}\frac{ e^{i\Pt\cdot \xt}}{\biggl [\Pt^2 + \overline{Q}^2\biggr ]}\Pt^i & = i\xt^i\left (\frac{\overline{Q}}{2\pi\vert\xt\vert}\right )^{d/2-1}K_{\frac{d}{2}-1}\left (\vert\xt\vert\overline{Q}\right ).
\end{split}
\end{equation}
In addition, we also need the integrals (see derivation in \cite{Beuf:2016wdz})
\begin{equation}
\begin{split}
\label{eq:FTintegralsextra1final}
\int \frac{\ud^{d-2}\Pt}{(2\pi)^{d-2}}\frac{e^{i\Pt\cdot \xt}}{\biggl [\Pt^2 + \overline{Q}^2\biggr ]}\log \left (\frac{\Pt^2 + \overline{Q}^2}{\overline{Q}^2}\right ) = \frac{1}{2\pi}\left (\frac{\overline{Q}}{2\pi\vert\xt\vert}\right )^{d/2-2}&K_{\frac{d}{2}-2}\left (\vert\xt\vert\overline{Q}\right ) 
\biggl \{\biggl [-\frac{1}{2}\log\left (\frac{\xt^2\overline{Q}^2}{4}\right ) + \Psi_{0}(1)\biggr ]\\
& + \mathcal{O}(d-4) \biggr \}
\end{split}
\end{equation}
\begin{equation}
\label{eq:FTintegralsextra2final}
\begin{split}
\int \frac{\ud^{d-2}\Pt}{(2\pi)^{d-2}}\frac{e^{i\Pt\cdot \xt}}{\biggl [\Pt^2 + \overline{Q}^2\biggr ]}\Pt^{i}\log \left (\frac{\Pt^2 + \overline{Q}^2}{\overline{Q}^2}\right ) = i\xt^i\left (\frac{\overline{Q}}{2\pi\vert\xt\vert}\right )^{d/2-1}& \biggl \{\biggl [-\frac{1}{2}\log\left (\frac{\xt^2\overline{Q}^2}{4}\right ) + \Psi_{0}(1)\biggr ]K_{\frac{d}{2}-1}\left (\vert\xt\vert\overline{Q}\right )\\
&  + \frac{1}{\vert \xt\vert\overline{Q}}K_{0}\left (\vert \xt\vert\overline{Q}\right ) +  \mathcal{O}(d-4) \biggr \},
\end{split}
\end{equation}
where $\Psi_{0}(x)$ is the digamma function with $\Psi_{0}(1) = -\gamma_{E}$, and
\begin{equation}
\label{eq:FTintegralsextra3final}
\begin{split}
\int \frac{\ud^{d-2}\Pt}{(2\pi)^{d-2}}\frac{e^{i\Pt\cdot \xt}}{\biggl [\Pt^2 + \overline{Q}^2\biggr ]}\Pt^{i}\frac{\left (\Pt^2 + \overline{Q}^2\right )}{\Pt^2}\log \left (\frac{\Pt^2 + \overline{Q}^2}{\overline{Q}^2}\right ) = 2i\xt^i\left (\frac{\overline{Q}}{2\pi\vert\xt\vert}\right )^{d/2-1} \biggl \{\frac{1}{\vert \xt\vert\overline{Q}}K_{0}\left (\vert\xt\vert\overline{Q}\right ) +  \mathcal{O}(d-4) \biggr \}.
\end{split}
\end{equation}

For the $q\bar{q}g$-component of the longitudinal virtual photon we need the following integral 
\begin{equation}
\mathcal{I}^i(\xt,\yt,\overline{Q}^2,\omega) = \mu^{2-\frac{d}{2}}\int \frac{\ud^2\Pt}{(2\pi)^2}\int\frac{\ud^{d-2}\Kt}{(2\pi)^{d-2}}\frac{\Kt^ie^{i\Pt\cdot \xt}e^{i\Kt\cdot \yt}}{\biggl [\Pt^2 + \overline{Q}^2\biggr ]\biggl [\Kt^2 + \omega\left (\Pt^2 + \overline{Q}^2\right )\biggr ]}
.
\end{equation}
Using \eqs\nr{eq:FTintegrals} and \nr{eq:SCHPARAM} we get
\begin{equation}
\mathcal{I}^i(\xt,\yt,\overline{Q}^2,\omega) = \mu^{2-\frac{d}{2}}\frac{i}{2}(4\pi)^{1-d/2}\yt^i\int_{0}^{\infty} \ud t t^{-d/2}e^{-\frac{-\yt^2}{4t}}\int_{0}^{\infty}\ud s e^{-(s + t\omega)\overline{Q}^2}\int \frac{\ud^2\Pt}{(2\pi)^2}e^{-(s+t\omega)\Pt^2}e^{i\Pt\cdot \xt}
,
\end{equation}
where the Gaussian integral over the transverse momentum $\Pt$ is 
\begin{equation}
\int \frac{\ud^2\Pt}{(2\pi)^2}e^{-(s+t\omega)\Pt^2}e^{i\Pt\cdot \xt} = (4\pi)^{-1}(s+t\omega)^{-1}e^{-\frac{\xt^2}{4(s+t\omega)}}.
\end{equation}
By making the change of variables $u = s + t\omega$,
\begin{equation}
\mathcal{I}^i(\xt,\yt,\overline{Q}^2,\omega) = \mu^{2-\frac{d}{2}}\frac{i}{2}(4\pi)^{-d/2}\yt^i\int_{0}^{\infty} \ud t t^{-d/2}e^{-\frac{-\yt^2}{4t}}\int_{t\omega}^{\infty}\frac{\ud u}{u}e^{-u\overline{Q}^2}e^{-\frac{\xt^2}{4u}}
\end{equation}
and changing the order of integration we obtain
\begin{equation}
\mathcal{I}^i(\xt,\yt,\overline{Q}^2,\omega) = \mu^{2-\frac{d}{2}}\frac{i}{2}(4\pi)^{-d/2}\yt^i\int_{0}^{\infty} \frac{\ud u}{u}e^{-u\overline{Q}^2}e^{-\frac{\xt^2}{4u}}  \int_{0}^{u/\omega} \ud t t^{-d/2}e^{-\frac{-\yt^2}{4t}}
.
\end{equation}
Finally, performing the outer integral with respect to $t$ we obtain the result 
\begin{equation}
\mathcal{I}^i(\xt,\yt,\overline{Q}^2,\omega) = \mu^{2-\frac{d}{2}}\frac{i}{8}\pi^{-d/2}\yt^i(\yt^2)^{1-d/2}\int_{0}^{\infty} \frac{\ud u}{u}e^{-u\overline{Q}^2}e^{-\frac{\xt^2}{4u}} \Gamma \left (\frac{d}{2}-1,\frac{\omega\yt^2}{4u} \right ),
\end{equation}
where $\Gamma(s,x)$ is the upper incomplete gamma function. For the case $d=4$,
\begin{equation}
\mathcal{I}^i(\xt,\yt,\overline{Q}^2,\omega) = \frac{i}{(2\pi)^2}\frac{\yt^i}{\yt^2}K_0\left (\overline{Q}\sqrt{\xt^2+\omega\yt^2}\right ).
\end{equation}

Similarly, for the $q\bar{q}g$-component of the transverse virtual photon we need the integrals 
\begin{equation}
\label{eq:Imasterrank2app}
\mathcal{I}^{ik}(\xt,\yt,\overline{Q}^2,\omega) = \mu^{2-\frac{d}{2}}\int \frac{\ud^2\Pt}{(2\pi)^2}\int\frac{\ud^{d-2}\Kt}{(2\pi)^{d-2}}\frac{\Pt^i\Kt^ke^{i\Pt\cdot \xt}e^{i\Kt\cdot \yt}}{\biggl [\Pt^2 + \overline{Q}^2\biggr ]\biggl [\Kt^2 + \omega\left (\Pt^2 + \overline{Q}^2\right )\biggr ]}
\end{equation}
and 
\begin{equation}
\label{eq:Imasterrank0app}
\mathcal{I}(\xt,\yt,\overline{Q}^2,\omega) = \mu^{2-\frac{d}{2}}\int \frac{\ud^2\Pt}{(2\pi)^2}\int\frac{\ud^{d-2}\Kt}{(2\pi)^{d-2}}\frac{e^{i\Pt\cdot \xt}e^{i\Kt\cdot \yt}}{\biggl [\Kt^2 + \omega\left (\Pt^2 + \overline{Q}^2\right )\biggr ]}.
\end{equation}
Following the same steps described previously we find 
\begin{equation}
\label{eq:Imasterrank2res}
\mathcal{I}^{ik}(\xt,\yt,\overline{Q}^2,\omega)  = -\mu^{2-d/2}\frac{\pi^{-d/2}}{16}\xt^i\yt^k(\yt^2)^{1-d/2}\int_{0}^{\infty} \frac{\ud u}{u^2}e^{-u\overline{Q}^2}e^{-\frac{\xt^2}{4u}}\Gamma \left (\frac{d}{2}-1,\frac{\omega\yt^2}{4u} \right )
\end{equation}
and
\begin{equation}
\label{eq:Imasterrank0res}
\mathcal{I}(\xt,\yt,\overline{Q}^2,\omega) = (2\pi)^{-d/2}\left (\frac{\mu}{\omega}\right )^{2-d/2}\left (\frac{\overline{Q}}{\sqrt{\xt^2 + \omega\yt^2}}\right )^{d/2-1}K_{\frac{d}{2}-1}\left (\overline{Q}\sqrt{\xt^2 + \omega\yt^2}\right ).
\end{equation}
For the case $d=4$
\begin{equation}
\label{Imasterrank0and2d4}
\begin{split}
\mathcal{I}^{ik}(\xt,\yt,\overline{Q}^2,\omega)  & = -\frac{1}{(2\pi)^2}\frac{\xt^i\yt^k}{\yt^2}\left (\frac{\overline{Q}}{\sqrt{\xt^2+\omega\yt^2}}\right )K_{1}\left (\overline{Q}\sqrt{\xt^2+\omega\yt^2}\right )\\
\mathcal{I}(\xt,\yt,\overline{Q}^2,\omega)  & =  \frac{1}{(2\pi)^2}\left (\frac{\overline{Q}}{\sqrt{\xt^2+\omega\yt^2}}\right )K_{1}\left (\overline{Q}\sqrt{\xt^2+\omega\yt^2}\right ).
\end{split}
\end{equation}

\section{Wilson line color algebra}
\label{app:color}

For the cross section we need the following $q\bar{q}$ and $q\bar{q}g$ matrix elements with eikonal operator $\hat{S}_E$:
\begin{equation}
\delta_{\alpha\beta}\delta_{\alpha'\beta'}\langle \bar{q}(\ell^+,\xt',h,\alpha')q(\ell'^+,\yt',-h,\beta')\vert1- \hat{S}_E\vert q(p^+,\xt,h,\alpha)\bar{q}(p'^+,\yt,-h,\beta)\rangle
\end{equation}
and
\begin{equation}
t^{a}_{\alpha\beta}t^{b}_{\beta'\alpha'}\langle \bar{q}(\ell^+,\xt',h,\alpha')q(\ell'^+,\yt',-h,\beta')g(w^+,\zt',\sigma',b)\vert1- \hat{S}_E\vert q(p^+,\xt,h,\alpha)\bar{q}(p'^+,\yt,-h,\beta)g(k^+,\zt,\sigma,a)\rangle.
\end{equation}
Using the definition of eikonal scattering operator \eq\nr{eq:eikonaloperatorqqbar} together with the normalization conditions in \eq\nr{eq:normalizationMS} one obtain 
\begin{equation}
\begin{split}
\langle &\bar{q}(\ell^+,\xt',h,\alpha')q(\ell'^+,\yt',-h,\beta')\vert 1- \hat{S}_E\vert q(p^+,\xt,h,\alpha)\bar{q}(p'^+,\yt,-h,\beta)\rangle\\
& = \biggl [\delta_{\alpha'\alpha}\delta_{\beta'\beta}-\sum_{\bar\alpha,\bar{\beta}}[U[A](\xt)]_{\bar{\alpha}\alpha}[U^{\dagger}[A](\yt)]_{\beta\bar{\beta}}  \delta_{\alpha'\bar\alpha}\delta_{\beta'\bar\beta} \biggr ] 4p^+p'^+(2\pi)^2\delta(p^+-\ell^+)\delta(p'^+-\ell'^+)\delta^{(2)}(\xt-\xt')\delta^{(2)}(\yt-\yt').
\end{split}
\end{equation}
On the cross section level this expression is multiplied with $\delta_{\alpha\beta}\delta_{\alpha'\beta'}$, and thus 
\begin{equation}
\label{eq:qqbarmatrixelement}
\begin{split}
\delta_{\alpha\beta}\delta_{\alpha'\beta'}\langle &\bar{q}(\ell^+,\xt',h,\alpha')q(\ell'^+,\yt',-h,\beta')\vert 1-\hat{S}_E\vert q(p^+,\xt,h,\alpha)\bar{q}(p'^+,\yt,-h,\beta)\rangle\\
& = \biggl [\nc - \mathrm{Tr}\left (U[A](\xt)U^{\dagger}[A](\yt)\right )\biggr ] 4p^+p'^+(2\pi)^2\delta(p^+-\ell^+)\delta(p'^+-\ell'^+)\delta^{(2)}(\xt-\xt')\delta^{(2)}(\yt-\yt').
\end{split}
\end{equation}
Similarly, for the $q\bar{q}g$-term one obtain 
\begin{equation}
\begin{split}
\langle & \bar{q}(\ell^+,\xt',h,\alpha')q(\ell'^+,\yt',-h,\beta')g(w^+,\zt',\sigma',b)\vert 1-\hat{S}_E\vert q(p^+,\xt,h,\alpha)\bar{q}(p'^+,\yt,-h,\beta)g(k^+,\zt,\sigma,a)\rangle\\
 = \biggl [\delta_{\alpha'\alpha}&\delta_{\beta'\beta}\delta_{ba}- \sum_{\bar{\alpha},\bar{\beta},c,a} [U[A](\xt)]_{\bar{\alpha}\alpha}[U^{\dagger}[A](\yt)]_{\beta\bar{\beta}}[V[A](\zt)]_{ca}(\delta_{\alpha'\bar{\alpha}}\delta_{\beta'\bar{\beta}}\delta_{bc})\biggr ] \\
&\times 8p^+p'^+k^+(2\pi)^3\delta(p^+-\ell^+)\delta(p'^+-\ell'^+)\delta(k^+-w^+)\delta^{(2)}(\xt-\xt')\delta^{(2)}(\yt-\yt')\delta^{(2)}(\zt-\zt')\delta_{\sigma,\sigma'}.
\end{split}
\end{equation}
On the cross section level this expression is multiplied with $t^{a}_{\alpha\beta}t^{b}_{\beta'\alpha'}$, and thus
\begin{equation}
\label{eq:qqbargmatrixelement}
\begin{split}
t^{a}_{\alpha\beta}t^{b}_{\beta'\alpha'}\langle & \bar{q}(\ell^+,\xt',h,\alpha')q(\ell'^+,\yt',-h,\beta')g(w^+,\zt',\sigma',b)\vert \hat{S}_E\vert q(p^+,\xt,h,\alpha)\bar{q}(p'^+,\yt,-h,\beta)g(k^+,\zt,\sigma,a)\rangle\\
 = \biggl [\nc\cf - &\sum_{b,a}\mathrm{Tr}\left (U[A](\xt)t^{a}U^{\dagger}[A](\yt)t^b\right )[V[A](\zt)]_{ba}\biggr ]\\
&\times 8p^+p'^+k^+(2\pi)^3\delta(p^+-\ell^+)\delta(p'^+-\ell'^+)\delta(k^+-w^+)\delta^{(2)}(\xt-\xt')\delta^{(2)}(\yt-\yt')\delta^{(2)}(\zt-\zt')\delta_{\sigma,\sigma'}.
\end{split}
\end{equation}
Rewriting the adjoint Wilson line as 
\begin{equation}
[V[A](\zt)]_{ba} = 2\mathrm{Tr}\left (U[A](\zt)t^{a}U^{\dagger}[A](\zt)t^{b}\right )
\end{equation}
and applying the Fierz identity 
\begin{equation}
t^{a}_{\alpha\beta}t^{a}_{\bar{\alpha}\bar{\beta}} = \frac{1}{2}\left (\delta_{\alpha\bar{\beta}}\delta_{\beta\bar{\alpha}}- \frac{1}{\nc}\delta_{\alpha\beta}\delta_{\bar{\alpha}\bar{\beta}}\right )
\end{equation}
together with the unitarity condition, $U[A](\zt)U^{\dagger}[A](\zt) = \mathbf{1}_{\nc}$, one finds the expression 
\begin{equation}
\begin{split}
\sum_{b,a}\mathrm{Tr}\left (U[A](\xt)t^{a}U^{\dagger}[A](\yt)t^b\right )[V[A](\zt)]_{ba} = &\frac{1}{2}\biggl [\mathrm{Tr}\left (U[A](\xt)U^{\dagger}[A](\zt)\right )\mathrm{Tr}\left (U[A](\zt)U^{\dagger}[A](\yt) \right )\\
& - \frac{1}{\nc}\mathrm{Tr}\left (U[A](\xt)U^{\dagger}[A](\yt) \right )\biggr ].
\end{split}
\end{equation}

\section{Subtraction procedures}
\label{app:sub}

The polynomial subtraction term in~\cite{Beuf:2017bpd} is taken as proportional to 
\begin{equation}
S_{\rm pol} = \frac{\Gamma(d/2-1)^2}{\pi^{d/2-1}}\int\ud^{d-2}\xt_2 (\xt_{20})^m(\xt_{20}^2)^{1-d/2}\biggl \{\xt_{20}^m(\xt_{20}^2)^{1-d/2} - \xt_{21}^m(\xt_{21}^2)^{1-d/2} \biggr \},
\end{equation}
where the first term corresponds to the desired UV divergence in the limit $\xt_{20}\to\ot$ and the second term is added in order to cancel the IR divergence introduced by the first term. 
We use here the notations of~\cite{Beuf:2017bpd}, which are related to ours by $\xt_0 \to \xt,$ $\xt_1\to \yt$, $\xt_2\to \zt$ and $\xt_{02}\to \rt_{xz}$ etc.
Using $\xt_{21} = \xt_{20}+\xt_{01}$ and $\xt_{21}^2 = \xt_{20}^2 + \xt_{01}^2 + 2\xt_{20}\cdot\xt_{01}$  we have
\begin{equation}
S_{\rm pol} = \frac{\Gamma(d/2-1)^2}{\pi^{d/2-1}}\int\ud^{d-2}\xt_2\biggl \{\xt_{20}^2(\xt_{20}^2)^{2-d} - \xt_{20}^2(\xt_{20}^2)^{1-d/2}(\xt_{21}^2)^{1-d/2} - \xt_{20}^m\xt_{01}^m(\xt_{20}^2)^{1-d/2}(\xt_{21}^2)^{1-d/2}  \biggr \}
\end{equation}
and performing the Feynman parametrization one finds the result
\begin{equation}
S_{\rm pol} = -(\xt_{01}^2)^{2-d/2}\Gamma(d/2-2).
\end{equation}
Note that this result is valid when $\varepsilon < 0$. Thus one has to analytically continue this to $\varepsilon > 0$ and the result is 
\begin{equation}
S_{\rm pol} = +\left (\frac{1}{\varepsilon} + \gamma_E  + \log(\xt_{01}^2) \right ). 
\end{equation}
Our subtraction term uses the integral
\begin{equation}
S_{\rm exp} = \frac{\Gamma(d/2-1)^2}{\pi^{d/2-1}} \int\ud^{d-2}\rt_{zx}(\rt_{zx}^2)^{3-d}e^{-\rt_{zx}^2/(\rt_{xy}^2\xi)},
\end{equation}
which has the same divergent behavior in the limit $\rt_{zx}\to \ot$, but moderated by an exponential function so that there is no IR divergence.  The constant  $\xi$ is taken as $\xi = e^{\gamma_E}$. This gives 
\begin{equation}
S_{\rm exp} = (\rt_{xy}^2\xi)^{2-d/2}\Gamma(d/2-1)\Gamma(2-d/2).
\end{equation}
This result is valid when $\varepsilon >0$ and we get 
\begin{equation}
S_{\rm exp} = +\left (\frac{1}{\varepsilon} + \gamma_E  + \log(\xt_{01}^2) \right ). 
\end{equation}

It can be illustrative to go to $d=4$ dimensions and perform the angular integral. Doing this one gets
\begin{eqnarray}\label{eq:gsubind4}
S_{\rm pol} &=& 2 \int \frac{\ud |\xt_{20}|}{|\xt_{20}|} \theta\left(|\xt_{10}|-|\xt_{20}|\right)
 \\
S_{\rm exp} &=& 2 \int \frac{\ud |\rt_{zx}|}{|\rt_{zx}|}e^{-\rt_{zx}^2/(\rt_{xy}^2\xi)}.
\end{eqnarray}
This shows that indeed both functions subtract the same UV divergence in the small daughter dipole limit, but at larger values of $|\xt_{20}|=|\rt_{zx}|$ the behavior is different. Although both choices lead to a perfectly finite final result, we believe that the discontinuous theta function in~\nr{eq:gsubind4} can be somewhat inconvenient from a numerical point of view in the multidimensional numerical integration required to evaluate the cross section in practice.

\section{Derivation of UV-finite terms for $\sigma^{\gamma^{\ast}_{\rm T}}$}
\label{app:finite}

Here we present the detailed computation of individual UV-finite contributions to the transverse virtual photon cross section appearing in \eq\nr{eq:Tfinitefromapp}.

The full cross term given in \eq\nr{eq:thetaT} is divided into three parts: The conribution coming from the instantaneus diagrams simply gives
\begin{equation}
2\Re e\biggl [\ref{diag:qgqbarinst}\ref{diag:qqbarginst}^{\ast} \biggr ] = 0.
\end{equation}
The interference terms between the radiative diagrams \ref{diag:qgqbarT}, \ref{diag:qqbargT}  and instantaneous diagrams \ref{diag:qgqbarinst}, \ref{diag:qqbarginst} simplifies to
\begin{equation}
\label{eq:cross1final}
\begin{split}
2\Re e\biggl [\left (\ref{diag:qgqbarT} + \ref{diag:qqbargT}\right )\left (\ref{diag:qgqbarinst} - \ref{diag:qqbarginst} \right )^{\ast} \biggr ] = \frac{2}{4(2\pi)^4}\frac{Q^2}{\Rt^2}&[K_1(Q\vert \Rt\vert)]^2\biggl [\frac{z_1^2z_2z_3^3}{z_1+z_2}\frac{\rt_{yxz}\cdot \rt_{zx}}{\rt_{zx}^2} +  \frac{z_1z_2z_3^2(z_1+z_2)^2}{z_2+z_3}\frac{\rt_{yxz}\cdot \rt_{zx}}{\rt_{zx}^2}\\
& + \frac{z_2z_3z_1^2(z_2+z_3)^2}{z_1+z_2}\frac{\rt_{xyz}\cdot \rt_{zy}}{\rt_{zy}^2} + \frac{z_2z_3^2z_1^3}{z_2+z_3}\frac{\rt_{xyz}\cdot \rt_{zy}}{\rt_{zy}^2}  \biggr ],
\end{split}
\end{equation}
where 
\begin{equation}
\begin{split}
\rt_{yxz}\cdot \rt_{zx} = \rt_{zx}^2\left (\frac{z_1}{z_1+z_2} \right ) - \rt_{zx}\cdot \rt_{zy}\\
\rt_{xyz}\cdot \rt_{zy} = \rt_{zy}^2\left (\frac{z_3}{z_2+z_3} \right ) - \rt_{zx}\cdot \rt_{zy}.
\end{split}
\end{equation}
Thanks to the above identities, \eq\nr{eq:cross1final} can be further simplified  to 
\begin{equation}
\label{eq:cross1final2}
\begin{split}
2\Re e\biggl [\left (\ref{diag:qgqbarT} + \ref{diag:qqbargT}\right )\left (\ref{diag:qgqbarinst} - \ref{diag:qqbarginst} \right )^{\ast} \biggr ] = \frac{2}{4(2\pi)^4}\frac{Q^2}{\Rt^2}&[K_1(Q\vert \Rt\vert)]^2 \frac{z_1z_2z_3}{(z_1+z_2)(z_2+z_3)}   \biggl [ \left (\frac{z_1}{z_1 + z_2}\right )A      \\
& +  \left (\frac{z_3}{z_2 + z_3}\right )B - A\frac{(\rt_{zx}\cdot \rt_{zy})}{\rt_{zx}^2} - B\frac{(\rt_{zx}\cdot \rt_{zy})}{\rt_{zy}^2}   \biggr ]
\end{split}
\end{equation}
with the coefficients 
\begin{equation}
\begin{split}
A & = z_3\biggl \{(z_1+z_2)^3 + z_1z_3(z_2 + z_3) \biggr \}\\
B & = z_1\biggl \{(z_2+z_3)^3 + z_1z_3(z_1 + z_2) \biggr \}. \\
\end{split}
\end{equation}
Finally, the interference term between radiative diagrams \ref{diag:qgqbarT} and \ref{diag:qqbargT} can be cast in the following form 
\begin{equation}
\label{eq:cross2final}
\begin{split}
2\Re e\biggl [\ref{diag:qgqbarT} \ref{diag:qqbargT}^{\ast} \biggr ] = \frac{2}{4(2\pi)^4}&\frac{Q^2}{\Rt^2}[K_1(Q\vert \Rt\vert)]^2\frac{z_1z_3}{(z_1+z_2)(z_2+z_3)}\biggl [z_2^2(z_1-z_3)^2\\
& - \biggl [z_1(z_1+z_2) + z_3(z_2 + z_3) \biggr ]\biggl [z_1(z_2+z_3) + z_3(z_1 + z_2)\biggr ]\frac{\Rt^2(\rt_{zx}\cdot \rt_{zy})}{\rt_{zx}^2\rt_{zy}^2}\\
& + 2z_1z_2z_3\biggl [(z_1+z_2)^2 + (z_2 + z_3)^2 \biggr ]\frac{(\rt_{zx}\cdot \rt_{zy})^2}{\rt_{zx}^2\rt_{zy}^2} \biggr ],
\end{split}
\end{equation}
where we have used the identity
\begin{equation}
\rt_{xyz}\cdot \rt_{yxz}  = -\frac{\Rt^2}{(z_1+z_2)(z_2 + z_3)}  + \frac{z_2}{(z_1+z_2)(z_2+z_3)}(\rt_{zx}\cdot \rt_{zy}).
\end{equation}
The term proportional to $(\rt_{zx}\cdot \rt_{zy})^2$ can be further simplified by noticing that 
\begin{equation}
\rt_{zx}\cdot \rt_{zy} = \frac{1}{2}\biggl [\left (\frac{z_2 + z_3}{z_3}\right )\rt_{zx}^2 + \left (\frac{z_1+z_2}{z_2}\right )\rt_{zy}^2 - \frac{\Rt^2}{z_1z_3} \biggr ].
\end{equation}
Straightforward algebra leads to
\begin{equation}
\label{eq:cross2final2}
\begin{split}
2\Re e\biggl [\ref{diag:qgqbarT} \ref{diag:qqbargT}^{\ast} \biggr ] = \frac{2}{4(2\pi)^4}&\frac{Q^2}{\Rt^2}[K_1(Q\vert \Rt\vert)]^2\frac{z_1z_2z_3}{(z_1+z_2)(z_2+z_3)}\biggl [- C\frac{\Rt^2(\rt_{zx}\cdot \rt_{zy})}{\rt_{zx}^2\rt_{zy}^2}\\
& + D\biggl [\left (\frac{z_1+z_2}{z_1} \right )\frac{(\rt_{zx}\cdot \rt_{zy})}{\rt_{zx}^2} + \left (\frac{z_2+z_3}{z_3} \right )\frac{(\rt_{zx}\cdot \rt_{zy})}{\rt_{zy}^2} \biggr ]  +  E\biggr ],
\end{split}
\end{equation}
where we have defined the coefficients
\begin{equation}
\begin{split}
C & = \frac{(z_1+z_2)(z_2+z_3)}{z_2}\biggl \{\biggl [(1-z_1)^2 + z_1^2 \biggr ] + \biggl [(1-z_3)^2 + z_3^2 \biggr ] \biggr \}\\
D & = z_1z_3\biggl \{ (z_1+z_2)^2 + (z_2 + z_3)^2 \biggr \}\\
E &= z_2(z_1-z_3)^2.
\end{split}
\end{equation}
Summing the contributions in \nr{eq:cross1final2} and \nr{eq:cross2final2} togehter we find for equation 
\begin{equation}
\label{eq:fullcross12final}
\begin{split}
\nr{eq:thetaT}  = \frac{2Q^2}{4(2\pi)^4}&\frac{z_1z_2z_3}{(z_1+z_2)(z_2+z_3)}\int_{\zt} \frac{[K_1(Q\vert \Rt\vert)]^2}{\Rt^2}\biggl [- C\frac{\Rt^2(\rt_{zx}\cdot \rt_{zy})}{\rt_{zx}^2\rt_{zy}^2} + \frac{(\rt_{zx}\cdot \rt_{zy})}{\rt_{zx}^2}\biggl \{D\left (\frac{z_1+z_2}{z_1} \right ) - A\biggr \}  \\
& + \frac{(\rt_{zx}\cdot \rt_{zy})}{\rt_{zy}^2}\biggr \{D\left (\frac{z_2+z_3}{z_3} \right ) -B\biggr \}  + \left (\frac{z_1}{z_1 + z_2}\right )A   +  \left (\frac{z_3}{z_2 + z_3}\right )B + E\biggr ]\left (1-\mathcal{S}_{xyz}\right ).
\end{split}
\end{equation}
The sum of contributions coming from the two instantaneus diagrams squared \eqs\nr{eq:diag:qgqbarinst:sqr} and \eqs\nr{eq:diag:qqbarginst:sqr} can be simplified to the following form
\begin{equation}
\label{eq:landmfinal}
\overline{\Theta}_{\ref{diag:qgqbarinst}} + \overline{\Theta}_{\ref{diag:qqbarginst}} = \frac{2Q^2}{4(2\pi)^4}\frac{z_1z_2z_3}{(z_1+z_2)(z_2+z_3)}H\int_{\zt} \frac{[K_1(Q\vert \Rt\vert)]^2}{\Rt^2}\left (1-\mathcal{S}_{xyz}\right ),
\end{equation}
where 
\begin{equation}
H = \frac{z_1z_2z_3}{2}\biggl \{\frac{(z_2+z_3)}{(z_1+z_2)} + \frac{(z_1+z_2)}{(z_2+z_3)} \biggr \}.
\end{equation}

Finally, combining the contributions in \eqs\nr{eq:fullcross12final} and \nr{eq:landmfinal} give the result shown in \eq\nr{eq:Tfinitefromapp}.

\bibliography{spires}
\bibliographystyle{JHEP-2modlong}

\end{document}